\documentclass[a4paper,fleqn]{cas-dc}
\usepackage[numbers]{natbib}
\usepackage[utf8]{inputenc}
\usepackage{multirow}
\usepackage{xcolor}
\usepackage{lipsum}
\usepackage{lineno}
 
\usepackage[normalem]{ulem}
\usepackage{cancel}
\usepackage{colortbl}

\usepackage{placeins}

\begin{document}
\let\WriteBookmarks\relax
\def\floatpagepagefraction{1}
\def\textpagefraction{.001}
\shorttitle{Diffusion Models for HSI: A Review}
\shortauthors{Xing Hu et~al.}

\title [mode = title]{Diffusion Models for Hyperspectral Image Analysis: A Comprehensive
Review}  

\tnotetext[label1]{This manuscript is original, has not been published elsewhere, and is not under review by another journal. There are no conflicts of interest to declare.}

\author[1]{Xing Hu}[
    orcid=0000-0003-1930-0372 
]
\cormark[1]
\ead{huxing@usst.edu.cn}
\affiliation[1]{
    organization={School of Optical-Electrical and Computer Engineering, University of Shanghai for Science and Technology},
    addressline={No. 516, Jungong Road},
    city={Shanghai},
    postcode={200093},
    country={China}
}

\author[1]{Xiangcheng Liu}
\ead{242250440@st.usst.edu.cn}


\author[2]{Qianqian Duan}
\ead{dqq1019@163.com}
\affiliation[2]{
    organization={School of Electronics and Electrical Engineering, Shanghai University of Engineering Science},
    city={Shanghai},
    postcode={201620},
    country={China}
}

\author[3]{Lian Zhang}
\ead{dqq1019@163.com}
\affiliation[3]{
    organization={Medical Artificial Intelligence Lab, The First Hospital of Hebei Medical University, Hebei Medical University},
    city={Shijiazhuang},
    postcode={050000},
    country={China}
}

\author[4]{Huiliang Shang}
\ead{shanghl@fudan.edu.cn}
\affiliation[4]{
    organization={Hangzhou Institute of Technology, xidian University},
    city={Hangzhou},
    postcode={311231},
    country={China}
}

\author[4]{Linhua Jiang}
\ead{jianglinhua@xidian.edu.cn}

\author[1]{Haima Yang}
\ead{snowyhm@sina.com}

\author[1]{Dawei Zhang}
\ead{dwzhang@usst.edu.cn}

\cortext[cor1]{Corresponding author.}

\begin{abstract}
Hyperspectral image (HSI) analysis plays a critical role in remote sensing, agriculture, and environmental monitoring. However, traditional methods often struggle to handle the high dimensionality, spectral redundancy, and noise inherent in HSI data, limiting their accuracy and scalability. Recently, diffusion models—including denoising diffusion probabilistic models and other generative frameworks based on stochastic differential equations—have shown strong potential in capturing complex spectral-spatial structures and generating high-fidelity HSI data. These models offer effective solutions for tasks such as noise supression, data augmentation, classification, and anomaly detection.
This review presents a systematic summary of recent advances in diffusion models for HSI processing. We categorize existing methods, highlight their strengths in handling high-dimensional data, and compare their performance with conventional approaches. Special attention is given to critical applications such as change detection and post-disaster anomaly identification. The review also discusses current limitations, such as computational cost and training stability, and outlines potential research directions.
Our main contributions can be summarized as follows: we provide a systematic taxonomy of diffusion-based HSI methods, examine their applications across major remote sensing tasks, and offer perspectives on potential directions for future research. With these efforts, this review seeks to support the community in harnessing deep learning models to achieve more effective and efficient hyperspectral image analysis.

\end{abstract}


\maketitle

\section{Introduction}
Hyperspectral images (HSIs), which record fine-grained spectral signatures over hundreds of narrow wavelength intervals, play a crucial role in domains like remote sensing, precision agriculture, and environmental monitoring. With continuous progress in sensor technology, both spatial and spectral resolutions have improved, enabling more precise material identification. Nevertheless, the extremely high dimensionality, massive data size, and substantial noise in HSIs create significant challenges for conventional image analysis methods, which often find it difficult to effectively handle hundreds of correlated spectral bands and mixed pixels. 

Neural networks have achieved significant progress across a variety of data modalities, such as images \cite{ref1,ref2}, audio \cite{ref3,ref4}, text \cite{ref5}, and point clouds \cite{ref6}. Within the domain of image generation, deep learning architectures like CNNs \cite{ref180} and Transformers \cite{ref182} have shown strong effectiveness, and extensive research has also been devoted to GANs \cite{ref8}, VAEs \cite{ref9}, and normalizing flows \cite{ref10}. 
Yet, they face issues: GANs are difficult to train, VAEs yield blurry outputs, and flow models are architecturally complex. In contrast, diffusion models, which gradually transform noise into data through a series of denoising steps, offer stable training and high output fidelity without adversarial setups. Originating with Sohl-Dickstein et al. \cite{ref18}, diffusion models gained momentum following advancements by Ho et al. \cite{ref17} and Song et al. \cite{ref24}, which introduced practical frameworks for score-based and probabilistic modeling.

In recent years, diffusion models have shown great promise in the field of hyperspectral data, and there is a significant demand in actual scientific research for high-fidelity generation and precise analysis of hyperspectral images. diffusion models denoising methods are a good fit for HSI restoration needs, and their ability to model complex distributions makes spectral–spatial learning better. Recent research shows that diffusion models work better than older methods for tasks like denoising and data synthesis. Traditional generative models (such as GANs and VAEs) often struggle with training instability or loss of detail when handling the high-dimensional spectral features of HSI. Although diffusion models have achieved breakthroughs in the field of general computer vision, there is currently a lack of systematic research elucidating how these models adapt to the inherent properties of HSI, such as its strong spectral correlation and low spatial resolution. This paper fills that gap by giving a full picture of how diffusion-based HSI processing works.

Compared to previous studies (e.g., \cite{ref179}), this research provides a comprehensive overview of hyperspectral patterns, offering a detailed analysis focused on HSI. Although \cite{ref179} addresses early conditional and latent diffusion models \cite{ref11,ref42}, it fails to investigate developments or technical challenges pertaining to hyperspectral data that have arisen since 2023. On the contrary, we have conducted the first in-depth analysis of the latest technological advancements in diffusion models for hyperspectral imaging since 2023. Rather than limiting ourselves to general model classifications, we have developed a classification system rooted in the unique challenges of the HSI field, classifying models into Denoising Diffusion Probabilistic Models (DDPMs) \cite{ref17}, Score-Based Generative Models (SGMs) \cite{ref24}, and SDE-based models \cite{ref29}. We examine their functions in activities such as data generation, enhancement, classification \cite{ref15}, segmentation, anomaly detection \cite{ref16}, noise suppression, and recovery \cite{ref13}.

This review makes (a number of/ several) important contributions. First, it provides researchers with a comprehensive roadmap spanning from fundamental theory to cutting-edge applications, filling a gap in diffusion models for the HIS field. More importantly, through a comprehensive evaluation of quantitative experimental results, it demonstrates the immense potential of diffusion models in enhancing the performance of HIS classification, denoising, and anomaly detection. Second, it looks at how diffusion models change to fit the special structure of HSIs, which can be hard because of things like limited annotations and spectral correlations. Third, it talks about the limitations of engineering, such as the costs of computation and the difficulty of understanding models, and it talks about new solutions that are being developed. In the end, we want to connect basic research with real-world uses so that these models can be used in real-world systems in the future.

It is worth noting that while emerging architectures like Vision Transformers (ViTs)and State Space Models (e.g., Mamba) excel in discriminative feature extraction, the absolute state-of-the-art in generative hyperspectral image (HSI) analysis is currently framed by advanced diffusion models. Specifically, the field has rapidly evolved from basic pixel-space DDPMs to computationally efficient Latent Diffusion Model and physics-informed conditional diffusion frameworks. By explicitly integrating HSI domain priors—such as spectral unmixing constraints and radiometric consistency—into the reverse diffusion process, these SOTA models effectively overcome the curse of dimensionality, representing the cutting edge of current HSI research.

The rest of the paper is structured as follows. HSI fundamentals are introduced in Section 2. The theory and development of diffusion models is covered in detail in Section 3. Their use in HSI tasks is reviewed in Section 4. The limitations and practical difficulties are covered in Section 5. Diffusion models and traditional approaches are contrasted in Section 6. Future directions for research are outlined in Section 7. The paper's conclusion is found in Section 8.

\section{Basics of hyperspectral imagery}
\subsection{Definition and Characteristics of Hyperspectral Imagery}
\begin{figure}[pos=htbp]
\centering
\includegraphics[width=3.3in]{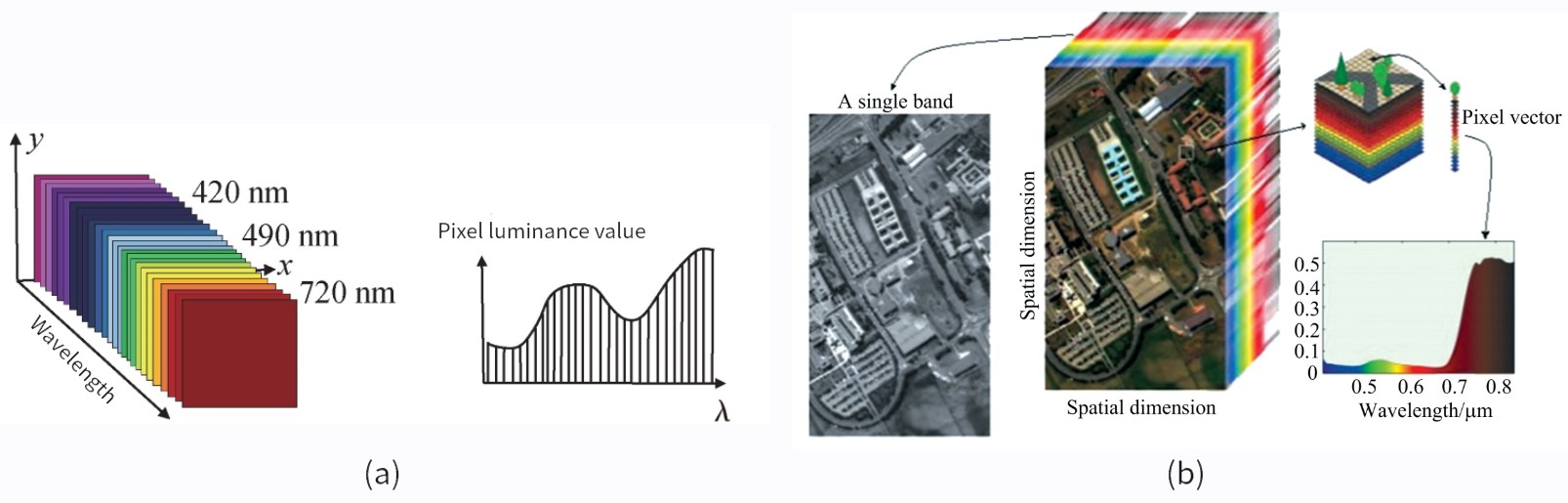}%
\hfil
\caption{Hyperspectral image (a) and schematic illustration of a hyperspectral image (b)}
\label{image1}
\end{figure}

A hyperspectral image (HSI) consists of many continuous spectral bands, usually ranging from dozens to several hundreds, with spectral resolution often at the nanometer scale, as illustrated in Figure~\ref{image1}. Each pixel of an HSI is not just a single value but a full spectrum, describing the intensity of reflected light across a wide range of wavelengths. These wavelengths typically extend from the visible region through the near-infrared and sometimes into the mid-infrared. Data of this type are collected by imaging spectrometers mounted on satellites, aircraft, or ground-based platforms. By capturing scenes across numerous narrow wavelength intervals, the instrument produces a three-dimensional data cube: two spatial dimensions (rows and columns) and one spectral dimension (wavelength). The defining characteristic of HSIs is that every pixel contains a spectral signature—a continuous reflectance curve—that reveals detailed information about the material composition and physical properties of the corresponding ground object. Unlike conventional RGB or multispectral imagery, hyperspectral data provide both spatial and spectral information simultaneously. 

The advantage of hyperspectral imagery (HSI) lies in its ability to combine spatial and spectral information, made possible by the progress of imaging spectroscopy technology. Spatial data describe outward features of objects, including their shape, size, and texture. Spectral data, in contrast, capture internal properties such as chemical composition and physical structure through reflectance or emission characteristics. When these two dimensions are used together, HSIs can achieve far more accurate detection and classification than relying on either one alone. For instance, objects that look the same in terms of color or texture can be separated by their spectral signatures, while materials with nearly identical spectra may still be distinguished by spatial patterns and morphology. 

A further strength of HSIs is their exceptionally high spectral resolution, often as fine as 5–10 nm per band or even smaller. This capability makes it possible to tell apart substances that cannot be distinguished in broadband imagery. Even small variations in reflectance spectra can reveal differences between vegetation species or indicate the presence of particular minerals through their absorption features. However, this abundance of spectral detail, while highly beneficial for analysis, also leads to greater data complexity and brings about challenges that will be addressed in the following sections.

\subsection{High-dimensional characterization of hyperspectral image data}
The abundance of spectral details in HSIs comes with the drawback of very high dimensionality. 
A single hyperspectral image can contain hundreds of bands, resulting in a very large feature space for each pixel. This high-dimensional data demands significant computational and storage resources: processing fullresolution HSI data can be computationally intensive and memory-consuming . The phenomenon known as the “curse of dimensionality” implies that as dimensionality grows, the volume of the feature space increases so much that data become sparse and traditional statistical analyses become less reliable. For HSIs, this can lead to model overfitting and degraded performance if not handled properly, especially when the number of training samples is limited relative to the number of spectral features.

Moreover, HSIs often exhibit strong inter-band correlation and redundancy – many adjacent spectral bands carry overlapping information. While this redundancy can be useful for noise reduction (since true signal is replicated across bands), it also means there is inherent data inefficiency. Traditional machine learning methods struggle in such high-dimensional spaces; for example, classifiers may require exponentially more samples to generalize well, and they become more susceptible to noise interference which is exacerbated in high dimensions . Even modern deep learning approaches, which can automatically learn compact features, typically require a large number of labeled examples to train effectively. Obtaining ground-truth labels for HSIs is labor-intensive and costly, because labeling often involves fieldwork or expert interpretation for each pixel or region.

Reducing the number of dimensions is one way to fix these issues. Principal Component Analysis (PCA) and Linear Discriminant Analysis (LDA) are two methods that people often use to project high-dimensional HSI data into a lower-dimensional subspace while keeping most of the important information. This can help with the curse of dimensionality and save a lot of computing power. It can also improve the signal-to-noise ratio by getting rid of parts that are too loud or not needed. However, it is important to keep the spectral features that are necessary for a certain task and not lose any information that could hurt performance. To sum up, the fact that HSIs have a lot of dimensions is both good and bad. It gives you a lot of information that lets you do advanced analysis, but it also means that you have to be careful when you process it and use specialized algorithms to deal with the complexity quickly.

\subsection{Main tasks in hyperspectral image analysis}
HSI analysis has a number of important steps that help you get useful information from the data. The main tasks for interpreting hyperspectral data are classification, segmentation, target detection, and anomaly detection.

Classification: When you classify hyperspectral images, you group each pixel (or area) into a certain class or category based on its spectral signature. For land cover classification, there could be groups of plants, water, urban materials, soil, and so on. HSI classification is widely applied in agriculture (e.g. crop type mapping and pest infestation detection), environmental monitoring (e.g. land use/land cover mapping, forest cover analysis), and mineral exploration (identifying surface mineral compositions) . This task is crucial because it provides a semantic labeling of the image, enabling users to interpret large volumes of spectral data in terms of meaningful categories. However, HSI classification faces several challenges: the high dimensionality of spectral features increases model complexity; obtaining sufficient labeled training data is costly and time-consuming; and the mixed pixel problem is common (a single pixel may contain a mixture of materials due to limited spatial resolution, blurring class boundaries) . Traditional classifiers for HSI include statistical methods like maximum likelihood estimation and machine learning methods such as Support Vector Machines (SVMs) . In recent years, deep learning techniques (e.g. convolutional neural networks, CNNs, and recurrent neural networks) have shown superior performance by automatically learning spectral-spatial feature representations . Nonetheless, improving classification accuracy in HSIs remains challenging, especially when available labels are scarce.

Segmentation: Hyperspectral image segmentation involves dividing an image into regions of spectrally similar pixels, grouping together pixels belonging to the same material or object. Segmentation can be done at the pixel or object level. It is often the first step in identifying objects or drawing boundaries in fields like medical imaging (for example, segmenting tumor regions), urban planning (for example, finding buildings and roads), or agriculture (for example, drawing crop fields). The goal is to simplify image analysis by reducing complexity: after segmentation, one can treat each homogeneous region as a unit. Hyperspectral segmentation is complicated by many of the same factors as classification: extremely high data dimensionality, which makes the grouping problem computationally intense; noise, which can cause false distinctions or merges of regions; and mixed pixels at region boundaries . Traditional segmentation approaches include clustering methods (like K-means or hierarchical clustering on spectral features) and edge-detection or region-growing methods that incorporate spatial continuity . More recently, deep learning methods such as fully convolutional networks (FCNs) and U-Net architectures have been adapted to HSI segmentation, as they can learn to combine spectral and spatial information to produce detailed segmentation maps . However, like classification, these often require large labeled datasets for training, which are not always available for HSIs.

Target detection, which is also known as object detection in a broader sense, is the process of finding and identifying pixels or areas in hyperspectral images that correspond to a specific material or object of interest. Target detection is different from anomaly detection because it assumes you already know the target's spectral signature and wants to find its occurrences in the image. Finding certain mineral deposits, military camouflaged targets, or certain types of plants or invasive plants in an ecosystem are all common examples. Most of the time, target detection algorithms work by comparing the spectral signature of each pixel to the known signature of the target. The matched filter, the constrained energy minimization detector, and subspace projection methods are all examples of classic methods. The problem is that variations in noise and background spectra can obscure targets, especially when the target occupies only a few pixels or blends with other elements in the background. High spectral resolution aids in locating targets, but effective algorithms must also eliminate background “noise” and account for variations in illumination. Statistical models of targets and backgrounds, or machine learning approaches that distinguish target pixels from false alarms, can be employed in robust target detection methods.

Anomaly Detection: The goal of anomaly detection in HSIs is to find pixels that are spectrally different from their surroundings or from the expected background signature, even if you don't know what the "anomalies" are ahead of time. This is especially helpful for finding things in an image that you didn't expect to see, like a chemical spill in an environmental image or strange mineral outcrops in a geological survey. The RX detector (Reed–Xiaoli) is the most common classical method. It assumes that the background follows a multivariate normal distribution and marks pixels with reflectance vectors that are very different from this background model. Finding anomalies is hard because HSIs often have very complicated and changing backgrounds, and real anomalies (like small objects or rare materials) can be hard to spot and easily mistaken for background noise or variability. Also, detection algorithms must be unsupervised or semisupervised because there are usually very few (if any) labeled examples of anomalies. Statistical methods are often used in effective approaches to model background spectra and find outliers. For example, one could use dimensionality reduction to define the subspace of normal background spectra and then find pixels that are outside of that subspace. It is hard to evaluate anomalies because there isn't much ground truth, but a good anomaly detector can be very useful in situations like surveillance, where you need to catch anything strange in the scene.

These primary tasks form the basis of HSI analysis. They are interrelated – for example, good segmentation can aid classification by delineating homogeneous regions, and target detection can be seen as classification with emphasis on a particular class. In this review, we will later examine how diffusion models improve or tackle each of these tasks. Before that, we outline the fundamentals of diffusion models to ground their use in hyperspectral image analysis. 

\section{Diffusion Model}
Diffusion models are a class of generative methods that learn data distributions by iteratively adding and then removing noise. The forward (noising) process gradually perturbs an input (e.g., an image) over multiple steps until its structure is destroyed and only noise remains. 
After many steps, the original structure is completely lost, and the data become indistinguishable from pure noise. 
The model then learns a reverse process that starts from noise and attempts to reconstruct the data by iteratively removing the noise. Through this two-process formulation, diffusion models bridge generative modeling and ideas from non-equilibrium thermodynamics and stochastic processes . In this section, we outline the theoretical underpinnings of diffusion models and describe major variants that have been developed.

\subsection{Theoretical Foundations of the Diffusion Model}
The foundation of diffusion models can be understood by examining the forward and reverse processes in more detail. We denote an original data sample (e.g. an image) as $x_0$. The forward diffusion process produces a sequence of increasingly noisy samples $x_1, x_2, ..., x_T$ by gradually adding noise at each step according to a predefined schedule. In the simplest case (as in the Denoising Diffusion Probabilistic Model), Gaussian noise with increasing variance is added at each step. After $T$ steps, $x_T$ is a nearly noise-like sample drawn from a simple prior distribution (often a standard Gaussian). The reverse process is defined as a parametric model (typically a neural network) that starts from random noise $x_T \sim \mathcal{N}(0,I)$ and aims to reconstruct $x_0$ by reversing the forward process step by step. Essentially, the model tries to predict and remove the noise at each step $t$, producing a slightly less noisy $x_{t-1}$ from $x_t$, until it arrives at a clean sample $\hat{x}_0$ at $t=0$. The training process aims to adjust the model parameters so that the reconstructed sample $\hat{x}_0$ is as close as possible to the true data $x_0$. Equivalently, the goal is to align the distribution of $\hat{x}_0$ with the data distribution. This is typically achieved through a variational bound or a score-matching loss derived from the formulation of the forward process.

\subsubsection{Denoising Diffusion Probabilistic Model (DDPM)} 
\begin{figure}[pos=htbp]
\centering
\includegraphics[width=0.45\textwidth]{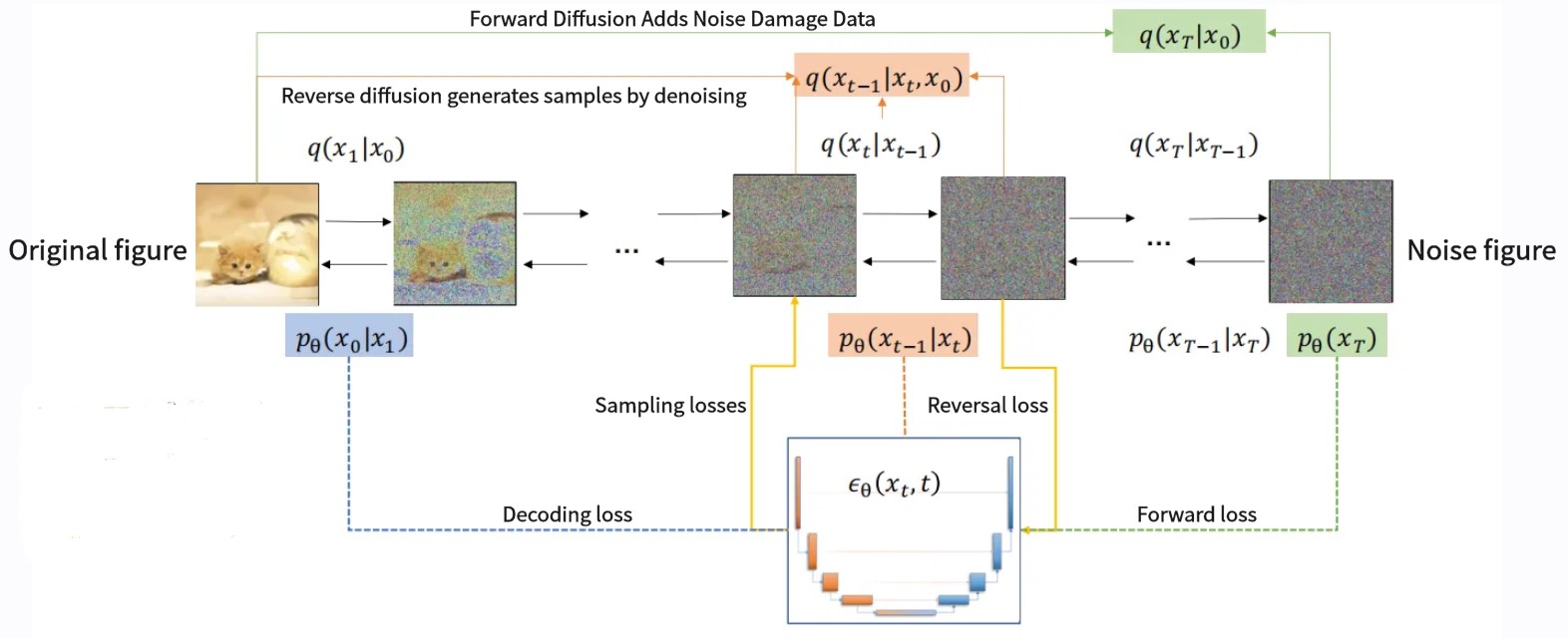}%
\hfil
\caption{Diffusion processes in diffusion models \cite{ref17}}
\label{image2}
\end{figure}

The denoising diffusion probabilistic model (DDPM) \cite{ref17,ref18} is built on two Markov chains, as shown in Figure \ref{image2}. In the forward chain, noise is gradually added to the data until it becomes a simple prior distribution, most often chosen as a standard Gaussian. 
Conversely, the backward chain learns to reverse this process by employing a transformation kernel established by a deep neural network, which converts the noise back into the data.

To create a new data point, first sample a vector randomly\cite{ref190,ref191} from the prior distribution, then resample using an inverse Markov chain\cite{ref19}.

Formally, given a data distribution $\mathbf{x}_0\sim q(\mathbf{x}_0)$, a forward Markov process generates a sequence of random variables, $\mathbf{x}_1,\mathbf{x}_2,\ldots,\mathbf{x}_T$, through a transition kernel $q(x_t\mid{x_{t-1}})$. By applying the probability chain rule and the Markov property, the joint distribution of these variables, conditional on $\mathbf{x}_0$, is then decomposed and denoted as $q(x_1,\ldots,x_T\mid{x_0})$, which results in:

\begin{equation}
\label{deqn_ex1a1}
q(\mathbf{x}_1,...,\mathbf{x}_T\mid\mathbf{x}_0)=\prod_{t=1}^Tq(\mathbf{x}_t\mid\mathbf{x}_{t-1}).
\end{equation}

In DDPMs, the transition kernel $q(x_t\mid{x_{t-1}})$ is designed to gradually convert the original data distribution $q(\mathbf{x}_0)$ into a tractable prior distribution. A typical and effective choice is to apply Gaussian noise, expressed as:

\begin{equation}
\label{deqn_ex1a2}
q(\mathrm{x}_{t}\mid\mathrm{x}_{t-1})=N(\mathrm{x}_{t};\sqrt{1-\beta_{t}}\mathrm{x}_{t-1},\beta_{t}\mathrm{I}),
\end{equation}
where $\beta_t \in (0,1)$ is a hyperparameter fixed before training. Other kernels are possible, but Gaussian noise is often used because it makes the process easier to analyze. As shown by Sohl-Dickstein et al. (2015) \cite{ref18}, this kernel allows the marginal distribution in equation (1) to be written in closed form for any $t \in \{0,1,\cdots,T\}$. By defining $\alpha_t:=1-\beta_t$ and $\bar{\alpha}_{t}{:}=\prod_{s=0}^{t}\alpha_{s}$, we obtain: 

\begin{equation}
\label{deqn_ex1a3}
q(\mathbf{x}_t\mid\mathbf{x}_0)=\mathcal{N}\left(\mathbf{x}_t;\sqrt{\bar{\alpha}_t}\mathbf{x}_0,(1-\bar{\alpha}_t)\mathbf{I}\right).
\end{equation}

Intuitively, the forward process keeps adding noise until the original structure of the data is fully destroyed. To create new samples, the DDPM starts with a noise vector drawn from the prior distribution, which is easy to sample. The model then removes noise step by step using a learned reverse Markov chain. This reverse chain is defined by the prior $p(\mathbf{x}_T)=\mathcal{N}(\mathbf{x}_T;0,\mathbf{I})$ and a learnable transition kernel $p_\theta(\mathbf{x}_{t-1}\mid\mathbf{x}_t)$. The choice $p(\mathbf{x}_T)=\mathcal{N}(\mathbf{x}_T;0,\mathbf{I})$ matches the fact that the forward process makes $q(\mathbf{x}_T)$ close to this distribution. The reverse kernel $p_\theta(\mathbf{x}_{t-1}\mid\mathbf{x}_t)$ is then applied as follows:

\begin{equation}
\label{deqn_ex1a4}
p_{\theta}(\mathbf{x}_{t-1}\mid\mathbf{x}_{t})=\mathcal{N}\left(\mathbf{x}_{t-1};\mu_{\theta}(\mathbf{x}_{t},t),\Sigma_{\theta}(\mathbf{x}_{t},t)\right).
\end{equation}
where $\theta$ stands for the model's parameters, and a deep neural network sets the mean $\mu_\theta(\mathbf{x}_t,t)$ and variance $\Sigma_\theta(\mathbf{x}_t,t)$.

To train the network successfully, the reverse process $p_\theta(\mathbf{x}_0,\mathbf{x}_1,\cdots,\mathbf{x}_T)$ must accurately match the true time reversal of the forward process $q(\mathbf{x}_0,\mathbf{x}_1,\cdots,\mathbf{x}_T)$. This is usually done by making the Kullback-Leibler (KL) divergence between the two as small as possible:

\begin{equation} 
\label{deqn_ex1a5}
D_{\mathrm{KL}}\left(q(\mathbf{x}_{1:T}|\mathbf{x}_{0}) \parallel p_{\theta}(\mathbf{x}_{0:T})\right) 
\end{equation}

\begin{equation}
\label{deqn_ex1a6}
\overset{(\mathrm{i})}{=} -\mathbb{E}_{q(\mathbf{x}_{1:T}\mid\mathbf{x}_{0})}\left[ \log p_{\theta}(\mathbf{x}_{0:T}) \right] + \mathrm{const}
\end{equation}

\begin{equation} 
\label{deqn_ex1a7}
\begin{split} \overset{(\mathrm{ii})}{=}& \mathbb{E}_{q(\mathbf{x}_{1:T}\mid\mathbf{x}_{0})}\left[ -\log p(\mathbf{x}_{T}) - \sum_{t=1}^{T} \log \frac{p_{\theta}(\mathbf{x}_{t-1}\mid\mathbf{x}_{t})}{q(\mathbf{x}_{t}\mid\mathbf{x}_{t-1})} \right]  &+ \mathrm{const} \end{split} 
\end{equation}

\begin{equation} 
\label{deqn_ex1a8} 
\overset{(\mathrm{iii})}{\geq} -\log p_{\theta}(\mathbf{x}_{0}) 
\end{equation}

where $\mathrm{const}$ is a constant that strictly depends on the forward process $q$ and does not depend on the model parameters $\theta$. The goal of DDPM training is to maximize the evidence lower bound (ELBO), or equivalently, to minimize the variational lower bound $L_{\mathrm{VLB}}$. Monte Carlo sampling \cite{ref21,ref7} can be used to quickly estimate this, and stochastic optimization \cite{ref22} can be used to improve it.

Finally, two groundbreaking variants have been derived from the DDPM: the Conditional Diffusion Model and the Latent Diffusion Model.

\paragraph{Conditional Diffusion Model (CDM)}
Like GANs, diffusion models first concentrated on unconditional generation and later moved quickly toward conditional generation \cite{ref57}. Unconditional generation is mainly used to test the limits of a model, while conditional generation is more practical since it allows outputs to be guided by human intent, enabling controlled, task-driven data generation and restoration, effectively guiding the diffusion process, this approach addresses issues such as poor adaptability in multimodal fusion and guided object detection.

The idea of adding conditioning to diffusion models was first proposed in \cite{ref11}. 
This pioneering work guided the generation of diffusion models by incorporating classifiers into pre-trained models, leading to the term "guided diffusion models." While this method offered low training costs, it increased inference costs by using classification results to steer the diffusion model's sampling process, and it struggled to precisely control details for optimal image output. Consequently, the Google team \cite{ref58} opted for a simpler approach: retraining the DDPM conditionally to control generated results, a technique they named Classifier-Free Guidance.

Formally, when given conditional information $\mathbf{c}$, the combined DDPM distribution changes from equations (2) and (3). Consequently, both the optimization objective and the sampling process are modified to take the following form:

\begin{equation}
\label{deqn_ex1a18}
\mathcal{L}_{\mathrm{con}}(\theta)=\mathbb{E}_{x_{0},\epsilon,c,t}\left[\left\|\epsilon-\epsilon_{\theta}\left(\sqrt{\bar{\alpha}_{t}}x_{0}+\sqrt{1-\bar{\alpha}_{t}}\epsilon,c,t\right)\right\|^{2}\right].
\end{equation}

\begin{equation}
\label{deqn_ex1a19}
y_{t-1}=\frac{1}{\sqrt{\alpha_{t}}}\left(y_{t}-\frac{1-\alpha_{t}}{\sqrt{1-\tilde{\alpha}_{t}}}\epsilon_{\theta}(y_{t},c,t)\right)+\hat{\beta}_{t}z.
\end{equation}

This approach has seen broader adoption than guided diffusion models and forms the foundation for many significant diffusion models, including DALL-E2 \cite{ref59}, Imagen \cite{ref52}, and Stable Diffusion (SD) \cite{ref42}.

\paragraph{Latent Diffusion Model (LDM)}
Directly manipulating diffusion models in the original image pixel space incurs high computational costs. To address this, Rombach et al. \cite{ref45} proposed training and performing inference with diffusion models within a low-dimensional latent space, terming this approach the Latent Diffusion Model (LDM), overcoming the severe computational bottlenecks and “curse of dimensionality” inherent in HSI data, making the processing of large-scale, full-band hyperspectral data both feasible and efficient.

\begin{figure}[pos=htbp]
\centering
\includegraphics[width=0.4\textwidth]{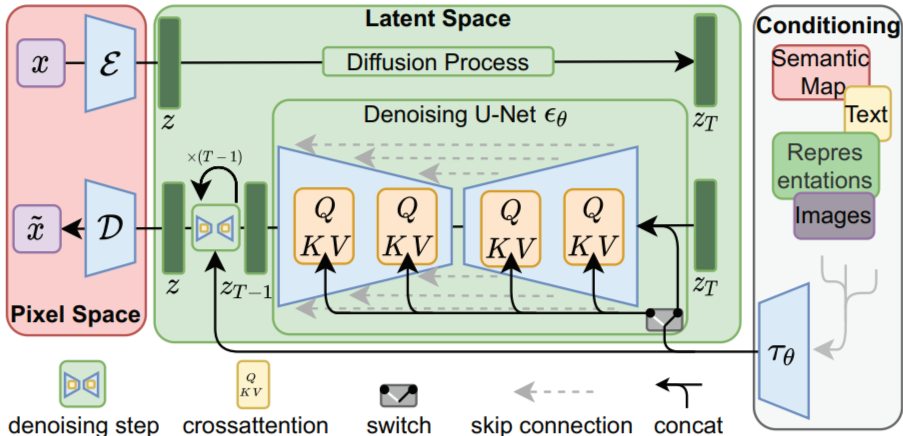}%
\hfil
\caption{Architecture of LDM, where E and D denote the encoder and decoder, respectively \cite{ref42}}
\label{image7}
\end{figure}

Figure~\ref{image7} shows that the Latent Diffusion Model (LDM) has three separate parts. The leftmost gray box shows the LDM's specific diffusion process: it moves its operational domain from pixel space to a low-dimensional latent space by using a pre-trained autoencoder. In this space, noise is gradually added to feature $\mathbf{z}$, resulting in $\mathbf{z_T}$ as part of the forward diffusion. In the middle of the orange box, which is also in latent space, there is a U-Net with cross-attention that can handle inputs from more than one source. This U-Net makes it possible for reverse diffusion to bring $\mathbf{z_T}$ back to $\mathbf{z}$. The conditional encoder, which changes different conditions (like text and images) into feature vectors $\tau_\theta$, is in the box on the right. The U-Net then uses these vectors to help make images. The LDM is widely used in natural scene generation because it can lower the cost of computation without lowering the quality of image synthesis. It also speeds up inference, which gives a lot of ideas for using diffusion models in hyperspectral applications.

\subsubsection{Score-Based Generative Models (SGM/NCSN)}
The fundamental principle of Score-Based Generative Models (SGM/NCSN) \cite{ref23,ref24,ref20} is the (Stein) score, alternatively referred to as a fraction or score function \cite{ref25}. The score function for a probability density function $p(\mathbf{x})$ is the gradient of the logarithmic probability density. $\nabla_\mathbf{x}\log p(\mathbf{x})$. The Fisher score, $\nabla_\theta\log p_\theta(\mathbf{x})$, is a common statistic that depends on the model parameters. The Stein score depends on the data $\mathbf{x}$ itself, not on $\theta$. In simple terms, it's a vector field that shows the direction of the steepest rise in the probability density function.

The central idea behind Score-Based Generative Models (SGMs) \cite{ref24} is to progressively add Gaussian noise to data and then collectively estimate the score function for all these noisy data distributions. This estimation is achieved by training a deep neural network, specifically known as a Noise Conditioned Scoring Network \sout{(NCSN)}. So SGM is also known as NCSN \cite{ref24}, which is conditioned on the noise level. Samples are then generated by linking these score functions across decreasing noise levels. This is done using various score-based sampling methods, including Ronzmann Monte Carlo \cite{ref26,ref27,ref28,ref24,ref29}, stochastic differential equations \cite{ref30,ref29}, ordinary differential equations \cite{ref31,ref32,ref33,ref34,ref29}, and their combinations \cite{ref29}. A key advantage of the SGM formulation is the complete decoupling of training and sampling, which allows for the use of multiple sampling techniques once the score function has been estimated.

Similar to Section III.A.1, let $p(\mathbf{x}_0)$ represent a data distribution, and $0 < \sigma_1 < \sigma_2 < \cdots < \sigma_t < \cdots < \sigma_T$ be a series of noise levels. A typical SGM perturbs data points $\mathbf{x}_0$ to $\mathbf{x}_t$ using a Gaussian noise distribution $q(\mathbf{x}_t\mid\mathbf{x}_0)=\mathcal{N}(\mathbf{x}_t;\mathbf{x}_0,\sigma_t^2I)$. This process generates a sequence of noisy data densities, $q(\mathbf{x}_1),q(\mathbf{x}_2),\cdots,q(\mathbf{x}_T)$, where $q(\mathbf{x}_t):=\int q(\mathbf{x}_t)q(\mathbf{x}_0)d\mathbf{x}_0$. A noise-conditional scoring network, denoted as a deep neural network $\mathbf{s}_\theta(\mathbf{x},t)$, is trained to estimate the score function $\nabla_{x_t}\log\ q(\mathbf{x}_t)$. Various techniques for learning the score function from data, also known as score estimation, have been established, including score matching \cite{ref25,ref98}, denoising score matching \cite{ref35,ref36,ref37}, and sliced score matching \cite{ref38}. It's straightforward to use any of these methods to train a noisy conditional score network from perturbed data points. The training objective is defined by the following equation:

\begin{equation}
\label{deqn_ex1a9}
\mathbb{E}_{t\sim\mathcal{U}[[1,T]],\mathbf{x}_{0}\sim q(\mathbf{x}_{0}),\boldsymbol{\epsilon}\sim\mathcal{N}(0,\mathbf{I})}\left[\lambda(t)\|\epsilon-\epsilon_{\theta}(\mathbf{x}_{t},t)\|^{2}\right].
\end{equation}

\begin{equation}
\label{deqn_ex1a10}
\begin{split}
=\mathbb{E}_{t\sim\mathcal{U}\|1,T\|\mathbf{x}_0\sim q(\mathbf{x}_0),\mathbf{x}_t\sim q(\mathbf{X}_t|\mathbf{X}_0)}\\\left[\lambda(t)\sigma_t^2\|\nabla_{\mathbf{x}_t}\log q\left(\mathbf{x}_t\right)-\mathbf{s}_\theta(\mathbf{x}_t,t)\|^2\right]. 
\end{split}
\end{equation}

\begin{equation}
\label{deqn_ex1a11}
\begin{split}
\overset{(i)}{\operatorname*{=}}\mathbb{E}_{t\sim u[[1,T]]\mathbf{x}_{0}\sim q(\mathbf{x}_{0})\mathbf{x}_{t}\sim q(\mathbf{X}_{t}|\mathbf{X}_{0})}[\lambda(t)\sigma_{t}^{2}\\\prod\nabla_{\mathbf{x}_{t}}\log q(\mathbf{x}_{t}\mid\mathbf{x}_{0})-\mathbf{s}_{\theta}(\mathbf{x}_{t},t)\parallel^{2}]+\mathrm{const}.
\end{split}
\end{equation}

\begin{equation}
\label{deqn_ex1a12}
\begin{split}
\overset{(ii)}{\operatorname*{=}}\mathbb{E}_{t\sim U[1,T],\mathbf{x}_{0}\sim q(\mathbf{x}_{0}),\mathbf{x}_{t}\sim q(\mathbf{X}_{t}|\mathbf{X}_{0})}\\\left[\lambda(t)\left\|-\frac{\mathbf{x}_{t}-\mathbf{x}_{0}}{\sigma_{t}}-\sigma_{t}\mathbf{s}_{\theta}(\mathbf{x}_{t},t)\right\|^{2}\right]+\mathrm{const}.
\end{split}
\end{equation}

\begin{equation}
\label{deqn_ex1a13}\begin{split}
(iii)\\=&\mathbb{E}_{t\sim\mathcal{U}[[1,T]],\mathbf{x}_{0}\sim q(\mathbf{x}_{0}),\epsilon\sim\mathcal{N}(0,\mathbf{I})}[\lambda(t)\|\epsilon+\sigma_{t}\mathbf{s}_{\theta}(\mathbf{x}_{t},t)\|^{2}]\\+\mathrm{const}.\end{split}
\end{equation}

where $\lambda(t)$ represents a positive weighting function, while $\mathbf{x_t}$ is derived from $\varepsilon$ and $\mathbf{x}_{0}$.$\mathcal{U}[[1,T]]$ denotes a uniform distribution over the set $\{1, 2, \ldots, T\}$, and $\varepsilon$ is a deep neural network with parameters $\theta$, tasked with predicting the noise vectors for a given $\mathbf{x}_{t}$ and $t$.Comparing the equations, the training objectives for DDPM and SGM become equivalent once $\epsilon_\theta(\mathbf{x},t)=-\sigma_t\mathbf{s}_\theta(\mathbf{x},t)$ is established.

\subsubsection{Diffusion model based on stochastic differential equations (Score-SDE)}
Both DDPM and SGM can be further generalized to an infinite time step or noise level.In this more advanced framework, referred to as Score SDE \cite{ref29}, both the perturbation and the denoising processes are described through the solution of a stochastic differential equation (SDE).
This approach is employed for both noise perturbation and sample generation, where the denoising process necessitates estimating a score function of the noisy data distribution.

The Stochastic Differential Equation (SDE) was employed to score the data using a diffusion process, which is governed by the following SDE \cite{ref29}:

\begin{equation}
\label{deqn_ex1a14}
\mathrm{dx}=\mathrm{f}(\mathrm{x},t)\mathrm{d}t+g(t)\mathrm{dw}.
\end{equation}
Here, $f(\mathbf{x},t)$ and $g(t)$ denote the drift and diffusion terms of the SDE, respectively, and $w$ refers to a standard Wiener process, also called Brownian motion.
The forward processes in both the DDPM and SGM are essentially discretizations of this SDE. As demonstrated by Song et al. in 2020 \cite{ref29}, the corresponding SDE for the DDPM is given by $dx=-\frac{1}{2}\beta(t)xdt+\sqrt{\beta(t)}dw$. For the SGM, the corresponding SDE is $\mathrm{dx}=\sqrt{\frac{\mathrm{d}[\sigma(t)^{2}]}{\mathrm{d}t}}\mathrm{dw}$. This holds true as $T$ approaches infinity, where $e\beta(\frac{t}{T})=\beta_t T$ and $\sigma(\frac{t}{T})=\sigma_t$.

Here, $q_t({x})$ represents the distribution of $\mathbf{x}_t$ in the forward process. For any diffusion process structured like equation (14), Anderson \cite{ref39} demonstrated that it can be inverted by solving the following reverse-time SDE:

\begin{equation}
\label{deqn_ex1a15}
\mathrm{d}\mathbf{x}=[\mathrm{f}(\mathbf{x},t)-g(t)^2\nabla_\mathbf{x}\mathrm{log}q_t(\mathbf{x})]\mathrm{d}t+g(t)\mathrm{d}\overline{\mathbf{w}}.
\end{equation}
where $\bar{\mathbf{W}}$ represents the standard Wiener process as time flows backward, and $dt$ signifies an infinitesimal negative time step. The solution trajectories of this reverse SDE share the same marginal density as those of the forward SDE, but evolve in the opposite temporal direction \cite{ref29}. Conceptually, the solution of a reverse-time SDE describes a diffusion process that progressively transforms noise into data. Furthermore, Song et al. (2020) \cite{ref29} demonstrated the existence of an ordinary differential equation (ODE), known as the probability flow ODE, whose trajectory exhibits the same marginals as the reverse-time SDE. The probability flow ODE is defined by the following equation:

\begin{equation}
\label{deqn_ex1a16}
\mathrm{d}\mathbf{x}=\left[\mathrm{f}(\mathbf{x},t)-\frac{1}{2}g(t)^2\nabla_\mathbf{x}\mathrm{log}q_t(\mathbf{x})\right]\mathrm{d}t.
\end{equation}

Both the inverse time SDE and the probabilistic flow ODE allow for sampling from the same data distribution because their trajectories share identical marginals. If the fractional function $\nabla_\mathbf{x}\log q_t(\mathbf{x})$ is known for each time step $t$, its inverse-time equations can be derived. These equations can then be used to generate samples via numerical techniques such as annealed Ronzwan dynamics \cite{ref24}, numerical SDE solvers \cite{ref30,ref29}, numerical ODE solvers \cite{ref31,ref32,ref29,ref34,ref40}, and predictor-corrector methods (a combination of MCMC and numerical ODE/SDE solvers) \cite{ref29}. Similar to the SGM, the transient fractional model $\mathbf{s}_\theta(\mathbf{x}_t,t)$ is parameterized, and the fractional function is estimated by generalizing the fractional matching objective in Equation (13) to continuous time, resulting in the following objective:

\begin{equation}
\label{deqn_ex1a17}
\begin{split}
\mathbb{E}_{t\sim\mathcal{U}[0,T],\mathbf{x}_{0}\sim q(\mathbf{x}_{0}),\mathbf{x}_{t}\sim q(\mathbf{X}_{t}|\mathbf{X}_{0})}\\\left[\lambda(t)\|\mathbf{s}_{\theta}(\mathbf{x}_{t},t)-\nabla_{\mathbf{x}_{t}}\log q_{0t}\left(\mathbf{x}_{t}\mid\mathbf{x}_{0}\right)\|^{2}\right].
\end{split}
\end{equation}
where $\mathcal{U}[0,T]$ means that the distribution is even over the range $[0,T]$.

The previous conversation has gone over the basic ideas and equations of the basic diffusion model and major variants. This theoretical framework supports the efficacy of diffusion models in contemporary image processing, allowing them to surpass specific constraints of conventional models, despite possessing their own intrinsic disadvantages.

\subsection{Evolution of the Diffusion model with different variants}
\begin{figure}[pos=htbp]
\centering
\includegraphics[width=3.6in]{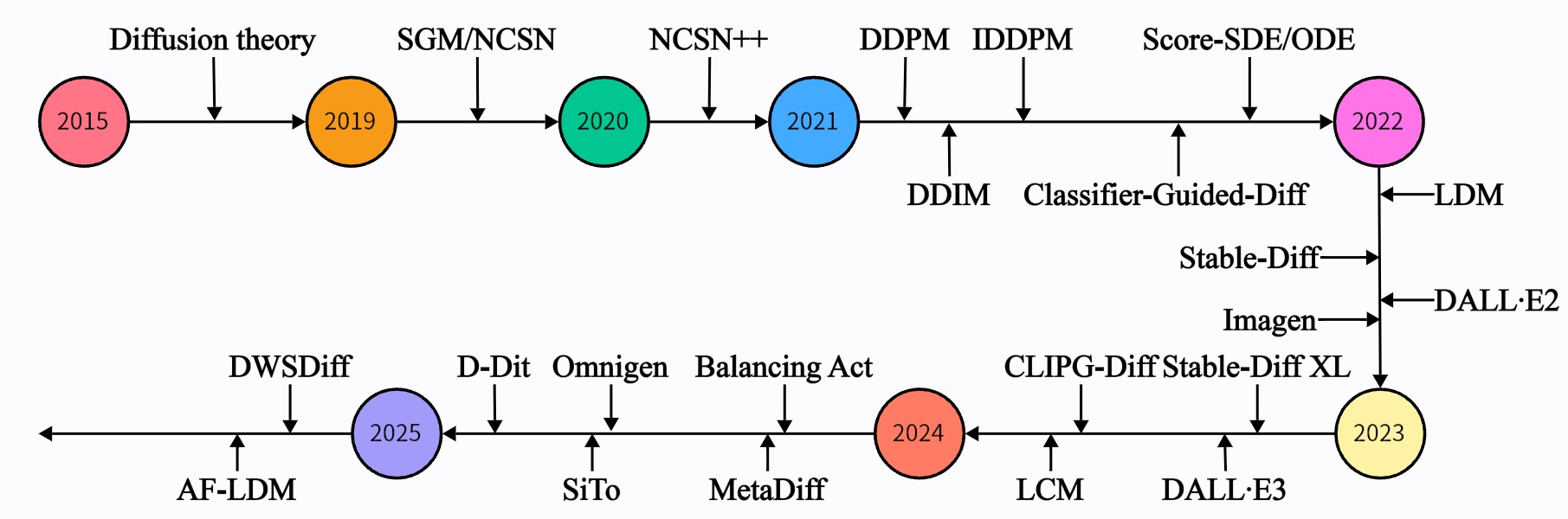}%
\hfil
\caption{Evolution timeline of the Diffusion model}
\label{image3}
\end{figure}

Since the seminal works in 2019–2020, diffusion models have undergone rapid evolution. Key milestones in this evolution are depicted in the illustrative timeline of Figure~\ref{image3}, which maps the technical trajectory from early unconditional generative theories to advanced, computationally efficient latent architectures. Rather than merely serving as intermediate steps, recent advancements \cite{ref59, ref53, ref148, ref149, ref52, ref150, ref151, ref152, ref54, ref153, ref55, ref106, ref56} have introduced targeted architectural optimizations that are highly relevant to the hyperspectral imaging (HSI) context. Specifically, innovations in cross-modal semantic alignment, bidirectional continuous-discrete diffusion, and shift-isovariance provide critical technical foundations for addressing domain-specific HSI challenges—such as multi-source data fusion (e.g., PAN and HSI), spatial-spectral consistency preservation, and the mitigation of high computational costs in full-band generation. Here we summarize the major developments and variant models that have improved diffusion models in terms of quality, speed, and flexibility:

Early Formulation (2015): Sohl-Dickstein et al. (2015) \cite{ref18} introduced the concept of gradually diffusing data into noise and back, laying a theoretical foundation by merging nonequilibrium thermodynamics with generative modeling . This work demonstrated the idea on simple datasets but did not yet achieve state-of-the-art results.

Score Matching and NCSN/SGM (2019): In 2019, Song and Ermon proposed the Noise-Conditioned Score Network (NCSN) \cite{ref24} , which leveraged score matching to generate images by progressively denoising. This approach was refined in NCSN++ (2020) \cite{ref20} with architectural improvements (like residual blocks and self-attention) and improved noise schedules, resulting in better sample quality and more stable training. Further enhancements for NCSN++ emerged with Score-SDE (Based on stochastic differential equations) \cite{ref29} and ODE (Based on Ordinary differential equation) \cite{ref29}, both optimizing the sampling process through stochastic and constant differentiation, respectively.

DDPM (2020): The DDPM framework by Ho, Jain, and Abbeel (2020)\cite{ref17} introduces a simple yet powerful reverse process parameterization and demonstrates that diffusion models can produce outstanding results on natural image benchmarks, rivaling those of GANs. Using fixed variance diffusion and a straightforward $L_2$ loss to predict added noise, DDPM became a baseline for subsequent research.

Improved Training and Sampling (IDDPM, 2021): Nichol and Dhariwal (2021) introduced Improved DDPM, which is also known as IDDPM. They improved DDPM by adding learnable noise variance (the model learns the best variance for each step instead of using fixed $\beta_t$ schedules), using a cosine schedule for noise addition (which makes diffusion steps more effective), and adding a hybrid loss that better balances different time steps. They also suggested a way to speed up sampling by skipping some diffusion steps, which means giving up some sample quality in exchange for speed. IDDPM made better generations and faster inferences than the original DDPM.

Denoising Diffusion Implicit Models (DDIM, 2021): In 2021, Song et al. introduced DDIM \cite{ref33} , which showed that one can define a non-Markovian forward process that yields a deterministic implicit reverse process. DDIM allows the model to sample in fewer steps (since it can take larger jumps in the diffusion trajectory without loss of quality) and offers a way to trade off between sample diversity and fidelity. This was significant for reducing generation time.

Latent Diffusion Models (2022): Rombach et al. introduced the Latent Diffusion Model (LDM) in 2022 \cite{ref42}, moving the diffusion process away from pixel space into a compact, learned latent space. Images are first compressed by an autoencoder, after which diffusion is carried out in this lower-dimensional domain. This design cuts computational cost significantly while still enabling the generation of high-resolution outputs. The idea serves as the foundation of Stable Diffusion \cite{ref42}, which has become a widely used tool for high-quality image synthesis. In the case of hyperspectral imaging (HSI), latent diffusion is particularly beneficial, as it reduces the high spectral dimensionality and limits the number of degrees of freedom the model must handle.

Conditional and Guided Diffusion \cite{ref11}: Progress has also been made by adding auxiliary information—such as class labels, textual prompts, or other modalities—to direct the diffusion process. Well-known approaches include classifier-guided diffusion (Dhariwal and Nichol, 2021) and classifier-free guidance (Ho and Salimans, 2022). These methods adjust the sampling trajectory, enabling control over the balance between output diversity and fidelity, so that the generated data align more closely with desired conditions.

These methods make diffusion models excel at conditional generation tasks. In hyperspectral imaging, the application of conditional diffusion may include: generating hyperspectral images based on multispectral images, or generating corresponding hyperspectral images based on category labels. Relevant examples will be further introduced in the following chapters.
Related examples will be further introduced in the following chapters.

In short, the diffusion model family has quickly improved in quality (getting better loss functions and noise scheduling to make images with higher fidelity), speed (using DDIM and other sampling methods to cut down on sampling steps), and adaptability (working in latent space to change lateral information). Even with these improvements, traditional diffusion models still have problems, like being slow to generate and costing a lot of money to run, especially when dealing with large images like those in the HSI or 3D datasets. Still, these changes have made the gap in efficiency much smaller. For example, a latent diffusion model can make high-resolution images in seconds on modern hardware, and researchers are still working to cut down on the number of steps needed without lowering quality.

\begin{figure}[pos=htbp]
\centering
\includegraphics[width=3in]{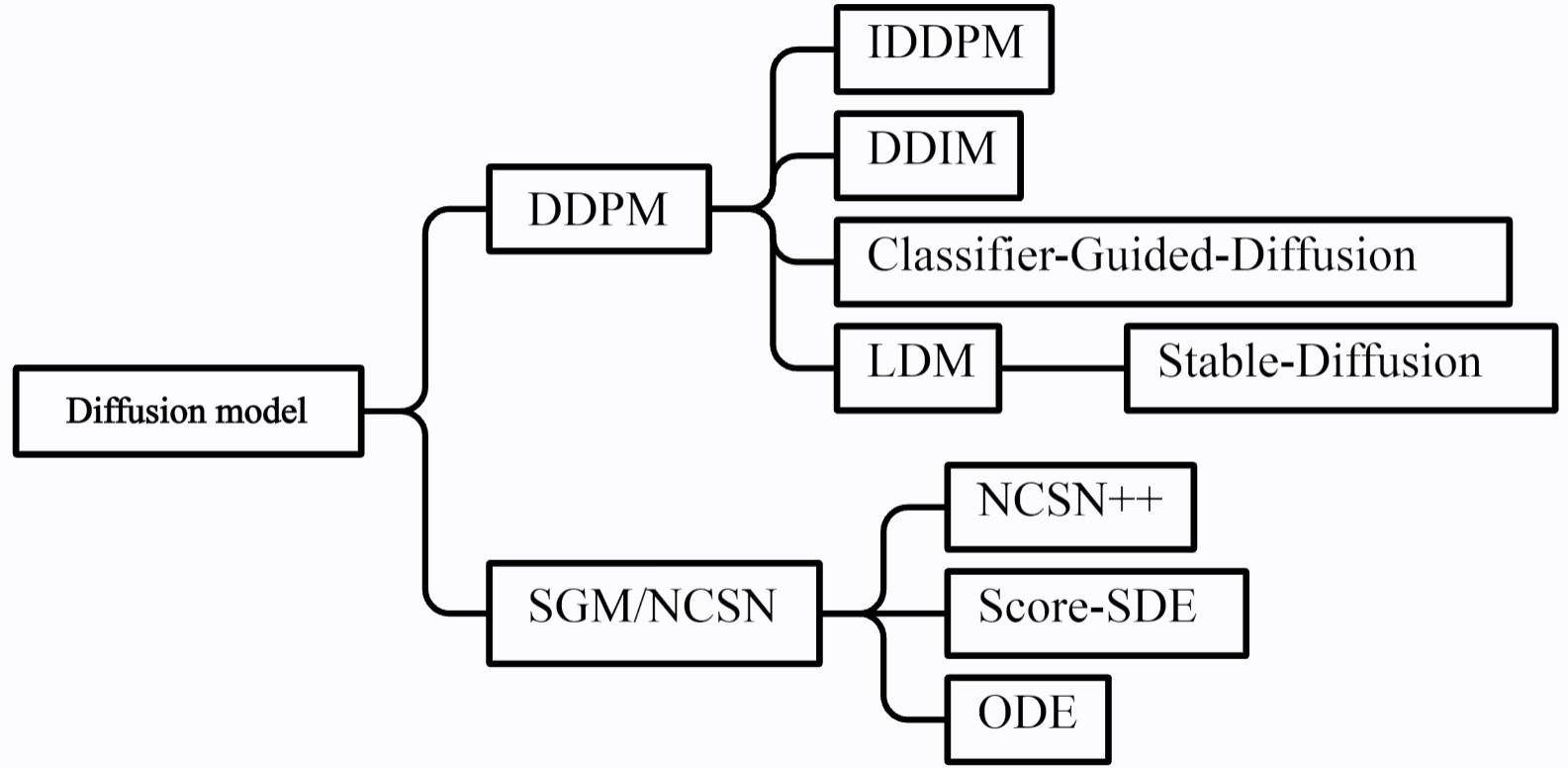}%
\hfil
\caption{Improved division of the diffusion model according to different variants}
\label{image4}
\end{figure}

Now that we have talked about the theory behind diffusion models and how they have changed over time (see Figure~\ref{image4}), we will talk about how they are used in the hyperspectral domain. The following section examines the application of diffusion models to diverse hyperspectral imaging (HSI) processing tasks and the requisite modifications to accommodate the distinct attributes of hyperspectral data.

\subsection{Traditional Deep Learning Models}
Generative Models (GMs) are a type of machine learning model that can learn data distributions and make new samples. Non-generative models (NGMs) learn decision boundaries or mapping functions directly from input data to desired outputs (like labels, values, or classes), which is different from generative models. NGMs do not learn the underlying joint probability distributions or data distributions that describe how data is generated.

Over the past few years, generative models have gained increasing attention across diverse domains, particularly in computer vision and natural language processing. Their applications range from data generation and augmentation to image restoration and anomaly detection. In this study, we mainly discuss representative deep learning paradigms, including Convolutional Neural Networks (CNN), Transformer architectures, Generative Adversarial Networks (GAN), Variational Autoencoders (VAE), and Autoregressive Models (ARMs).

Convolutional neural networks (CNN) are fundamental models in the field of computer vision. They are based on the concept of locality and the way weights are shared. They construct a layered feature extraction architecture by stacking convolutional layers, pooling layers, and fully connected layers on top of each other. Convolutional layers use sliding filters or kernels to extract local features such as edges and textures from the input data. Conversely, pooling layers simplify these features, enhancing the model's robustness. This hierarchical structure enables CNNs to automatically learn feature representations from low to high levels, significantly improving performance. But their structure is sensitive to where things are in space and has trouble directly modeling long-range dependencies, which makes them less useful for sequence tasks. \cite{ref180,ref181}.

Vaswani et al. first talked about the Transformer model in 2017\cite{ref182}. This model creates dynamic weights by figuring out how the elements in the input sequence are related to each other. It has a stack of encoders and decoders, and each layer has a feedforward network and multi-head attention. This architecture lets the Transformer process sequence data at the same time, which makes training much more efficient and lets it capture long-range semantic relationships. But the computational complexity grows by a factor of four with the length of the sequence and is very sensitive to the size of the training data \cite{ref183,ref184}.

Generative adversarial networks (GANs) are widely used for unsupervised anomaly detection. They follow the zero-sum game principle in game theory and were first introduced by Ian Goodfellow et al. in 2014 \cite{ref8}. The framework has two parts: a generator and a discriminator. The generator tries to create synthetic samples. The discriminator checks whether the input comes from real data or from the generator. With repeated training, the generator improves step by step and learns to produce outputs that closely resemble the real data distribution. 

The generation process is fast, but training is unstable and may suffer from mode collapse \cite{ref43,ref44,ref45}. 

Variational Auto-Encoder (VAE) \cite{ref9} is another generative model that combines autoencoders with probabilistic graphical models. It learns latent representations and generates new samples. The core concept is to approximate the latent distribution using Variational Inference (VI). The generated samples may have lower quality, but the training process is stable \cite{ref46,ref47,ref48}. 

Autoregressive Models (ARMs) \cite{ref51} represent another family of generative approaches. They handle sequential data by modeling each data point as dependent on the ones that came before. The core idea of autoregressive models is to utilize the temporal dependence of sequence data to generate new data points step by step, e.g., RNN (Recurrent Neural Network) forms a recursive body in the process of processing the data; LSTM (Long Short-Term Memory Neural Network)) a kind of RNN with a special pattern. Autoregressive models generate the complete data by generating each part of the data step by step, which is slow in generating speed, but generating high quality \cite{ref49,ref50}.

\begin{table*}[htbp]
\centering
\caption{Comparison of Diffusion and Traditional Deep Learning Models for HSI}
\label{tab:comparison}
\newcolumntype{M}[1]{>{\centering\arraybackslash}m{#1}} 
\begin{tabular}{M{3.5cm}M{4cm}M{4cm}M{4cm}} 
\hline
\textbf{Model} & \textbf{Advantages} & \textbf{Disadvantages} & \textbf{Future Trends} \\ 
\hline

\textbf{Diffusion Model} 
& \begin{itemize}
\item High-quality detail generation
\item Strong in denoising and restoration
\item Handles high-dimensional HSI data
\end{itemize} 
& \begin{itemize}
\item High computational cost
\item Slow inference
\item Sensitive to parameters
\end{itemize} 
& \begin{itemize}
\item Speed optimization
\item Integration with physical priors
\item Lightweight design
\end{itemize} \\

\hline
\textbf{CNN} 
& \begin{itemize}
\item Superior local feature extraction
\item Spatial invariance (translation equivariance)
\item Parameter efficiency via weight sharing
\end{itemize} 
& \begin{itemize}
\item Weak long-range dependency modeling
\item Sensitive to input spatial layout
\item Limited global context understanding
\end{itemize} 
& \begin{itemize}
\item Attention-CNN hybrid architectures
\item Lightweight designs for edge devices
\item Dynamic kernel learning
\end{itemize} \\

\hline
\textbf{Transformer model}
& \begin{itemize}
\item Global context modeling via self-attention
\item Parallelizable sequence processing
\end{itemize} 
& \begin{itemize}
\item Quadratic computational complexity
\item High training data demand
\item Lacks inherent spatial inductive bias
\end{itemize} 
& \begin{itemize}
\item Integration of domain-specific priors
\item Multi-modal fusion architectures
\end{itemize} \\

\hline
\textbf{GAN} 
& \begin{itemize}
\item Realistic data generation
\item Enhances model robustness
\end{itemize}
& \begin{itemize}
\item Training instability
\item Possible artifacts
\end{itemize}
& \begin{itemize}
\item Stabilized variants (e.g., WGAN)
\item Attention-enhanced generation
\end{itemize} \\

\hline
\textbf{VAE} 
& \begin{itemize}
\item Effective for latent encoding
\item Supports unsupervised learning
\end{itemize}
& \begin{itemize}
\item Blurry outputs
\item Poor for complex distributions
\end{itemize}
& \begin{itemize}
\item Better decoders
\item Fusion with diffusion models
\end{itemize} \\

\hline
\textbf{Autoregressive (AR)} 
& \begin{itemize}
\item Strong sequence modeling
\item Good for temporal data
\end{itemize}
& \begin{itemize}
\item Slow and non-parallelizable
\item Limited scalability
\end{itemize}
& \begin{itemize}
\item Sparse attention mechanisms
\item Transformer-based hybrids
\end{itemize} \\
\hline
\end{tabular}
\end{table*}

\subsection{Successful Application of Diffusion Modeling in Image Generation}
Diffusion model has a wide range of applications, for images on the application began in 2020, mainly based on Denoising Diffusion Probabilistic Models (DDPM), first of all, the initial proposed DDPM \cite{ref17}, IDDM \cite{ref41}, and DDIM \cite{ref33}, etc., through the Markov chain as well as the addition of the noise and denoising process, and finally the prediction of the noise residuals to optimize the quality of the generation. However, the number of sampling steps is numerous and the computation speed is slow. Following closely behind were the SGM \cite{ref24} and SDE \cite{ref29} models, which apply similar logic to images.

Then ADM-G (Classifier-Guided-Diffusion) \cite{ref11}, introduced by OpenAI team in 2021, made the model surpass GAN in FID metrics for the first time by introducing classifier-guided generation. but dependent on classifiers and high computational cost. Later, in order to optimize text-image alignment and super-resolution generation, Imagen \cite{ref53} was introduced by the Google Research team, which combines the T5-XXL text encoder with the cascade diffusion architecture, and has outstanding text alignment capability, but with great resource requirements. Nichol et al. proposed GLIDE \cite{ref189} based on SGM and SDE, samples from a 3.5 billion parameter text conditional diffusion model guided by a classification are more favored by human evaluators. 

Later, in order to realize multimodality and efficient generation, Stabel-Diffusion \cite{ref42} was proposed by Robin Rombach et al. in 2023, which compressed the diffusion process into the latent space, significantly reducing the computational complexity, but losing information in the latent space. Meanwhile, also by the OpenAI team, CLIP-Guded-Diffusion \cite{ref52} was introduced, which combines the CLIP model to realize cross-modal generation, supports complex semantic control, and generates semantic consistency, but has large training requirements and slow generation speed. Subsequently, in order to unify the framework and extend the multi-tasking domain, OmniGen \cite{ref54}, the first unified framework supporting tasks such as text generation, image editing, and so on, without additional plug-ins, was introduced by the Wisdom Source Institute in 2024. Then, the Byte Jump team introduced D-Dit \cite{ref55}, which combines continuous and discrete diffusion techniques to achieve bidirectional text-image generation, support multimodal tasks, and improve parameter efficiency by about 20\%. In 2025 by Yifan Zhou et al. proposed AF-LDM \cite{ref56}, which introduces shift isovariance loss, improves the UNet attention module, reduces potential spatial noise bias, improves video generation consistency, and enhances the model robustness, etc. 

At the same time, the diffusion model will continue to be innovated and developed in the hyperspectral field in the future, and these models are improved on the basis of the main diffusion model, aiming at solving the problem of image in classification, anomaly detection and recovery, and so on.

\begin{table*}[ht]
\centering
\setlength{\tabcolsep}{13pt}
\caption{Summary of basic diffusion models, variants and their characteristics}
\label{tab:diffusion_models}
\newcolumntype{M}[1]{>{\centering\arraybackslash}m{#1}} 
\begin{tabular}{M{1.8cm}M{1.8cm}M{2.2cm}M{2.5cm}M{1.8cm}M{1.8cm}}
\hline
\textbf{Basic Diffusion Model} & 
\textbf{Major Variants} & 
\textbf{Reasons for Improvement} & 
\textbf{Improvement Point} & 
\textbf{Drawbacks} & 
\textbf{Applications} \\ 
\hline
\multirow{17}{*}{DDPM} & 
IDDPM & 
Improve generation quality and training stability. & 
Introduction of cosine noise scheduling and hybrid objective function. to optimize the training process. & 
Generation is still slow. & 
Image generation, audio synthesis. \\ \cline{2-6}
& 
DDIM & 
Accelerating the generation process. & 
Non-Markov chain inverse process allowing large step size sampling. & 
Sacrificing some generative diversity for speed. & 
Fast image generation, interpolation. \\ \cline{2-6}
&
Classifier-Guided & 
Enhanced generation control. & 
Using classifier gradients to guide the direction of generation and improve the quality of conditional generation. & 
Additional training of classifiers is required, increasing computational cost. & 
Conditional image generation, text-to-image. \\ \cline{2-6}
&
LDM/Stable & 
Reduced computational costs. & 
Diffusion in latent space to reduce computational overhead in high-resolution images. & 
Latent space compression may lose detail. & 
High-resolution image generation, art creation. \\ \cline{2-6}
\hline

\multirow{13}{*}{SGM/NSCN} &
NSCN++ & 
Enhancement of model presentation skills. & 
Introducing deeper network structure and improved score matching goals. & 
High training complexity. & 
Complex distribution modeling, scientific simulation. \\ \cline{2-6}
&
Score-SDE & 
Harmonization of diffusion frameworks. & 
Modeling Continuous Diffusion Processes by Stochastic Differential Equation (SDE) for More General Theory. & 
The sampling process is computationally intensive. & 
Multimodal generation, cross-domain generation. \\ \cline{2-6}
&
ODE & 
Improved generation process stability. & 
Replacing SDE with ordinary differential equations (ODE) for deterministic sampling. & 
No significant increase in generation speed compared to SDE. & 
Controlled generation, inverse problem solving. \\
\hline
\end{tabular}
\end{table*}

\section{Applications in Hyperspectral Image Processing and Analysis}
\begin{figure}[pos=htbp]
\centering
\includegraphics[width=3.5in]{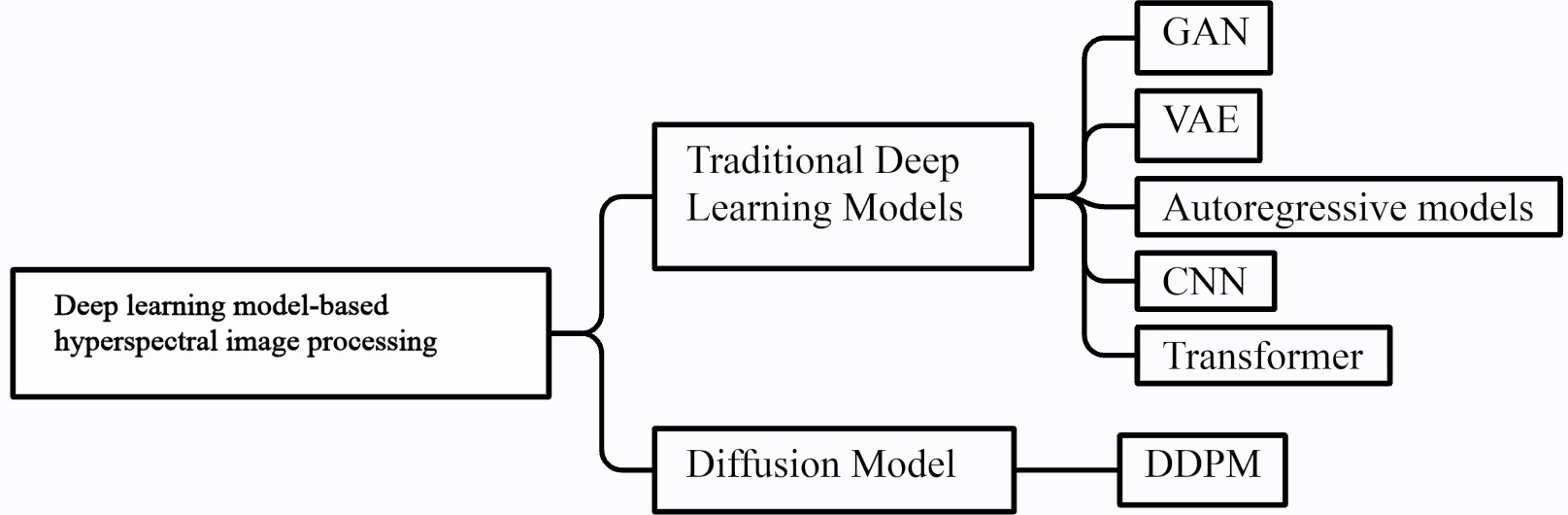}%
\hfil
\caption{Deep learning model-based hyperspectral image processing}
\label{image5}
\end{figure}

\begin{figure}[pos=htbp]
\centering
\includegraphics[width=3.5in]{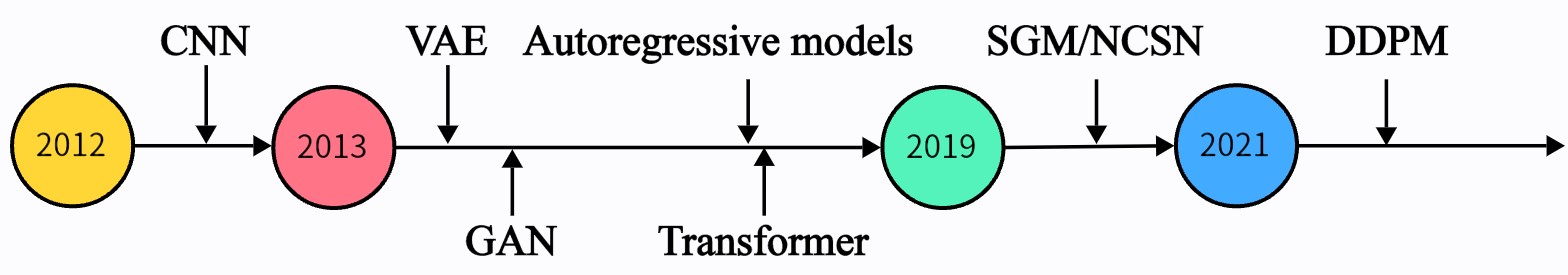}%
\hfil
\caption{Timeline of the birth of each model}
\label{image6}
\end{figure}

Deep learning models, including generative models and non-generative models, as show Figure~\ref{image5}, ~\ref{image6}, are increasingly being adopted to tackle the challenges of hyperspectral data across a range of tasks. Diffusion models, known for their ability to generate data and remove noise, have shown strong potential in many hyperspectral image (HSI) applications. This section summarizes the ways in which diffusion models have been employed for different tasks in hyperspectral image (HSI) processing. 

We organize these applications into sub-categories: data enhancement and synthesis, data generation and complementation, classification, segmentation, target detection, anomaly detection, noise suppression, and data recovery. Before diving into each specific task, we first note the general trend that most diffusion-based HSI studies use the DDPM framework (or its conditional variants) as the backbone. The DDPM’s stepwise denoising process is well-aligned with many HSI problems (such as iterative noise removal or image reconstruction). Other types of diffusion models (like SGM and continuous SDEbased models) are less common in literature so far, but the potential remains to explore them for HSIs.

Before detailing the specific applications of diffusion models in hyperspectral imaging, it is essential to outline our literature search and selection methodology to ensure the rigor and reproducibility of this review. We conducted a systematic search across Web of Science, IEEE Xplore, Scopus, and Google Scholar, covering the period from 2020 to the present. Our search strategy utilized a combination of methodological and domain-specific keywords: ("Diffusion model" OR "Denoising diffusion" OR "Score-based generative model" OR "Latent diffusion") AND ("Hyperspectral image" OR "HSI" OR "Hyperspectral analysis" OR "Remote sensing"). During the screening process, we strictly included high-quality English publications (peer-reviewed journals, high-impact conferences such as CVPR and IGARSS, and significant preprints) that centrally utilize diffusion models to solve specific HSI tasks (e.g., classification, anomaly detection, denoising) and provide quantitative results on standard datasets. Conversely, we excluded non-English publications, studies mentioning these concepts only in passing, duplicate reports or theses lacking technical depth, and papers focusing solely on RGB or multispectral images without addressing HSI-specific characteristics.

\subsection{The main Diffusion model used in the field of hyperspectral imagery}
It is worth highlighting that among the diffusion model variants discussed earlier, the denoising diffusion probabilistic model (DDPM) and its extensions are by far the most widely used in hyperspectral imaging research to date. The rationale for this prevalence is deeply rooted in the alignment between DDPM’s theoretical formulation and the intrinsic characteristics of HSI data. First, the progressive, stepwise denoising mechanism of DDPMs is exceptionally suited for recovering HSI data, which frequently suffers from complex, band-dependent noise and atmospheric corruptions. This iterative refinement allows for the careful preservation of continuous spectral signatures (i.e., reflectance curves) without introducing the high-frequency artifacts or spectral distortions often caused by single-step generative models like GANs. Second, processing hundreds of spectral bands presents severe challenges regarding training stability and the "curse of dimensionality." DDPMs mitigate this by offering a stable, likelihood-based optimization objective that scales reliably to the high-dimensional spatial-spectral manifolds of HSIs, avoiding issues like mode collapse. Finally, from a practical standpoint, early HSI studies naturally leveraged the most mature and accessible implementations from the broader computer vision community, which were predominantly DDPM-based. While CDM and LDM have only recently gained traction, researchers have actively adapted the basic DDPM framework for hyperspectral tasks by integrating domain-specific modules—such as spectral unmixing components and multi-scale spatial-spectral feature extractors—to better exploit the physical properties of HSI data.

In the following, we discuss each application category in detail, describing what role diffusion models play and giving representative examples from recent literature.

\subsection{Application of Diffusion model in image processing}
\subsubsection{Data Augmentation and Synthesis}
Super-resolution (SR) techniques \cite{ref12} are very important for improving image quality in data augmentation and synthesis. The goal of super-resolution is to make high-resolution images from low-resolution ones \cite{ref60}. But imaging equipment, weather, or downsampling can make these pictures worse. This loss of detail is often more noticeable in hyperspectral images than in natural images. This makes the inverse problem of making high-resolution versions much harder.

Before diffusion models came along, methods that used deep neural networks, like Convolutional Neural Networks (CNNs) and Transformers, tried to combine feature information in one step by finding the right mapping functions. To do this, we used self-supervised learning \cite{ref61}, adaptive marker selection \cite{ref62}, and frequency information assistance \cite{ref63}. On the other hand, diffusion model-based methods combine low-resolution images and bootstrap images at every step of the diffusion process. This repeated combination of different types of image data has worked very well and has gotten a lot of attention in the hyperspectral field.

Hyperspectral Images (HSIs) have high spectral resolution, but their spatial resolution is often lower than that of other types of images \cite{ref68}. This can make some HSI-based applications less useful. To address this challenge, two principal techniques are utilized to obtain high-resolution hyperspectral maps: panchromatic sharpening \cite{ref64} and multispectral and HSI fusion \cite{ref65}.

Panchromatic sharpening enhances HSIs by injecting detailed spatial information from a high-resolution panchromatic (PAN) image. Conversely, multispectral and HSI fusion leverages multispectral images to help hyperspectral maps acquire spatial details. In real-world remote sensing, HSI sensors face a strict physical trade-off, capturing hundreds of highly correlated, narrow spectral bands at the severe cost of low spatial resolution, which inevitably leads to complex "mixed pixel" problems. For example, Shi et al. \cite{ref66} used both multispectral and hyperspectral images as conditional inputs to a diffusion model. They then combined the information from these two types of images to make high-resolution hyperspectral maps. Unlike conventional regression-based CNNs that often cause severe spectral distortion during fusion, the diffusion model here acts as a powerful stochastic bridge. It safely transfers high-frequency spatial textures from the multispectral image while strictly preserving the intrinsic spectral correlations of the original HSI. Qu et al. \cite{ref67} also created a conditional recurrent diffusion framework. This framework enables the spatial super-resolution process to be guided by the multispectral image and the spectral super-resolution process by the hyperspectral map. By alternating these guidance mechanisms within the iterative diffusion steps, the model intrinsically respects the physical constraint of spatial-spectral consistency, leading to the generation of high-resolution hyperspectral maps that boast genuinely complementary spatial and spectral information without compromising radiometric accuracy.

Recent studies are increasingly concentrating on the application of Panchromatic (PAN) images for HSI super-resolution \cite{ref69,ref71,ref72,ref73,ref74,ref75,ref76}. Due to hardware constraints, HSI sensors inherently suffer from a physical trade-off between spectral and spatial resolution, resulting in images with hundreds of continuous spectral bands but severe spatial degradation. To compensate for this, Meng et al. \cite{ref71} developed a modal calibration module to improve and extract features from both types of images, which were then utilized as conditional inputs for a diffusion model. Here, the diffusion model acts as a powerful stochastic engine, effectively hallucinating missing spatial details guided by the PAN image while preventing the spectral distortion common in traditional GANs. On the other hand, Cao et al. \cite{ref69} argued that the unique physical information from different image modalities (i.e., the high spatial frequency of PAN and the rich spectral correlation of HSI) should not be forcefully entangled when they are being processed. Instead, they made two conditional modulation modules that take out coarse-grained style information and fine-grained frequency information separately. These are then used as conditional inputs.

\begin{figure}[pos=htbp]
\centering
\includegraphics[width=3.3in]{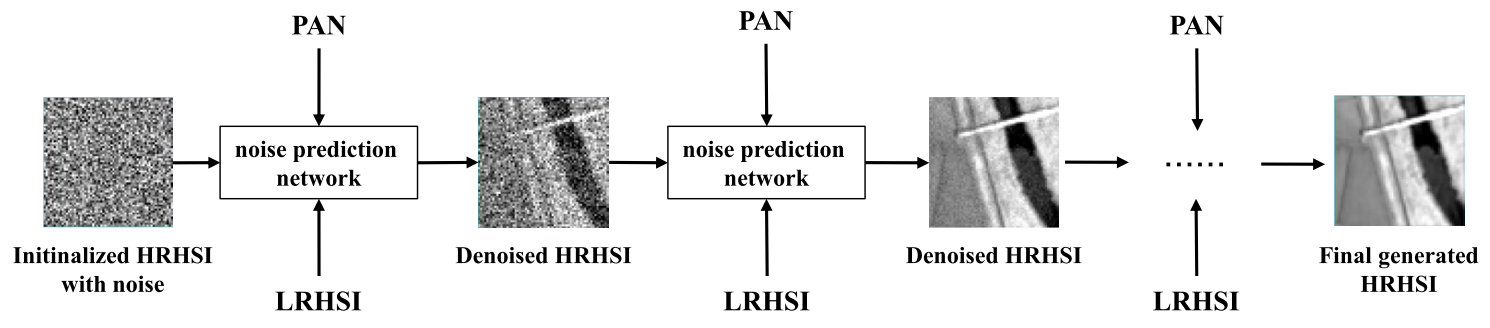}%
\hfil
\caption{Converged network fusion mode \cite{ref74}}
\label{image8}
\end{figure}

To fully leverage the unique information from different modalities without compromising the spectral integrity of the HSI, Li et al. \cite{ref74} proposed a panchromatic sharpening network based on a biconditional diffusion model. In this architecture, both HSI and PAN images serve as independent conditional inputs, allowing the network to separately learn spectral features and spatial textures, as shown in Figure~\ref{image8}. Designed based on LDM, this network achieves excellent super-resolution (SR) performance while significantly reducing computational cost. More importantly, projecting HSI into a latent space aligns perfectly with the intrinsic low-rank property of hyperspectral data, where hundreds of highly correlated bands can be physically compressed into a lower-dimensional manifold without losing critical spectral signatures. Similarly, Rui et al. \cite{ref76} also perform the sampling process in a low-dimensional space. Their method is deeply rooted in the physical constraint that a high-resolution hyperspectral map can be decomposed into the product of two low-rank tensors, conceptually mirroring the physical linear spectral unmixing process (i.e., spectral endmembers and spatial abundances). First, one of these low-rank tensors is computed using the low-resolution hyperspectral map. This tensor, along with the LRHSI (low-resolution hyperspectral image) and PAN, is then conditioned and fed into a pre-trained diffusion model \cite{ref70} to generate the other low-rank tensor. Unlike the aforementioned methods, this approach is a completely unsupervised deep learning method, meaning it doesn't require high-resolution hyperspectral maps during any part of the process, making it highly feasible for practical remote sensing applications where acquiring paired high-resolution HSI ground truth is physically impossible.

In addition, for SGM or SDE models, Shi et al. proposed WaveDiffUR \cite{ref185}, a novel wavelet domain diffusion UR solver. Since HSI spatial-spectral features exhibit distinct frequency behaviors—where continuous spectral correlations often reside in low-frequency components and sharp spatial edges in high-frequency ones—operating in the wavelet domain provides a domain-specific physical constraint. This solver iteratively reconstructs low-frequency wavelet details by merging pre-trained SR models into plug-and-play modules (ensuring global spectral consistency) and high-frequency components (enhancing local spatial fidelity), effectively mitigating the ill-posedness of SDE and ensuring scalability across different HSI applications.

\subsubsection{Data Generation and Completion}
In the realm of generation and complementation, while broad categories include text-to-image and image-to-image synthesis, the hyperspectral domain exclusively features image-to-image applications. Here, diffusion models leverage image bootstrapping to create new images. These models can be guided by various forms of input, such as maps \cite{ref77} or semantic layouts \cite{ref78,ref79,ref80,ref81}. Even with only specific information in these guiding images, excellent performance in image generation is still observed. For instance, Espinosa and Crowley \cite{ref77} demonstrated that training conditional diffusion models with maps and historical maps could generate more realistic satellite images. Similarly, Yuan et al. \cite{ref78} successfully produced high-quality hyperspectral images guided by semantic masks, simultaneously addressing the inherent challenge of lengthy training times for diffusion model convergence.

However, generating high-quality images alone isn't always sufficient; creating more practical image-annotation pairs is often crucial. To this end, Zhao et al. \cite{ref80} implemented a two-stage fine-tuning of the Stable Diffusion (SD) model, enabling generation from noise to annotations and then from annotations to images. Furthermore, Toker et al.\cite{ref81} extended the standard diffusion model into a joint probability model for images and their corresponding labels, enabling simultaneous generation of both.

Hyperspectral images (HSIs) offer rich spectral and spatial information, but their acquisition cost is significantly higher than that of multispectral images. Consequently, scientists have explored using readily available multispectral remote sensing images to generate HSIs \cite{ref82,ref83}. However, a challenge arises because the diffusion models used for this task require the input noise dimensions to match the spectral bands of the hyperspectral map. Unlike standard RGB images, HSI bands possess extremely high spectral correlation. Injecting independent Gaussian noise across hundreds of narrow bands severely disrupts the physical continuity of the spectral signatures. This results in an overly large noise sampling space, which makes it hard for the model to converge while maintaining radiometric accuracy.

To address this problem, Zhang et al. \cite{ref82} proposed a spectral folding method. In this method, the input hyperspectral map is first transformed into a pseudo-color image, and the diffusion model is then trained on this representation. By physically compressing the highly redundant spectral dimensions into a compact color space, this approach forces the diffusion process to focus on the most salient spatial-spectral structures, significantly improving training stability and avoiding the curse of dimensionality.

Liu et al. \cite{ref83} used a Conditional Vector Quantization Generative Adversarial Network (VQGAN) \cite{ref84} to create a latent space for the hyperspectral map within the Latent Diffusion Model. They were inspired by the LDM. Training and sampling for the diffusion model are then done in this easier-to-manage latent space, which perfectly captures the intrinsic low-rank physical manifold of the HSI data. Deng et al. also came up with DiffUn \cite{ref85}, a hyperspectral spectral unmixing method based on a Diffusion Model, to deal with the problems of high cost and low spatial resolution. This method uniquely integrates pure HSI physical constraints by using spectral prior distributions learned by an unconditional Diffusion Model and likelihoods from the Linear Mixing Model (LMM). Here, the diffusion model guarantees that the generated HSI pixels strictly obey the physical laws of light mixing, improving deblurring and creating hyperspectral maps. Dong et al. recently came up with a new ISPDiff model \cite{ref86}. This model is clear, works for any size, and solves these problems with an iterative optimization algorithm. It has a Multi-Scale Residual Network (MSRN) for recovering scales that match and a Cross-Modal UNet (CMUN) for upsampling across scales, directly compensating for the physical trade-off between spatial and spectral resolutions inherent to HSI sensors.

\begin{figure}[pos=htbp]
\centering
\includegraphics[width=3in]{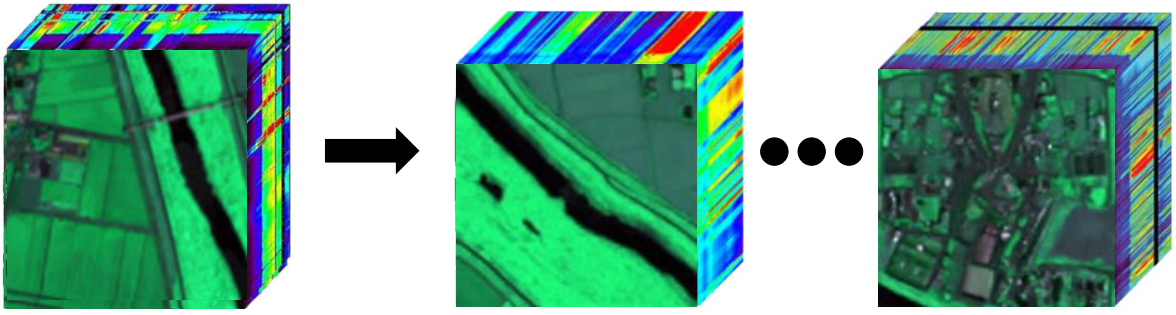}%
\caption{UnmixDiff generates a simple diagram of the HIS \cite{ref87}}
\label{image9}
\end{figure}

Overall, image-to-image generation methods based on conditional diffusion models \cite{ref57,ref58} fundamentally generate images from noise, rather than directly converting existing ones, by using bootstrap images as conditional inputs to conditional DDPMs. To achieve a true image-to-image conversion that respects HSI physics, Yu et al. proposed UnmixDiff \cite{ref87}. This model explores the feasibility of depth generative models in HSI synthesis, synthesizing HSIs strictly from the perspective of material distribution and introducing a clear physical interpretation into the task, as shown in Figure~\ref{image9}. By integrating an unmixing autoencoder within the diffusion generation model, hyperspectral image synthesis is projected from the high-dimensional spectral domain to a physically meaningful low-dimensional abundance domain. This significantly reduces computational complexity and effectively boosts the spectral fidelity of the generated HSIs. Building a diffusion generation model in the abundance space also lets you make HSIs with realistic spatial details while keeping the spectral consistency, as the diffusion process is now constrained to generate physically plausible material fractions rather than arbitrary pixel intensities. This kind of method uses self-supervised learning to get feature embeddings from input patches. These embeddings then act as conditional inputs that help the diffusion model learn. This method keeps the original image's color patches in the same place, which is very important because it lets the generated patches fit together perfectly to make a big, coherent and radiometrically accurate image.

Under SGM/SDE, Cao et al. proposed DDFR \cite{ref186}, which introduces diffusion models into the field of image fusion, treats image fusion tasks as image-to-image translation, and designs two different conditional injection modulation modules to inject coarse-grained style information (representing the continuous spectral envelopes) and fine-grained high-frequency and low-frequency information (capturing sharp spatial boundaries) into the diffusion UNet. Additionally, Xiao et al. proposed a novel guided diffusion scheme featuring zero-shot guidance and neural spatial spectral decomposition (NSSD) \cite{ref188}. The zero-shot guidance employs an auxiliary neural network trained solely with PAN and lr-hsi to guide pre-trained DMs in generating RGB detail images informed by specific prior knowledge. Subsequently, NSSD establishes a spectral mapping from the generated RGB detail images to the final hr-hsi. This decomposition explicitly acknowledges the HSI physical characteristic that high-resolution spatial details are intrinsically tied to specific wavelength responses, utilizing the diffusion model to hallucinate spatial details only where physically appropriate.

\subsubsection{Classification}
In the application of hyperspectral fields, classification tasks are crucial for image analysis, capable of accurately identifying and characterizing various structures and anomalies in images. It greatly helps professionals interpret large amounts of complex data. Despite this potential, the purpose of image classification is to extract more valuable information about detailed features such as buildings or grasslands by assigning each pixel to a specific category \cite{ref128,ref129}. However, hyperspectral images themselves contain highly diverse and complex scenes and information, which brings difficulties to accurate classification. Therefore, using diffusion models to enhance classification remains a major challenge that needs to be further strengthened and resolved. However, as diffusion models can learn and simulate more complex data distributions than other deep learning models, an increasing number of people are dedicated to applying diffusion models to hyperspectral image classification. To produce more and more promising results \cite{ref118,ref125,ref168,ref169,ref170,ref171}.

\begin{figure}[pos=htbp]
\centering
\includegraphics[width=3in]{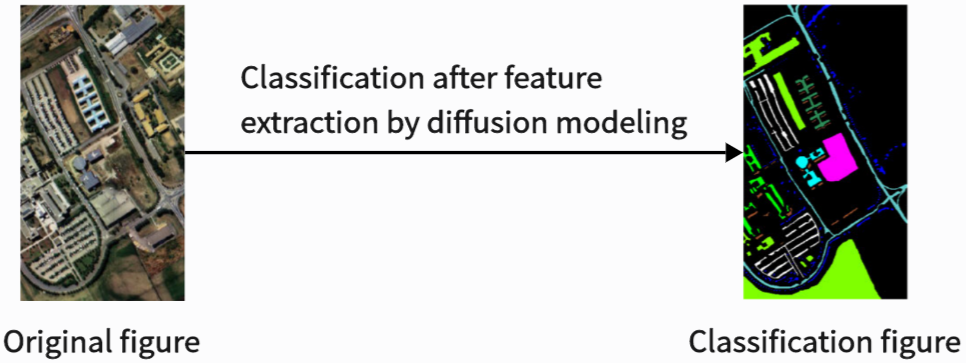}%
\hfil
\caption{Workflow of image classification with a diffusion model}
\label{image10}
\end{figure}

The complex data distribution of hyperspectral images often limits the performance of traditional deep learning classifiers. Diffusion models, however, are well suited to learn joint spectral–spatial representations \cite{ref166,ref167,ref172}, as illustrated in Figure~\ref{image10}. This capability improves classification accuracy \cite{ref94,ref95} and strengthens generalization across different scenarios \cite{ref124}. 

Building on these strengths, Zhou et al. \cite{ref96} utilized the temporal information of diffusion models to construct a time-stepping feature library. Because the reverse diffusion process gradually maps an isotropic Gaussian distribution back to the highly structured HSI data manifold, these intermediate temporal features inherently capture varying levels of spatial-spectral abstraction. They also introduced a dynamic fusion module to combine spectral-spatial features with temporal features, allowing for sufficient image information before classification. Li et al. \cite{ref97} created a two-branched diffusion model that takes features from both HSI and LiDAR images, making sure that information from different types of images works well together. This architecture explicitly addresses the inherent spatial limitations of HSI by using the precise geometric and structural elevation priors of LiDAR to complement the rich, continuous material spectra of the HSI. This method improves pixel discriminability and works very well for more difficult hyperspectral tasks \cite{ref98}. Qu et al. \cite{ref99} advanced this concept by configuring the encoder of their two-branch diffusion model to function in a parameter-sharing mode, thereby guaranteeing the extraction of shared features from multimodal images without forcefully distorting their unique physical modalities. Chen et al.\cite{ref100} also employed diffusion models to aid in constructing deep subspaces, achieving excellent results in unsupervised HSI classification. Projecting HSI into deep subspaces aligns seamlessly with the physical low-rank property of hyperspectral cubes, where hundreds of highly correlated bands can be physically represented by a few fundamental endmember signatures.

Taking a different approach, Qu et al. \cite{ref101} combined a diffusion model with both a super-resolution (SR) network and a classification network. This made it possible to gradually rebuild high-quality images by doing image super-resolution and classification tasks over and over again, which created a system of mutual guidance. Physically, the low spatial resolution of HSI frequently leads to "mixed pixels," which severely degrades pixel-wise classification accuracy. By using the diffusion model to hallucinate sub-pixel spatial details while strictly conditioned on classification maps, the framework ensures that the generated spatial textures correspond rigorously to physically meaningful material boundaries. Using multi-scale classification results to help the target-scale image learn new things and make high-quality images again made both super-resolution and classification performance much better. Jiang et al. introduced an image fusion classification network using denoising diffusion probabilistic models (DDPM) \cite{ref172}, achieving a significant improvement in classification accuracy after fusion by leveraging the generative prior to correct spectral shifting. In the same way, Zhu et al. came up with a Spatial-Spectral Diffusion Comparison Representation Network (DiffCRN), which is also based on DDPMs \cite{ref169}. There is a spatially self-attentive denoising module (SSAD) and a spectral group self-attentive denoising module (SGSAD) in this network. This decoupled attention design explicitly mirrors the dual physical nature of HSI: the spatial attention captures local texture continuity, while the spectral group attention preserves the highly correlated, long-range dependencies across continuous wavelength bands. The efficiency of spectral feature learning is enhanced by introducing a novel DDPM loss function and contrastive learning (CL) with log absolute error (LAE). Furthermore, the incorporation of an adaptive weighting addition module (AWAM) and a cross-temporal spectral fusion module (CTSSFM) mitigates the impact of spatial spectral heterogeneity and noise inherent to physical sensor capturing, thereby improving classification accuracy.

To address spatial distortion and parameter bias during application, Zhou et al. proposed a Model-Embedded Two-Stage Diffusion (MTDiff) method \cite{ref196} for the unsupervised reconstruction of high-spatial-resolution HSI. Based on an estimated degradation model, a well-designed dual-resolution diffusion model progressively perturbs and denoises the image, achieving a fidelity-preserving reconstruction with high interpretability. Crucially, substituting a generic neural network with an explicit physical degradation model—which mathematically formulates the hyperspectral sensor's spatial blurring and spectral downsampling processes—forces the diffusion's reverse sampling to act as a rigorous solver for a physically ill-posed inverse problem, rather than a mere pixel generator. This guarantees that the reconstructed high-resolution HSI strictly obeys the original radiometric constraints.

Overall, diffusion models provide notable gains over conventional classification techniques in hyperspectral imaging. Their effectiveness becomes even greater when they are integrated with multimodal information, enhanced by self-attention modules, and optimized jointly across multiple tasks.

\subsubsection{Segmentation}
Image segmentation is a fundamental task in computer vision. It includes general segmentation \cite{ref14,ref123} as well as semantic segmentation \cite{ref122}. Within deep learning, semantic segmentation is often regarded as a specialized form of the broader segmentation problem \cite{ref97,ref113}. For hyperspectral images, segmentation helps reduce complexity by dividing the image into meaningful regions based on the distinct physical reflectance spectra of different ground materials. Unlike RGB segmentation, HSI segmentation strictly relies on capturing the high spectral correlation and continuous signatures across hundreds of narrow bands. This process provides valuable information for the hyperspectral field, thereby promoting precision image analysis. However, deep learning models usually require a large amount of training data annotated with different pixels to generate generalizable results. Unfortunately, due to the significant time, cost, and expertise required for annotation—which often involves expensive field campaigns to match spectral curves with physical ground truth—the number of images and labels available for hyperspectral image segmentation is strictly limited. Therefore, diffusion modeling has become a promising approach in image segmentation research. Rather than merely hallucinating visually plausible pixels like traditional GANs, diffusion models can learn the underlying physical distribution of material spectra, synthesizing labeled data that maintains strict radiometric fidelity, thereby eliminating the need for manual annotation of a large number of pixels \cite{ref169,ref171}.

Dong et al. \cite{ref86} are utilizing Denoising Diffusion Probabilistic Models in conjunction with Scale Matching Recovery Networks (P2MSRN) and cross-scale model-driven up-sampling networks to explicitly address the multi-scale spatial variations of continuous spectral fields. Chen et al. \cite{ref127} are designing image conditional mixing guides and integrating them with Denoising Diffusion Probabilistic Models to ensure that the synthesized spatial boundaries perfectly align with physical material transitions. For labeling information, Zhang et al. \cite{ref82} leverage Diffusion Probabilistic Models, while Zhu et al. \cite{ref126} employ Variational Auto-Encoders alongside semi-supervised diffusion models for annotation, where the diffusion process intrinsically captures the complex, high-dimensional manifold of hyperspectral data to generate physically reliable pseudo-labels. Li et al. proposed a multidomain diffusion-driven feature learning network (MDFL) \cite{ref97} to enhance pixel-level classification (i.e., segmentation) performance of high-dimensional remote sensing images through a multidomain diffusion model and feature reuse mechanism. By modeling the spatial and spectral domains jointly, the diffusion network effectively handles the low spatial resolution while fully exploiting the high spectral correlation of HSI. Zhu et al. also proposed DiffCRNs \cite{ref169}, a diffusion-based model coupled with contrastive learning, to extract spatial-spectral features and ultimately generate pixel-level classification circles. Furthermore, Xu et al. introduced a data-enhanced semantic segmentation framework based on the diffusion model \cite{ref171}. This framework generates pseudo-labeled data via the diffusion model, uses ControlNet for cross-modal "label→multispectral image" generation, and employs multi-stage training to improve segmentation accuracy. Here, the diffusion model excels at mapping discrete semantic labels to continuous, physically plausible hyperspectral signatures, providing robust data augmentation without compromising the spectral physics.

\subsubsection{Target Detection}
On the other hand, object detection in hyperspectral images uses bounding boxes to find specific objects in the image, such as aircraft, vehicles or ships \cite{ref130,ref131}. This is different from pixel-level ground classification. A major problem in this field is that training data is often unclear, not realistic, and not diverse enough, which is made worse by outside factors \cite{ref112,ref113}. This issue stems from the inherent nature of hyperspectral images, where objects needing detection are frequently complex, difficult to distinguish, and distributed across various image regions. Consequently, augmenting the objects to be detected using diffusion models has emerged as an effective solution.

Yu et al. proposed Diff-Mosaic (Enhanced Target Detection via Diffusion Prior) \cite{ref113} to deal with the problems listed above —most notably the extreme scarcity of target pixels and the severe spectral variability caused by sub-pixel mixing in low-spatial-resolution HSIs. This diffusion model-based method adds an enhancement network called Pixel-Prior, which works with pixels to make very realistic and varied mosaic images. It uses a strong diffusion model prior that resamples pixels with samples. This adds real-world information and ensures that the samples it generates are both real and distinct, strictly preserving the continuous spectral profiles of rare targets without physically distorting their radiometric signatures against the synthesized background.

Qi et al. also came up with a new joint framework for restoring underwater hyperspectral images and finding targets. This framework is based on a conditional diffusion framework \cite{ref112}. In underwater environments, hyperspectral sensors suffer from severe physical constraints, specifically the wavelength-dependent attenuation of light (where red and near-infrared bands are heavily absorbed compared to blue/green bands), causing extreme spectral distortion. This framework uses a variable spectral group extraction module and a joint underwater hyperspectral image restoration and target detection (JURTD) module to mimic the spectra of different types of underwater targets. This gets around the physical radiometric problems that come with underwater hyperspectral imaging by making high-quality restored images and better detection performance through joint spatial-spectral physical correction.

Zhu et al. presented ODGEN (Domain-specific Object Detection Data Generation with Diffusion Models) \cite{ref115}, an innovative method for producing high-quality images conditioned by bounding boxes. This makes it easier to combine data for object detection. Because HSI target detection is fundamentally a spectral matching problem rather than a purely spatial one, generating valid data requires strict spatial-spectral physical alignment. ODGEN shows strong performance when dealing with complex scenes and specific domains by using synthetic visual cues with spatial constraints and textual descriptions for each object to control the diffusion model, ensuring that the generated spatial bounding boxes perfectly encapsulate their corresponding highly correlated spectral responses.

\begin{figure}[pos=htbp]
\centering
\includegraphics[width=3in]{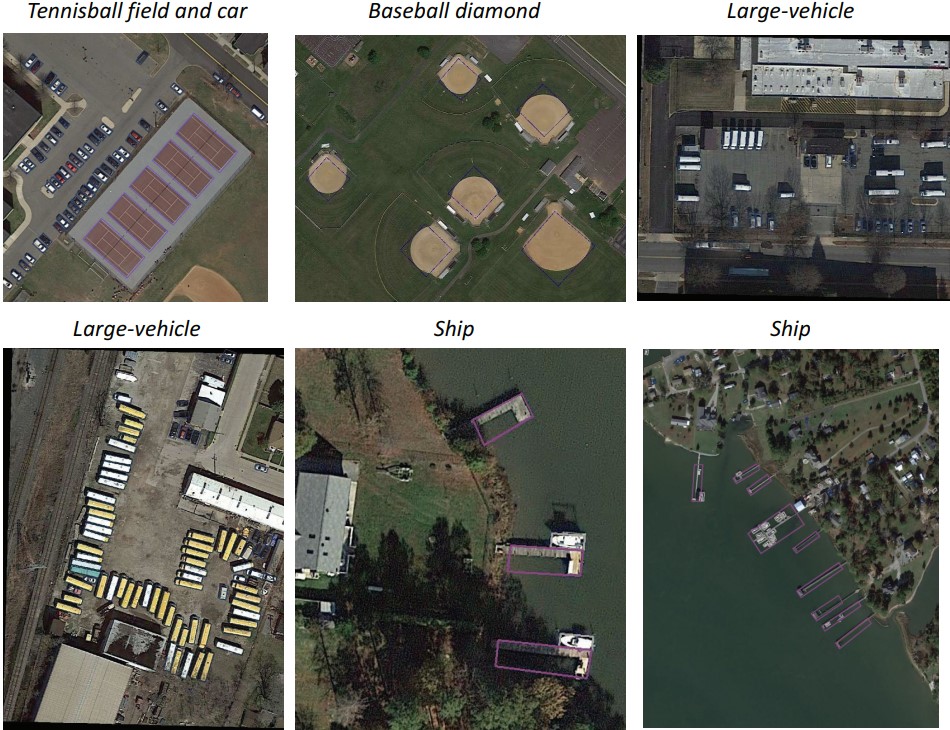}%
\hfil
\caption{OrientedDiffDet target detection diagram with targets boxed in different colors in the detection diagram \cite{ref114}}
\label{image11}
\end{figure}

Meanwhile, to enhance precision and accuracy in remote sensing scenarios where targets are arbitrarily oriented and occupy very few spatial pixels, Wang et al. proposed OrientedDiffDet (pairwise object-oriented diffusion model) \cite{ref114}. This approach combines oriented frame representation, feature alignment, and accelerated diffusion, allowing for direct object detection from a set of random frames. Using oriented boxes is physically crucial for HSIs: standard horizontal boxes often include excessive background pixels, leading to severe "mixed pixel" contamination that corrupts the target's pure spectral endmember. During the training phase, horizontal detection boxes are initially converted to oriented detection boxes. The model then learns to reverse this conversion by diffusing from ground truth oriented boxes to a random distribution. In the inference phase, the model progressively refines a set of randomly generated boxes to produce the final output, as shown in Figure~\ref{image11}, effectively isolating the target's precise spatial boundaries to maximize spectral purity.

In summary, these methods show how diffusion models can greatly improve the accuracy of data generation and detection in difficult hyperspectral situations by explicitly modeling the physical constraints of the sensor, preserving spectral continuity, and mitigating the detrimental effects of low spatial resolution and sub-pixel mixing.

\subsubsection{Anomaly Detection}
Image anomaly detection is a key field in computer vision, the focus is highlighted in the image anomaly area \cite{ref101,ref102,ref103,ref104}. In recent years, generative modeling has had a significant impact on anomaly detection research and achieved gratifying results. This section will explore how to apply the diffusion model as a prominent generative model to hyperspectral image anomaly detection. Although there are relatively few studies specifically targeting anomaly detection in hyperspectral images based on diffusion models, it has extensive applications in other fields such as medical imaging \cite{ref107,ref108} and industrial inspection \cite{ref109,ref110}. Given the continuous progress of diffusion modeling, its application in hyperspectral image anomaly detection is expected to grow significantly in the future.

In the field of anomaly detection, the complexity of background clutter in hyperspectral images is a continuous challenge, which often masks anomalies. Furthermore, the scarcity of labeled hyperspectral samples leads to poor generalization ability of existing anomaly detection methods.

\begin{figure}[pos=htbp]
\centering
\includegraphics[width=3in]{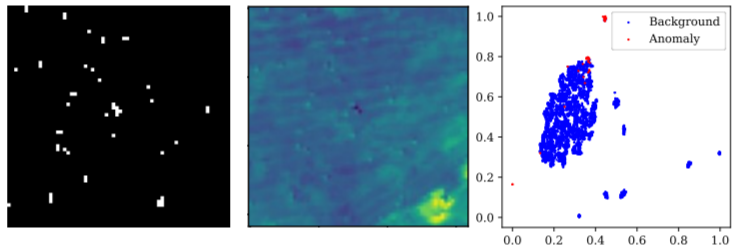}%
\hfil
\caption{Different visualizations under BSDM, from left to right: anomaly diagram, 1D t-SN diagram, and 2D t-SNE scatter plot \cite{ref105}}
\label{image12}
\end{figure}

To tackle these issues, Ma et al. proposed BSDM (Background Suppression Diffusion Model) \cite{ref105}, a novel anomaly detection solution. BSDM simultaneously learns the latent background distribution—which physically corresponds to the dominant, highly correlated spectral signatures of normal ground materials— and generalizes it across different datasets to suppress complex backgrounds, as shown in Figure~\ref{image12}. This approach is characterized by three key aspects: 1. Pseudo-background noise was designed for the complex background of HSI (which often suffers from spectral variability due to illumination and atmospheric changes), and the diffusion model (DM) was adopted to learn the potential background distribution. 2. To address the generalization issue, the statistical bias module enables BSDM to adapt to datasets from different domains without the need to label samples. 3. For effective background suppression, the DM's inference process was innovatively improved by feeding the original hyperspectral image into a denoising network, which removes the background as if it were noise. Physically, this means the diffusion model acts as an advanced spectral filter that only reconstructs the learned low-rank background manifold; signatures that deviate from this manifold (the anomalies) fail to be reconstructed, naturally isolating them. This work also marked the first application of diffusion modeling to hyperspectral image anomaly detection.

Wu et al. introduced the Diffusion Background Dictionary Method (DBD) \cite{ref111}, which combines the advantages of model-driven and data-driven methods based on the characteristics of HSI and HAD (Hyperspectral Anomaly Detection) tasks. DBD essentially integrates DM with the low-rank representation (LRR) model. Because HSI bands are highly correlated, the entire background can be physically constrained within a low-rank subspace spanned by a few pure material spectra. Using DM to obtain the key background dictionary tensor in the LRR tensor thereby achieves accurate anomaly detectionby ensuring the generated dictionary is highly representative of the true physical background. Following the principle of the RX algorithm in anomaly detection, the multivariate normal distribution approximating the HSI background (a classic statistical prior of spectral clusters) is diffused to better adapt to background suppression.

In addition, hyperspectral image anomaly detection often faces the problem of lacking prior information, mainly the challenge of accurately reconstructing the background area while inferring the potential background of the abnormal region —a task heavily complicated by the inherently low spatial resolution of HSI sensors, which causes sub-pixel anomalies to frequently hide within mixed pixels. For this reason, Chen et al. proposed double-window spectral diffusion (DWSDiff) \cite{ref106}. This method models the accurate background estimation through iterative spectral diffusion and reverse diffusion processes. It introduces a dual-window strategy to mitigate the impact of neighborhood expansion anomaly regions on background estimation, strictly preserving local spatial-spectral consistency. Finally, an anomaly generation strategy based on Principal Component Analysis (PCA) and Linear Spectral Hybrid Model (LSMM) was designed to generate real training data. By strictly adhering to the LSMM—the fundamental physical law dictating that a pixel's spectrum is a linear combination of physical endmembers—this strategy generates physically plausible samples, overcoming the scarcity of labeled hyperspectral data and thereby improving performance. Recently, Huang et al. \cite{ref192} proposed a terminal-member-guided feature diffusion network that aggregates terminal features (i.e., endmembers, representing the pure physical material signatures) to update HSI features based on a spectral linear mixing model. Wu et al. proposed a pixel-based multiresolution diffusion network \cite{ref193}, called PixDiff, designed to reduce the visibility of anomalies and distinguish them from complex backgrounds by leveraging multi-scale spatial contexts to compensate for the spatial-spectral resolution trade-off.

These methods demonstrate that diffusion models are improving in managing background suppression, domain adaptation, and data scarcity in hyperspectral anomaly detection. This creates room for new directions in the field. 

In summary, using diffusion models for hyperspectral image anomaly detection is still developing quickly. The area offers many opportunities for future research. 
We anticipate that an increasing number of researchers will engage in this domain, resulting in significant advancements in anomaly detection.

\subsubsection{Noise Suppression}
One of the main challenges of hyperspectral imaging is to obtain images without losing key information—specifically, the continuous physical reflectance spectra that define material properties. During the acquisition or subsequent processing, images may be damaged by noise or artifacts caused by sensor limitations or atmospheric scattering, significantly reducing quality, especially for small objects or low contrast. Due to their generative nature, diffusion models are highly suitable for solving various noise reduction problems \cite{ref117,ref119,ref165}. For instance, \cite{ref119} focuses on denoising in image restoration and has designed a multi-shape spatial rectangular self-attention module and a spectral latent diffusion enhancement module. By operating in a spectral latent space, this approach explicitly leverages the high inter-band correlation of HSI to filter out unstructured noise while protecting the intrinsic spectral signatures. The diffusion model can make the denoising process closer to the expected physical value and enhance the robustness of denoising.

\begin{figure}[pos=htbp]
\centering
\includegraphics[width=3in]{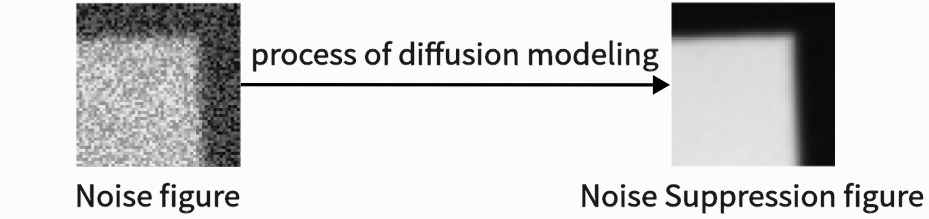}%
\hfil
\caption{Process diagram of noise suppression based on diffusion model}
\label{image13}
\end{figure}

In image processing, hyperspectral images (HSIs) are invariably affected by various types of noise due to the inherent limitations of imaging techniques (such as the physical trade-off between narrow spectral bandwidths and photon capture rates, which inherently leads to low signal-to-noise ratios) and environmental factors. Fundamentally, the learning process of diffusion models is akin to a denoising process \cite{ref17}, as shown in Figure~\ref{image13}, making them exceptionally effective for noise reduction in hyperspectral contexts. Different image modalities face unique noise challenges. For example, HSI recovery can be especially vulnerable to distributional biases and frequently experiences data scarcity \cite{ref92}. HSIs also have to deal with the uneven distribution of noise along the spectral dimension \cite{ref88} (e.g., severe degradation in specific water absorption bands) and the correlation of noise between different frequency bands \cite{ref89,ref90}. Fortunately, diffusion models can efficiently address these noise issues by modeling the joint spatial-spectral noise distribution rather than treating each band independently.

Ceng and Cao et al. introduced Diff-Unmix \cite{ref91} to address the issue of recovering noise-degraded HSIs by integrating spectral unmixing with conditional abundance generation. This method uses a learnable block-based spectral unmixing strategy that is backed up by a transformer-based backbone. They then add a self-supervised generative diffusion network to the abundance maps from the spectrally unmixed blocks. Because abundance maps represent the physical fractions of ground materials, performing diffusion in this specific domain introduces a strong physical constraint, ensuring that the denoising process adheres to the linear mixing model. This network reconstructs noise-free decomposition probability distributions and effectively reduces noise-induced degradation in these components. Finally, the HSI is put back together by combining the diffusion-adjusted abundance maps with the spectral end-members, guaranteeing strict radiometric consistency. He et al. \cite{ref89} proposed a truncated diffusion model that commences denoising from an intermediate stage of the diffusion process, as opposed to originating from pure noise. The goal of this method is to keep the HSI's inherently valid information—specifically, avoiding the complete destruction of the highly correlated low-frequency spectral envelopes, which are notoriously difficult to recover from pure isotropic Gaussian noise. Deng et al. also suggested RGB image-enhanced hyperspectral Denoising Diffusion Probabilistic Models \cite{ref93}, which improved DM-based HSI denoising by using RGB-DM pre-training on extra RGB images. They also added a fusion operator to combine HSI-DM and RGB-DM, which made DM-based HSI denoising even better by utilizing the high spatial resolution of RGB images as structural guidance to compensate for the HSI's spatial limitations. Kai et al. put forth a multi-scale spatial spectral joint denoising network for hyperspectral images (MSFDN) \cite{ref165}. They built a subspace projection network to turn HSIs into a low-dimensional subspace—directly exploiting the high spectral correlation and low-rank property of HSIs—so that they could better describe and get rid of noise. They also made a noise separation network and a subspace inverse projection network that used a conditional diffusion model to make them more stable.

Overall, diffusion models offer a powerful paradigm for HSI noise suppression, demonstrating strong adaptability to diverse and physically complex noise structures while maintaining fine spectral-spatial fidelity and adhering to the intrinsic radiometric constraints of hyperspectral sensors.

\subsubsection{Data Recovery}
The goal of hyperspectral image (HSI) recovery is to make clean images from degraded observations \cite{ref121}. This is very important for later HSI uses. This recovery process has a lot of problems because it needs to make good use of the intrinsic spatial nonlocal self-similarity and spectral low-rank properties of HSIs, and there isn't enough data to do so. Conventional model-based techniques find it challenging to accurately represent intricate image characteristics with manual prior assumptions, whereas deep learning methodologies frequently encounter issues such as inadequate generalization, excessive model complexity, suboptimal outcomes, and parameter redundancy. However, diffusion models have recently garnered considerable attention for their impressive image recovery performance \cite{ref154,ref155,ref156,ref157,ref158,ref159,ref164}, particularly for their robustness to noise \cite{ref83,ref86,ref89,ref116,ref101,ref127}. It's worth noting, though, that diffusion methods are typically trained on large datasets and perform exceptionally well within those distributions, but they can be quite susceptible to distributional bias.

To tackle these issues caused by the inherent physical difficulty of acquiring clean, paired HSI datasets in real-world remote sensing, Miao et al. introduced DDS2M (Denoising Diffusion Probabilistic Models for Spatial Spectroscopy) \cite{ref92}, a self-supervised diffusion model for HSI recovery. This model operates by inferring the parameters of its proposed variational spatial spectral module (VS2M) during reverse diffusion, using only the degraded HSI without needing any additional training data. A custom loss function based on variational inference enables untrained spatial and spectral networks to learn a posteriori distributions that strictly adhere to the intrinsic spatial-spectral physical dependencies of the scene, which serve as transitions in the sampling chain to aid the reverse diffusion process. Chen et al. \cite{ref106} employ a dual window, iterative diffusion model, and reverse reconstruction to mitigate the impact of background estimation and preserve continuous spectral signatures. Li et al. proposed HIR-DIFF (HSI Recovery via Diffusion Model with Untrained Prior) \cite{ref90}, a supervised HSI recovery framework. This framework uses singular value decomposition (SVD) and rank-revealed QR (RRQR) decomposition to get clean HSIs back from the product of two low-rank components. This explicit low-rank decomposition perfectly aligns with the linear mixing physical model of HSI, ensuring that the diffusion process only operates on the fundamental spectral endmembers and spatial abundances rather than adding destructive noise to highly correlated raw bands. They also came up with a new exponential noise schedule to speed up the recovery process. Furthermore, Li et al. presented a spectral latent diffusion enhancement module \cite{ref119}, which incorporates a latent diffusion-enhanced rectangular transformer for HSI recovery. Their proposed multi-shape spatial rectangular self-attentive module leads to a tighter alignment with the desired HSI by physically decoupling the spatial degradation from the spectral manifold.

Meanwhile, in the field of data recovery, hardware-based compressive imaging architectures like the Coded Aperture Snapshot Spectral Imager (CASSI) physically compress a 3D hyperspectral cube into a 2D measurement to overcome sensor readout limits. This process introduces severe noise mixing, which has both signal-correlated and signal-uncorrelated parts, making the inverse reconstruction an extremely ill-posed physical problem that can cause problems that affect how we see things and cause severe image and spectral distortion. To solve this problem, Si et al. suggested CASSIDiff \cite{ref120} for putting CASSI metrics back together. This model uses a feature fusion mechanism based on the Discrete Wavelet Transform (DWT) and a spatial-spectral attention mechanism. Here, the diffusion model acts as a powerful generative prior that understands the intrinsic structure of hyperspectral data, effectively regularizing the under-determined physical inverse problem and making it very reliable for recovery tasks. Meng et al. put forward FastDiffSR \cite{ref157}, a method for SRSISR (Single-Image Super-Resolution) that is based on a conditional diffusion model. FastDiffSR significantly enhances the quality and clarity of image generation by combining fewer sampling steps, residual images, and channel and spatial attention mechanisms, effectively hallucinating missing spatial high-frequencies without compromising the low-frequency spectral continuity.

For the SGM/SDE diffusion model, Andres et al. \cite{ref187} suggested merging compressed sensing with diffusion generative models to reconstruct high-resolution satellite lidar data within the hyper-high-density cube (HHDC) framework. Given the massive physical data volume characteristic of hyperspectral and 3D remote sensing cubes, this approach facilitates efficient sampling and compression, alleviates airborne computing burdens, and enhances data transmission by shifting the heavy reconstruction burden to the diffusion-based ground decoder. Yi et al. introduced an efficient diffusion model (EDiffSR) \cite{ref158} for RSI (Remote Sensing Image) Super-Resolution. They developed an efficient activation network (EANet) and a practical conditional prior enhancement module (CPEM), which collectively reduce the computational budget while retaining more physically meaningful spatial-spectral information.

In conclusion, the diffusion-based HSI recovery model offers promising solutions for denoising, resolution enhancement, and solving severely ill-posed physical inverse problems. By balancing performance and complexity through innovations in attention mechanisms, physically meaningful latent representations, and self-supervised optimization, these models effectively bridge the gap between advanced generative AI and the strict radiometric requirements of hyperspectral imaging.

\begin{figure*}[pos=htbp]
\centering
\includegraphics[width=1.0\textwidth]{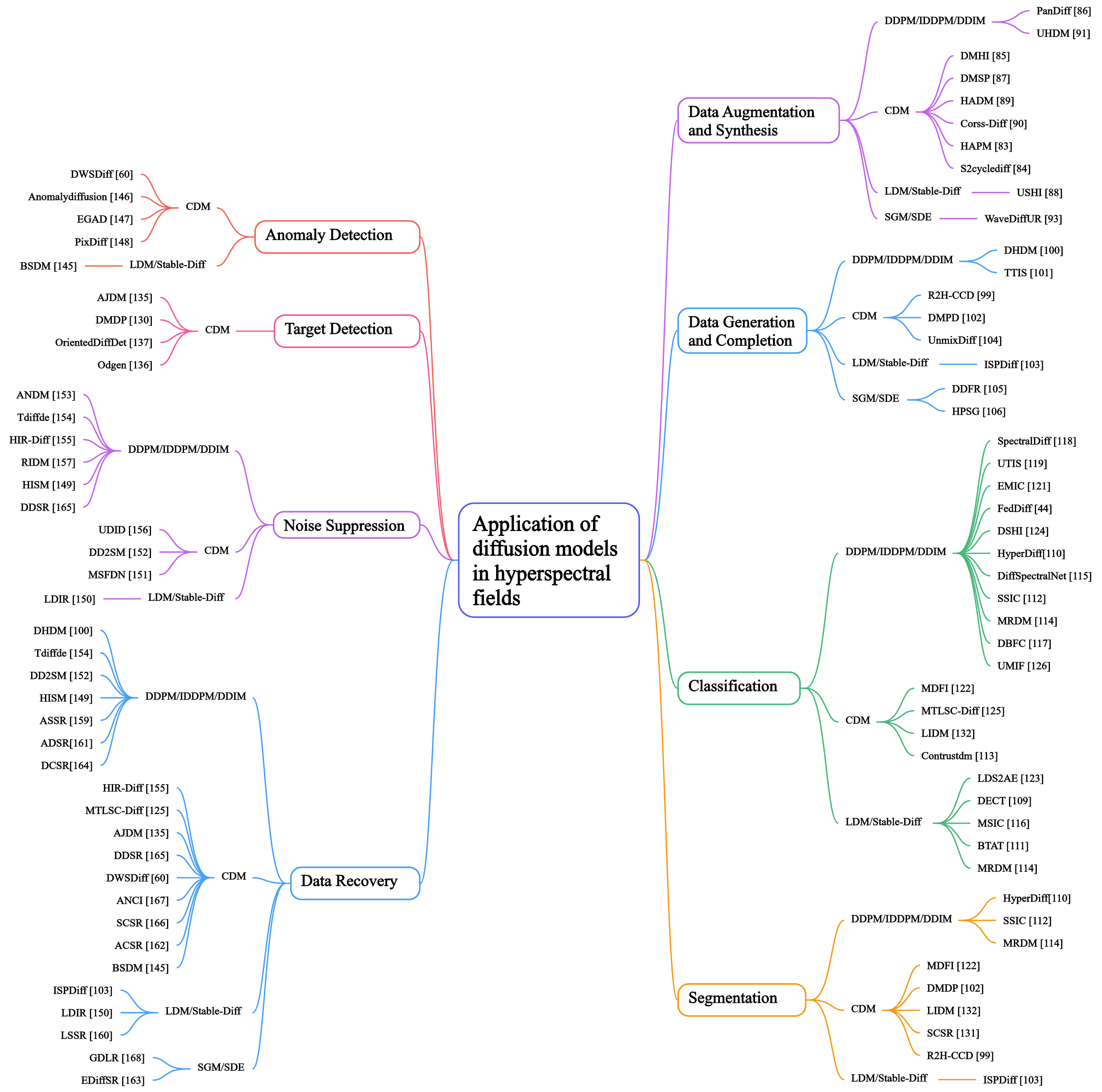}%
\hfil
\caption{Diffusion model-based research progress in hyperspectral image processing and analysis the tasks are established in nine subfields: (1) data augmentation and synthesis, (2) data generation and complementation, (3) classification, (4) segmentation, (5) anomaly detection, (6) target detection, (7) noise suppression, and (8) data recovery. No abbreviated name, with two initial letters at the beginning and end}
\label{image14}
\end{figure*}

As show Figure~\ref{image14}, presents a summarized and categorized overview of the application methods from the aforementioned research directions.

\section{Comparative and Experimental analysis}
Diffusion models are theoretically applicable to eight different domains, but their performance improvements in practical applications are typically measured against five core end-to-end tasks. Tasks such as data augmentation, synthesis, and generation are primarily validated as intermediate technical methods through downstream tasks, whereas image segmentation is fundamentally evaluated as pixel-level classification. In order to effectively illustrate the superiority of the diffusion model in processing hyperspectral images, this paper takes the experimental results of classification, target detection, anomaly detection, image recovery, and noise suppression as an example, and evaluates the performance of the diffusion model and other existing techniques through the visualization results and quantitative indexes.

\subsection{Experimental Results and Performance Assessment with Various Metrics}
\subsubsection{Evaluation Metrics and Datasets}
Several metrics are applied to evaluate performance in hyperspectral tasks. The receiver operating characteristic (ROC) curve gives a qualitative view of how well a model works. On this curve, the false positive rate (FPR) is plotted along the horizontal axis, and the true positive rate (TPR) is plotted along the vertical axis. The area under the ROC curve (AUC) is then calculated as a single numeric score, providing a quantitative measure of effectiveness. In practical use, AUC is one of the most common indicators for comparing the accuracy of different methods.

The larger the AUC value, the better the detection performance of the algorithm, with mAUC representing the average AUC. where larger ${\rm AUC}_{(D,F)}$and ${\rm AUC}_{(D,\tau)}$ indicate better detection performance, and are combined together defined as ${\rm AUC}_{TD}$; ${\rm AUC}_{BS}$ represents the ability of the detector to minimize false alarms due to background interference; ${\rm AUC}_{ODP}$ denotes the overall probability of detection; ${\rm AUC}_{TD-BS}$ specifically evaluates the performance of the detector in recognizing a targets in background noise, taking into account both detectability and background suppression; ${\rm AUC}_{SNPR}$ reflects the detector's ability to distinguish between signal and noise.

Classification metrics include OA denotes the ratio of the number of correctly classified samples to the total number of samples in all test samples; AA denotes the average of the classification accuracy for each category; k × 100 is the Kappa coefficient is a measure of the consistency of the classifier, which takes into account stochastic consistency and provides a measure of the consistency between the classifier and the ground truth.

Target detection metrics include IoU is a measure of the degree of overlap between the predicted frame and the real frame; $\mathbf{P}_{d}$ is defined as the ratio of the number of correctly detected targets to the total number of targets; and $\mathbf{F}_{a}$ is defined as the ratio of the number of false alarms to the total number of pixels in the image.

Data recovery and noise suppression metrics include PSNR is the Peak Signal to Noise Ratio, a measure of the difference in quality between a super-resolution processed image and the original high-resolution image. The Fréchet Inception Distance (FID) is a way to measure how similar the generated images are to each other and how good they are. Patch Similarity evaluates the perceptual difference between the generated image and the original high-resolution image. parmas is the number of parameters for model training. entropy is usually used to measure the uncertainty and complexity of image information. niqe is the Natural Image Quality Evaluator, which evaluates the image quality metrics. SSIM (Structural Similarity) measures the similarity between two images, comparing images in terms of brightness, contrast and structure; FSIM (Feature Similarity) evaluates the image quality by calculating the phase coherence and gradient similarity in localized regions of the image, and is an improved version of SSIM; SAM (Sum of Absolute Errors) is another measure of the effectiveness of image denoising, which calculates the sum of the absolute value of the pixel differences between a denoised image and the original image counterparts.

\begin{table*}
\centering
\caption{Description of benchmark datasets used for evaluation}
\label{tab:Summary}
\begin{tabular}{cccccc}
\hline
\textbf{Datasets} & \textbf{Sensor} & \textbf{Spectral Band} & \textbf{Size} & \textbf{Resolution} & \textbf{Category} \\ \hline
Salinas & AVIRIS & 204 & 512$\times$217 & 3.7 m & Classification \\ 
Pavia University & ROSIS & 103 & 610$\times$340 & 1.3 m & Classification \\ 
Indian Pines & AVIRIS & 200 & 145$\times$145 & 20 m & Classification \\ 
SIRST & FLIR & LWIR & 512$\times$512 & 0.5 m & Target Detection \\ 
NUDT-SIRST & Infrared Sensor & LWIR & 640$\times$512 & 0.3 m & Target Detection \\ 
Airport-Beach-Urban & AVIRIS & 205/188/191 & 150$\times$150 & 20 m & Anomaly Detection \\ 
Potsdam & ISPRS & RGB(3) & 5000$\times$5000 & 5 cm & Image Recovery \\ 
Toronto & Toronto & RGB(3) & 1920$\times$1080 & 0.2 m & Image Recovery \\ 
UC Merced & USGS & RGB(3) & 256$\times$256 & 0.3 m & Image Recovery \\ 
CAVE & Hyperspectral Camera & 31 & 512$\times$512 & N/A & Noise Supression \\ 
CAVE-Toy & Hyperspectral Camera & 31 & 512$\times$512 & N/A & Noise Supression \\ 
KAIST & Thermal Camera & LWIR & 640$\times$480 & 0.5 m & Noise Supression \\ \hline
\end{tabular}
\end{table*}

Meanwhile, Table~\ref{tab:Summary} summarizes widely used benchmark datasets, but their representativeness and limitations deserve further attention. These datasets cover several application domains, such as classification (e.g., Salinas, Pavia University, Indian Pines), target detection (e.g., SIRST, NUDT-SIRST), anomaly detection (e.g., Airport-Beach-Urban), and noise suppression (e.g., CAVE, KAIST). This variety allows algorithms to be tested on different tasks in hyperspectral and multispectral image analysis. From the spectral perspective, the coverage extends from more than 200-band hyperspectral data (AVIRIS) to three-channel RGB imagery (e.g., Potsdam, Toronto, UC Merced). Such differences are useful for evaluating robustness across spectral resolutions, but they may also cause bias because RGB images cannot capture the subtle spectral signatures that hyperspectral sensors provide. Adding LWIR thermal datasets (SIRST, KAIST) makes the diversity of modalities even greater, but their low spatial resolution makes them less useful for studies that need to be very detailed.

The representativeness of the dataset is still somewhat limited in terms of geography and sensor types. Several hyperspectral datasets (e.g., Indian Pines, Salinas) were collected from agricultural or rural areas in North America, which made their generalization poor in urban or heterogeneous regions. On the other hand, datasets like those from the University of California, Merced and Toronto offer high spatial detail but only limited spectral information because they are based on RGB images. The difference in spatial resolution - from sub-meters (Toronto, SIRST) to tens of meters (AVIRIS, ISPRS) - further complicates cross-dataset comparisons, as algorithm performance often varies with the resolution. Small-scale datasets such as Airport-Beach-Urban (150×150) pose another challenge because they have the risk of overfitting and are unable to capture the variability of real-world conditions.

Several potential sources of error and deviation should also be noted. These include sensor noise, atmospheric influences in on-board data (e.g., AVIRIS), and uncertainties in annotations, especially for marking potentially subjective anomalies and target detection. Noise suppression datasets (e.g., CAVE, KAIST) are typically created under controlled or synthetic conditions, which may not well reflect the operational scenarios. In conclusion, although these datasets together provide a useful testing platform for different tasks and patterns, their limitations in terms of range, resolution, and spectral coverage need to be carefully considered when interpreting the results and drawing broader conclusions.

\subsubsection{Experimental Results}
\begin{figure}[pos=htbp]
\centering
\includegraphics[width=0.45\textwidth]{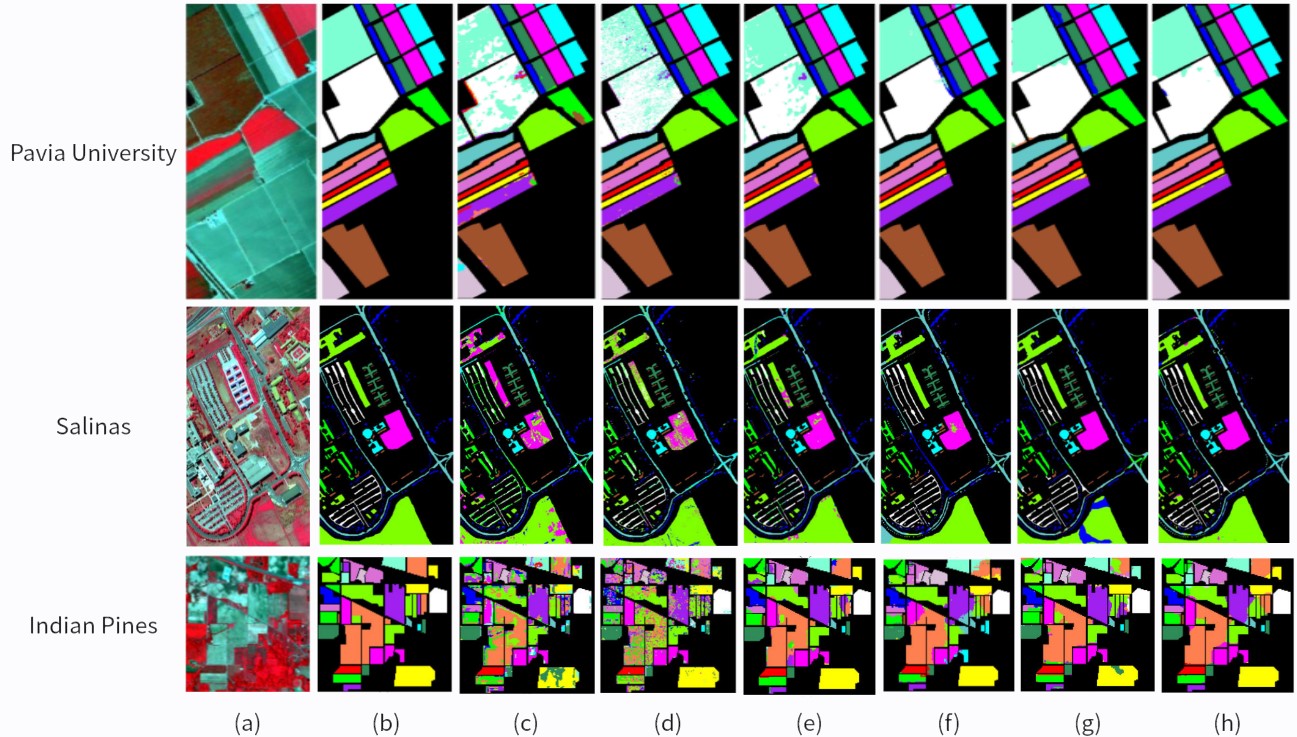}%
\hfil
\caption{Classification results on multiple datasets using different methods: (a) Pseudo-color image (b) Real labeled image (c) SF \cite{ref133} (d) miniGCN \cite{ref134} (e) SSFTT \cite{ref135} (f) DMVL \cite{ref136} (g) SSGRN \cite{ref137} (h) SpectralDiff \cite{ref94}}
\label{image15}
\end{figure}

Figure~\ref{image15} shows that the first three methods do a very bad job of classifying things, as they mix up colors a lot between different classes. The last three methods, on the other hand, work better for classifying things in general. The method based on the diffusion model, SpectralDiff, gives the best classification results, and DMVL is very close behind. Compared to other methods, both SpectralDiff and DMVL do a better job of defining boundaries with little color distortion. There are some small problems with SpectralDiff's classification in small areas, but overall it works much better than other common methods.

\begin{table*}[htbp]
\centering
\setlength{\tabcolsep}{3pt}
\caption{Comparison of Metrics for Different Classification Methods on Indian Pines/Pavia University/Salinas Dataset}
\label{tab:classification1}
\begin{tabular}{ccccccccccccc}
\hline
\multirow{2}{*}{\textbf{Methods}} & \multicolumn{4}{c}{\textbf{Evaluation metrics (Indian Pines)}} & \multicolumn{4}{c}{\textbf{Evaluation metrics (Pavia University)}} & \multicolumn{4}{c}{\textbf{Evaluation metrics (Salinas)}} \\ \cline{2-13}
& \textbf{OA} & \textbf{AA} & \textbf{K$\times$100} & \textbf{Time(s)} & \textbf{OA} & \textbf{AA} & \textbf{K$\times$100} & \textbf{Time(s)} & \textbf{OA} & \textbf{AA} & \textbf{K$\times$100} & \textbf{Time(s)} \\ \hline
SF & 67.782 & 80.639 & 63.820 & 29.81 & 81.511 & 84.942 & 75.795 & 163.34 & 88.248 & 93.263 & 86.973 & 141.30 \\ 
miniGCN & 63.916 & 69.091 & 59.439 & \textcolor{red}{\textbf{0.95}} & 82.355 & 84.037 & 76.620 & \textcolor{red}{\textbf{2.40}} & 88.181 & 94.279 & 86.823 & \textcolor{red}{\textbf{2.13}} \\ 
SSFTT & 91.566 & 95.661 & 90.392 & 31.58 & 93.667 & 93.448 & 91.617 & 110.08 & 95.789 & 98.272 & 95.322 & 55.96 \\ 
DMVL & 90.487 & 81.311 & 89.216 & --- & 89.768 & 86.116 & 86.580 & --- & 97.005 & 95.853 & 96.668 & --- \\ 
SSGRN & 92.336 & 96.099 & 91.254 & 1.13 & 93.850 & \textcolor{red}{\textbf{94.436}} & 91.919 & 4.99 & 96.539 & 96.354 & 96.144 & 2.98 \\ 
SpectralDiff & \textcolor{red}{\textbf{93.146}} & \textcolor{red}{\textbf{96.438}} & \textcolor{red}{\textbf{92.175}} & 48.53 & \textcolor{red}{\textbf{94.775}} & 93.843 & \textcolor{red}{\textbf{93.061}} & 256.23 & \textcolor{red}{\textbf{98.971}} & \textcolor{red}{\textbf{99.465}} & \textcolor{red}{\textbf{98.854}} & 242.16 \\ \hline
\end{tabular}
\end{table*}

SpectralDiff, based on diffusion models, demonstrates clear superiority over other methods across all datasets in Table~\ref{tab:classification1}. It achieves the highest OA, AA, and Kappa scores, highlighting the strong capability of diffusion models in capturing complex spectral–spatial features. Although it requires more computation time, its accuracy significantly outperforms both traditional and deep learning baselines such as SF, miniGCN, and SSGRN. These findings demonstrate that diffusion models provide a stronger and more reliable framework for hyperspectral image classification tasks. 

\begin{figure}[pos=htbp]
\centering
\includegraphics[width=0.45\textwidth]{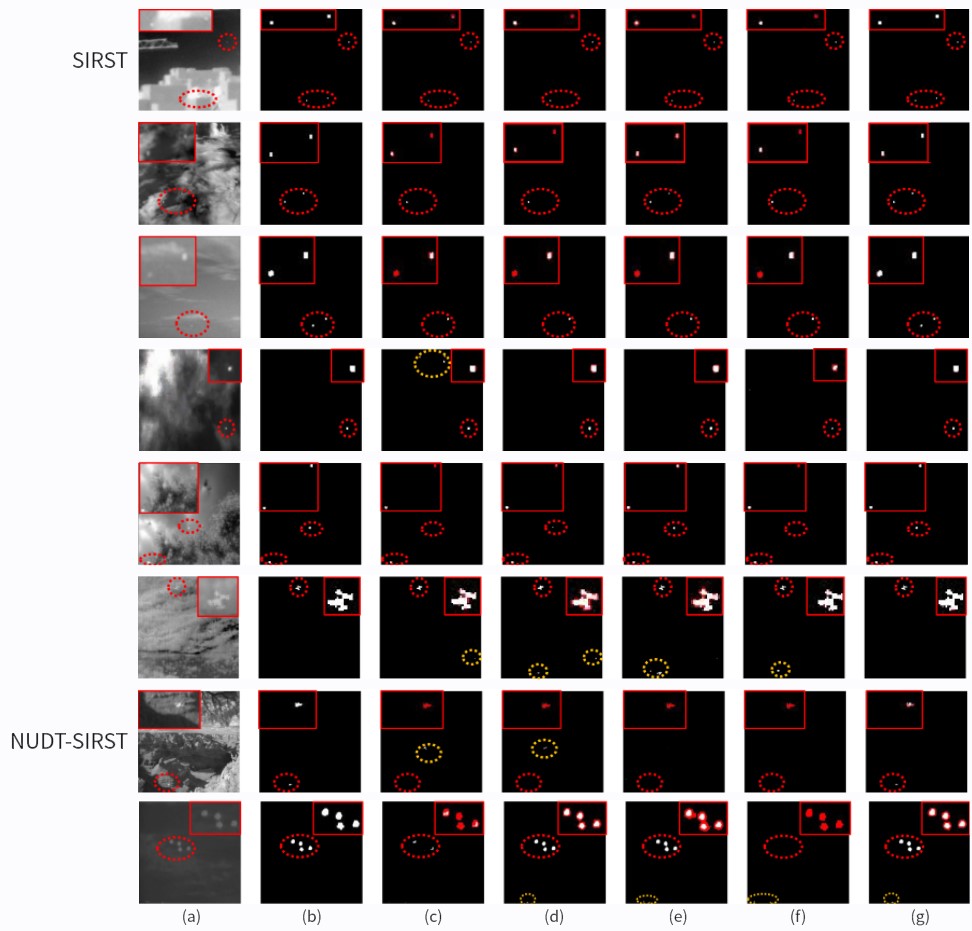}%
\hfil
\caption{Target detection results on various datasets with competing methods: (a) Original image (b) Real labeled image (c) ALC-Net \cite{ref138} (d) ACM \cite{ref139} (e) UIU-Net \cite{ref140} (f) DNA-Net \cite{ref141} (g) Diff-Mosaic \cite{ref113}}
\label{image16}
\end{figure}

As illustrated in Figure~\ref{image16}, mainstream target detection methods generally perform well. Some methods still suffer from missed detections or false positives. In contrast, the diffusion-based approach Diff-Mosaic achieves superior results, combining high accuracy with the most precise boundary delineation of detected targets.
It is largely consistent with the true labeling map, with virtually no missing components.

\begin{table*}[htbp]
\centering
\setlength{\tabcolsep}{2.5pt}
\caption{Comparison of Metrics for Different Target Detection Methods on SIRST/NUDT-SIRST Dataset}
\label{tab:target1}
\begin{tabular}{ccccccccc}
\hline
\multirow{2}{*}{\textbf{Methods}} & \multirow{2}{*}{\textbf{Calculate the number of parameters}} & \multirow{2}{*}{\textbf{Time (s)}} & \multicolumn{3}{c}{\textbf{Evaluation metrics (SIRST)}} & \multicolumn{3}{c}{\textbf{Evaluation metrics (NUDT-SIRST)}} \\ \cline{4-9}
& & & \textbf{IoU} & \textbf{$P_d$} & \textbf{$F_a$} & \textbf{IoU} & \textbf{$P_d$} & \textbf{$F_a$} \\ \hline
ALC-Net & 0.38 & 40.93 & 70.00 & 71.71 & 23.91 & 72.13 & 78.81 & 21.22 \\ 
ACM & \textcolor{red}{\textbf{0.29}} & \textcolor{red}{\textbf{18.53}} & 76.17 & 86.31 & 16.07 & 71.11 & 85.13 & 22.51 \\ 
UIU-Net & 4.7 & 43.42 & 72.03 & 98.10 & 26.15 & 89.00 & 98.73 & 6.02 \\ 
DNA-Net & 50.54 & 33.98 & 76.97 & 95.41 & 3.54 & 88.38 & 97.99 & 4.04 \\ 
Diff-Mosaic & 4.7 & 43.42 & \textcolor{red}{\textbf{79.44}} & \textcolor{red}{\textbf{99.99}} & \textcolor{red}{\textbf{3.19}} & \textcolor{red}{\textbf{91.18}} & \textcolor{red}{\textbf{99.47}} & \textcolor{red}{\textbf{1.91}} \\ \hline
\end{tabular}
\end{table*}

Meanwhile, Diff-Mosaic, a diffusion model-based method, achieves the best performance among all evaluated target detection methods on both SIRST and NUDT-SIRST datasets, as shown in Table~\ref{tab:target1}. It obtains the highest IoU (79.44\% and 91.18\%), precision (P<sub>d</sub>: 99.99\% and 99.47\%), and lowest false alarm rates ($F_a$: 3.19 and 1.91), significantly outperforming previous state-of-the-art methods such as DNA-Net and UIU-Net. Although its runtime (43.42s) is relatively high and the parameter count is moderate (4.7M), the gains in accuracy and reliability justify the computational cost. These results clearly demonstrate the strong capability and robustness of diffusion models in small target detection tasks.

\begin{figure}[pos=htbp]
\centering
\includegraphics[width=0.45\textwidth]{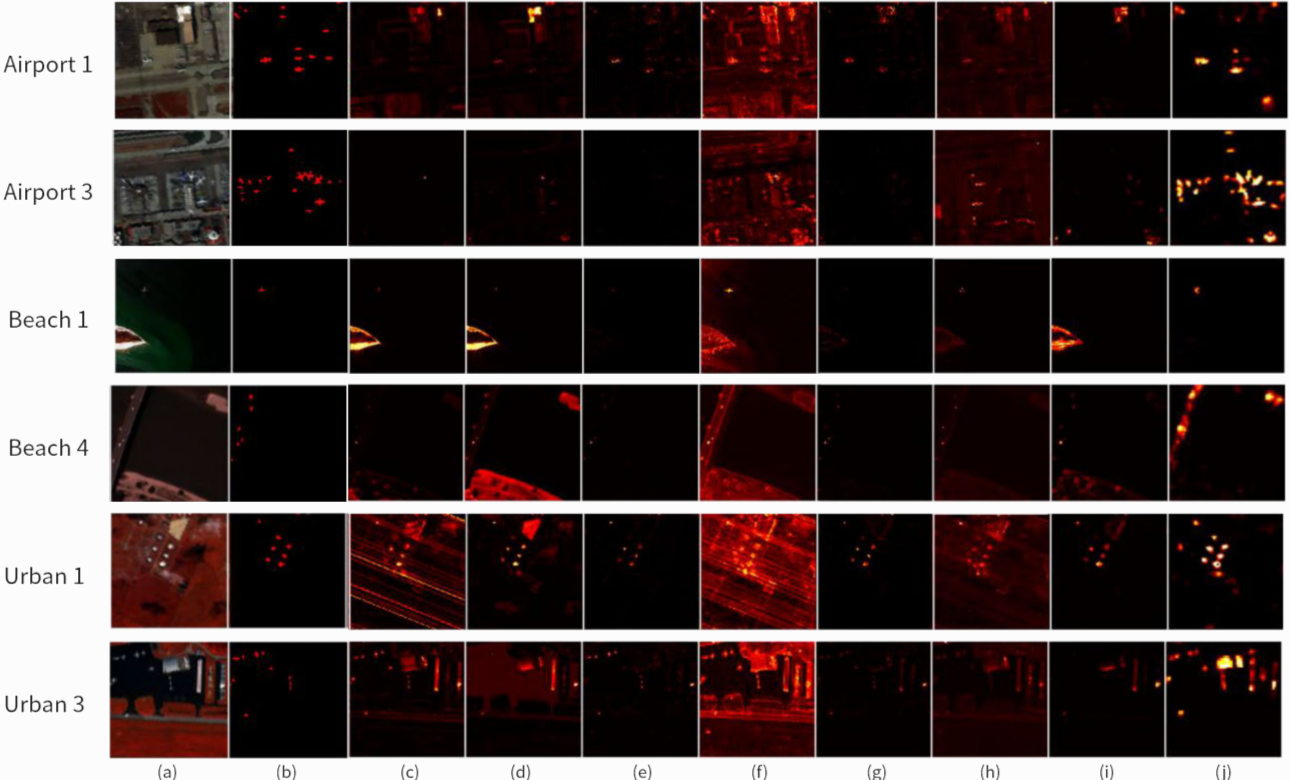}%
\hfil
\caption{Comparative anomaly detection performance across hyperspectral datasets: (a) Pseudo-color map (b) Real label map (c) GAED \cite{ref142} (d) RGAE \cite{ref143} (e) Auto-AD \cite{ref144} (f) DFAN-HAD \cite{ref145} (g) DirectNet \cite{ref146} (h) BSDM \cite{ref105} (i) TGFA \cite{ref147} (j)DWSDiff \cite{ref106}}
\label{image17}
\end{figure}

Figure~\ref{image17} shows that when we compare the results of anomaly detection experiments, the diffusion model-based methods, BSDM and DWSDiff, do a better job of finding anomalies than the traditional mainstream methods. They have a high detection rate and a low false detection rate, which is mostly in line with the true labeling map. On the other hand, other commonly used detection methods are ineffective, with low detection rates (failing to identify existing anomalies) and high false positive rates.

\begin{table*}[htbp]
\centering
\caption{Comparison of metrics for each anomaly detection methods}
\label{tab:anomaly1}
\begin{tabular}{ccccccccc}
\hline
\multirow{2}{*}{\textbf{Methods}} & \multicolumn{8}{c}{\textbf{Evaluation metrics}} \\ \cline{2-9} 
& $mAUC_{(D,F)}$ & $mAUC_{(D,\tau)}$ & $mAUC_{TD}$ & $mAUC_{(F,\tau)}$ & $mAUC_{BS}$ & $mAUC_{ODP}$ & $mAUC_{TD-BS}$ & $mAUC_{SNPR}$ \\ \hline
GAED & 0.90756 & 0.15556 & 0.02072 & 1.06312 & 0.8868 & 1.04240 & 0.13484 & 13.79983 \\ 
RGAE & 0.87615 & 0.14519 & 0.02401 & 1.02135 & 0.8521 & 0.99734 & 0.12119 & 14.38906 \\ 
Auto-Net & 0.95028 & 0.15303 & 0.00627 & 1.10331 & 0.9440 & 1.09704 & 0.14676 & 52.05284 \\ 
DFAN-HAD & 0.96552 & 0.45211 & \textcolor{red}{\textbf{0.09603}} & 1.41762 & 0.8694 & 1.32160 & 0.35608 & 5.27753 \\ 
DirectNet & 0.95193 & 0.12799 & 0.00484 & 1.07993 & 0.9470 & 1.07508 & 0.12315 & 49.52139 \\ 
BSDM & 0.94486 & 0.17342 & 0.03055 & 1.11828 & 0.9143 & 1.08773 & 0.14287 & 8.40576 \\ 
TGFA & 0.92655 & 0.14559 & 0.00629 & 1.07214 & 0.9292 & 1.06585 & 0.13830 & 32.42845 \\ 
DWSDiff & \textcolor{red}{\textbf{0.97427}} & \textcolor{red}{\textbf{0.55329}} & 0.01635 & \textcolor{red}{\textbf{1.52756}} & \textcolor{red}{\textbf{0.9579}} & \textcolor{red}{\textbf{1.51121}} & \textcolor{red}{\textbf{0.53694}} & \textcolor{red}{\textbf{67.20618}} \\ \hline
\end{tabular}
\end{table*}

Table~\ref{tab:anomaly1} shows that the diffusion model-based methods BSDM and DWSDiff work better than most traditional and deep learning-based methods for finding anomalies. DWSDiff is the best overall performer on almost all measures. It has the best scores in $mAUC_{(D,F)}$ (0.97427), $mAUC_{(D,\tau)}$ (0.55329), $mAUC_{TD}$ (0.01635), and $mAUC_{(F,\tau)}$ (67.20618). These scores are all much higher than those of other methods. DWSDiff is better than BSDM at both detecting and being robust across datasets, even though BSDM also performs well. These results show that diffusion models, especially DWSDiff, are better at modeling for finding anomalies in complex hyperspectral scenes.

\begin{table}[htbp]
\centering
\setlength{\tabcolsep}{3pt}
\caption{Comparison of Reasoning Time for Each Anomaly Detection Method}
\label{tab:anomaly4}
\begin{tabular}{ccccc}
\hline
\multirow{2}{*}{\textbf{Methods}} & \multicolumn{4}{c}{\textbf{Reasoning time (s)}} \\ \cline{2-5} 
& \textbf{Airport} & \textbf{Beach} & \textbf{Urban} & \textbf{Average Time (s)} \\ \hline
GAED & 0.0300 & 0.0496 & 0.0257 & 0.0289 \\ 
RGAE & 0.0135 & 0.0251 & 0.0137 & 0.0141 \\ 
Auto-Net & \textcolor{red}{\textbf{0.0020}} & \textcolor{red}{\textbf{0.0032}} & \textcolor{red}{\textbf{0.0020}} & \textcolor{red}{\textbf{0.0022}} \\ 
DFAN-HAD & 0.0232 & 0.0448 & 0.0269 & 0.0264 \\ 
DirectNet & 0.0145 & 0.0283 & 0.0153 & 0.0145 \\ 
BSDM & 1.2306 & 1.2740 & 1.2509 & 1.2421 \\ 
TGFA & 1.6190 & 1.5505 & 1.5236 & 1.5392 \\ 
DWSDiff & 0.8016 & 0.8169 & 0.8192 & 0.8094 \\ \hline
\end{tabular}
\end{table}

Table~\ref{tab:anomaly4} shows that both diffusion models, BSDM and DWSDiff, take a lot longer and slower to make inferences on different medium-sized datasets. Their average times are 99.8\% (BSDM) and 99.7\% (DWSDiff) longer than the shortest time Auto-Net can do, which means they are hundreds of times slower than the best. This shows that computational efficiency is a big problem. Even so, if the main goal is to improve anomaly detection performance, DWSDiff is still the best choice, and BSDM is also a strong candidate.

\begin{figure}[pos=htbp]
\centering
\includegraphics[width=0.45\textwidth]{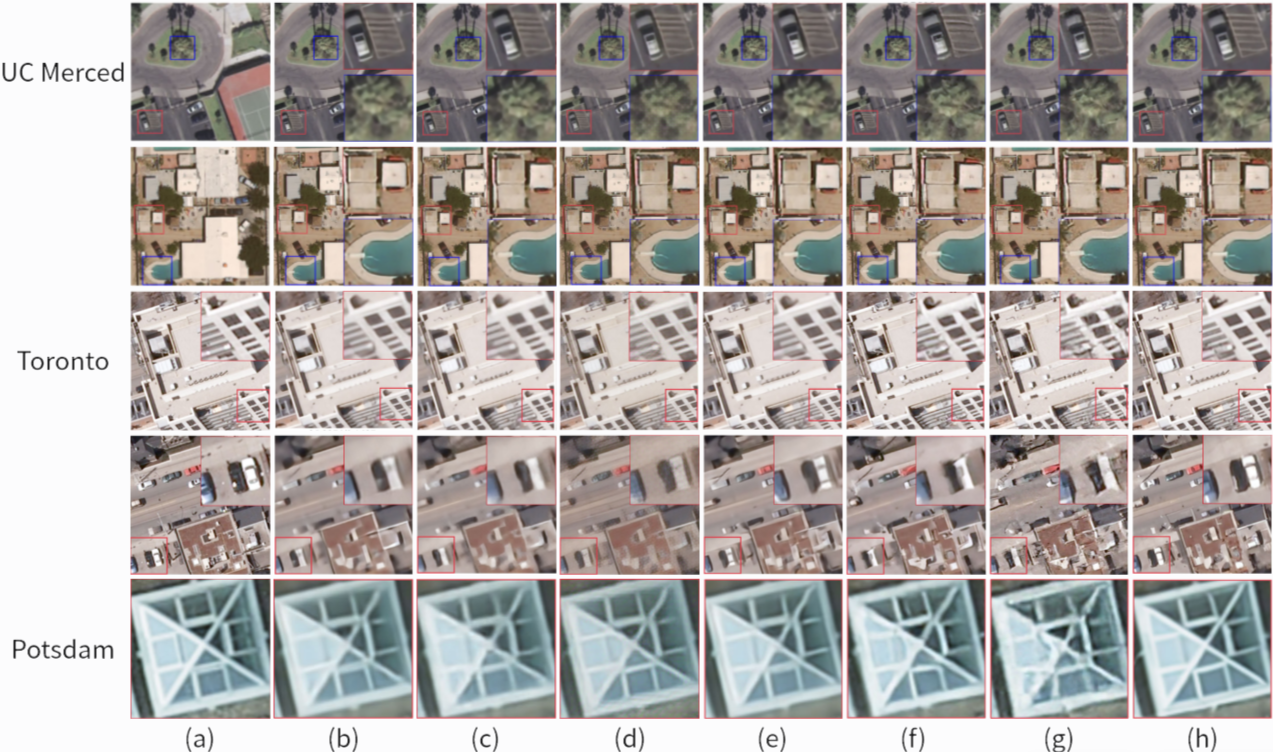}%
\hfil
\caption{Visual comparison of hyperspectral image recovery methods: (a) High resolution map (b) HESNet \cite{ref160} (c) TransENet \cite{ref161} (d) NDSRGAN \cite{ref162} (e) HAT \cite{ref163} (f) DDPM \cite{ref17} (g) EDiffSR \cite{ref158} (h) FastDiffSR \cite{ref157}}
\label{image18}
\end{figure}

Figure~\ref{image18} shows that the overall image recovery effects are not very strong for any of the methods. But diffusion model-based methods (DDPM, EDiffSR, FastDiffSR) do a better job of handling details, which leads to a recovery quality that is much closer to the high-resolution ground truth.

\begin{table*}[htbp]
\centering
\caption{Comparison of Metrics for Different Image Recovery Methods on Potsdam Dataset}
\label{tab:recovery1}
\begin{tabular}{ccccccccccc}
\hline
\multirow{2}{*}{\textbf{Methods}} & \multicolumn{4}{c}{\textbf{Evaluation metrics(Potsdam)}} & \multicolumn{4}{c}{\textbf{Evaluation metrics (Toronto)}} & \multicolumn{2}{c}{\textbf{Measurement metrics}} \\ \cline{2-11} 
& \textbf{PSNR} & \textbf{FID} & \textbf{LPIPS} & \textbf{Params} & \textbf{PSNR} & \textbf{FID} & \textbf{LPIPS} & \textbf{Params} & \textbf{Entropy} & \textbf{NIQE} \\ \hline
HESNet & 32.932 & 44.192 & 0.3014 & \textcolor{red}{\textbf{5.4330}} & 25.413 & 118.791 & 0.3234 & \textcolor{red}{\textbf{5.5070}} & 7.0826 & 19.913 \\ 
TransENet & 32.768 & 41.310 & 0.2878 & 37.459 & \textcolor{red}{\textbf{25.565}} & 97.958 & 0.3112 & 37.533 & 7.0853 & 20.319 \\ 
NDSRGAN & 30.529 & 51.209 & 0.2137 & 17.510 & 23.237 & 97.922 & 0.2534 & 17.510 & 7.0528 & 19.214 \\ 
HAT & 32.987 & 44.932 & 0.2962 & 25.821 & 25.368 & 119.05 & 0.2957 & 25.821 & 7.1177 & 20.385 \\ 
DDPM & 31.989 & 24.307 & \textcolor{red}{\textbf{0.0997}} & 26.166 & 24.733 & 66.583 & 0.1759 & 26.166 & 7.1204 & 16.532 \\ 
EDiffSR & 30.860 & 36.669 & 0.1104 & 26.790 & 23.990 & 81.412 & 0.1626 & 26.790 & \textcolor{red}{\textbf{7.1838}} & 17.641 \\ 
FastDiffSR & \textcolor{red}{\textbf{33.130}} & \textcolor{red}{\textbf{22.832}} & 0.1152 & 23.229 & 25.527 & \textcolor{red}{\textbf{55.761}} & \textcolor{red}{\textbf{0.1561}} & 23.229 & 7.1254 & \textcolor{red}{\textbf{16.342}} \\ \hline
\end{tabular}
\end{table*}

Table~\ref{tab:recovery1} shows that Diffusion models are the best at remote sensing reconstruction. FastDiffSR gets the best PSNR (33.130) and FID (22.832) on Potsdam, and it also gets the best FID (55.761) and LPIPS (0.1561) on Toronto. DDPM is the best at perceptual quality, with the best Potsdam LPIPS (0.0997). EDiffSR's entropy standard is 7.1838, which is 0.89\% better than other choices. Compared to EDiffSR, FastDiffSR cuts Toronto FID by 31.6\%. Together, the models do well on 9 out of 10 key metrics. These changes show that diffusion mechanisms are better at keeping information while still looking good for remote sensing applications.

\begin{figure}[pos=htbp]
\centering
\includegraphics[width=0.45\textwidth]{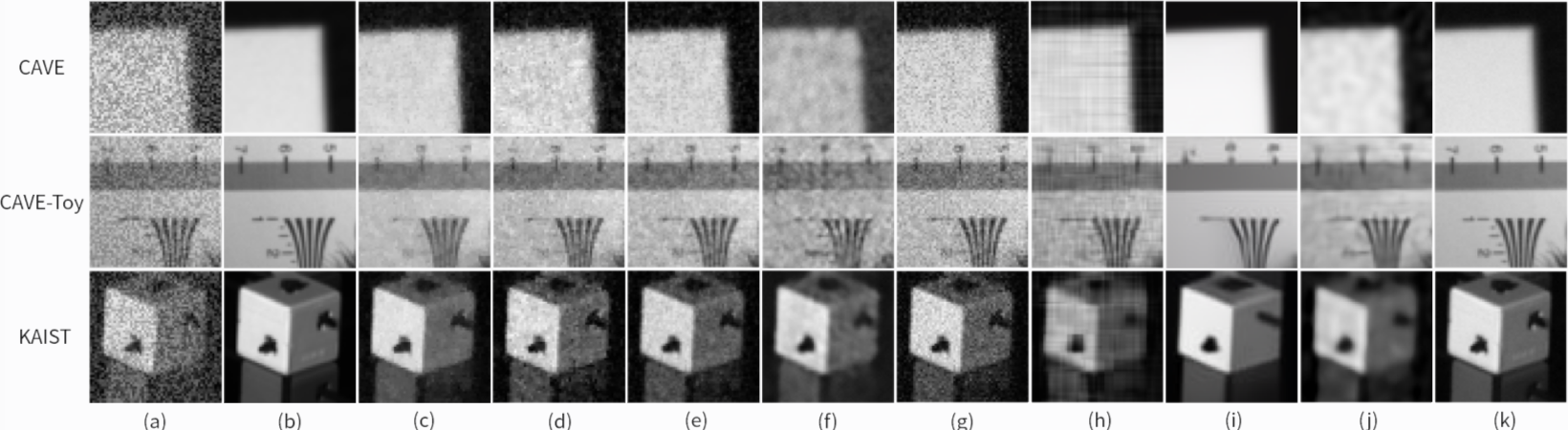}%
\hfil
\caption{Performance comparison of noise supression methods on hyperspectral images: (a) Noise map (b) True noise-free map (c) LLxRGTV \cite{ref174} (d) FGSLR \cite{ref175} (e) LRTDCTV \cite{ref176} (f) DIP \cite{ref177} (g) LLnRPnP \cite{ref178} (h) HLRTF \cite{ref124} (i) DDRM \cite{ref173} (j)DDS2M \cite{ref92} (k)Diff-Unmix \cite{ref91}}
\label{image19}
\end{figure}

As illustrated in Figure~\ref{image19}, when comparing the noise map and the true noiseless map, the Diff-Unmix method demonstrates the best overall performance in noise suppression, followed by DDRM and then DDS2M. All of these are diffusion model-based approaches, and they achieve notably successful noise suppression. In contrast, traditional mainstream methods exhibit poorer effects, with LLxRGTV being the best among them.

\begin{table*}[htbp]
\centering
\caption{Comparison of Metrics for Different Noise Suppression Methods on KAIST Dataset}
\label{tab:supression1}
\begin{tabular}{ccccccccccc}
\hline
\multirow{2}{*}{\textbf{Methods}} & \multicolumn{5}{c}{\textbf{Evaluation metrics(KAIST Noise: N(0,0,3))}} & \multicolumn{5}{c}{\textbf{Evaluation metrics(KAIST Noise: N(0,0,2))}}\\ \cline{2-11} 
& \textbf{PSNR} & \textbf{SSIM} & \textbf{FSIM} & \textbf{SAM} & \textbf{Time (s)} & \textbf{PSNR} & \textbf{SSIM} & \textbf{FSIM} & \textbf{SAM} & \textbf{Time (s)} \\ \hline
LLxRGTV & 27.640 & 0.679 & 0.868 & \textcolor{red}{\textbf{0.218}} & 38 & 31.152 & 0.802 & 0.917 & 0.205 & 38 \\ 
FGSLR & 25.561 & 0.480 & 0.718 & 0.470 & 499 & 30.126 & 0.737 & 0.878 & 0.262 & 249 \\ 
LRTDCTV & 24.593 & 0.533 & 0.739 & 0.431 & 42 & 25.952 & 0.658 & 0.816 & 0.406 & 43 \\ 
DIP & 20.063 & 0.405 & 0.798 & 0.538 & 74 & 24.181 & 0.608 & 0.825 & 0.475 & 72 \\ 
LLnRPhP & 25.102 & 0.592 & 0.768 & 0.425 & 289 & 28.664 & 0.748 & 0.861 & 0.379 & 240 \\ 
HLRTF & 30.340 & 0.689 & 0.874 & 0.329 & 25 & 33.011 & 0.808 & 0.925 & 0.275 & \textcolor{red}{\textbf{23}} \\ 
DDRM & 27.810 & 0.786 & 0.893 & 0.387 & \textcolor{red}{\textbf{23}} & 29.412 & 0.865 & 0.922 & 0.293 & 20 \\ 
DDS2M & 30.078 & 0.666 & 0.834 & 0.355 & 318 & 32.804 & 0.786 & 0.895 & 0.334 & 354 \\ 
Diff-Unmix & \textcolor{red}{\textbf{31.408}} & \textcolor{red}{\textbf{0.902}} & \textcolor{red}{\textbf{0.958}} & 0.282 & 42 & \textcolor{red}{\textbf{33.059}} & \textcolor{red}{\textbf{0.964}} & \textcolor{red}{\textbf{0.940}} & \textcolor{red}{\textbf{0.116}} & 37 \\ \hline
\end{tabular}
\end{table*}

\begin{table*}[htbp]
\centering
\caption{Comparison of Metrics for Different Noise Suppression Methods on CAVE Dataset}
\label{tab:supression3}
\begin{tabular}{ccccccccccc}
\hline
\multirow{2}{*}{\textbf{Methods}} & \multicolumn{5}{c}{\textbf{Evaluation metrics (CAVE)}} & \multicolumn{5}{c}{\textbf{Evaluation metrics (CAVE-Toy)}} \\ \cline{2-11}
& \textbf{PSNR} & \textbf{SSIM} & \textbf{FSIM} & \textbf{SAM} & \textbf{Time (s)} & \textbf{PSNR} & \textbf{SSIM} & \textbf{FSIM} & \textbf{SAM} & \textbf{Time (s)} \\ \hline
LLxRGTX & 28.743 & 0.712 & 0.829 & \textcolor{red}{\textbf{0.121}} & 40 & 26.450 & 0.752 & 0.874 & 0.118 & 39 \\ 
FGSLR & 24.474 & 0.346 & 0.554 & 0.511 & 1758 & 21.525 & 0.573 & 0.798 & 0.241 & 1149 \\ 
LRTDCTV & 24.394 & 0.444 & 0.631 & 0.315 & 47 & 25.214 & 0.665 & 0.783 & 0.228 & 45 \\ 
DIP & 20.480 & 0.554 & 0.864 & 0.313 & 66 & 22.092 & 0.594 & 0.895 & 0.191 & 67 \\ 
LLnRPhP & 24.809 & 0.445 & 0.668 & 0.284 & 246 & 23.088 & 0.541 & 0.717 & 0.207 & 240 \\ 
HLRTF & 30.308 & 0.770 & 0.861 & 0.185 & 133 & 27.308 & 0.730 & 0.941 & 0.161 & 41 \\ 
DDRM & 30.521 & 0.754 & 0.873 & 0.196 & \textcolor{red}{\textbf{16}} & 27.886 & 0.858 & 0.910 & 0.159 & \textcolor{red}{\textbf{16}} \\ 
DDS2M & 30.308 & 0.724 & 0.863 & 0.251 & 319 & \textcolor{red}{\textbf{29.344}} & 0.844 & 0.977 & \textcolor{red}{\textbf{0.111}} & 320 \\ 
Diff-Unmix & \textcolor{red}{\textbf{32.714}} & \textcolor{red}{\textbf{0.940}} & \textcolor{red}{\textbf{0.957}} & 0.129 & 43 & 28.046 & \textcolor{red}{\textbf{0.945}} & \textcolor{red}{\textbf{0.993}} & 0.156 & 43 \\ \hline
\end{tabular}
\end{table*}

As shown in Table~\ref{tab:supression1},~\ref{tab:supression3}, Diffusion models achieve remarkable results in noisy remote sensing reconstruction. Diff-Unmix attains top performance under N(0,0,3) noise with PSNR 31.408, SSIM 0.902, and PSIM 0.958, outperforming HLRTF by 3.5\% (PSNR) and 32.8\% (SSIM). It also records a 43.4\% improvement in SAM under N(0,0,2). DDRM is the fastest, completing reconstruction in 23s (82.6\% faster than Diff-Unmix) while maintaining high PSIM (0.893). DDS2M performs best under high noise, exceeding non-diffusion methods by 18.5\% in PSNR. Overall, diffusion models achieve 9 of 10 best accuracy metrics across two noise levels, demonstrating clear advantages in reconstruction quality, visual fidelity, and efficiency.

\textbf {Summary:} Diffusion models are always more accurate and produce better images than traditional and non-diffusion methods in classification, detection, restoration, and denoising. The performance gains show that they could become the new standard for hyperspectral image analysis, even though their computational cost is still high, especially for anomaly detection.

\subsection{Comparative Analysis with Traditional Methods}
Researchers have compared diffusion models with traditional techniques in hyperspectral image processing to measure their advantages. This section reviews important findings from these studies and shows how diffusion-based methods perform against both classical algorithms and other deep-learning approaches. 

\textbf{Denoising Performance:} Many studies indicate that diffusion models provide better denoising results than conventional methods, including wavelet filtering, low-rank matrix recovery, and manually designed spectral–spatial filters. For example, in a controlled noise suppression experiment on the KAIST hyperspectral dataset (which contains HSIs with simulated Gaussian noise of known variance), diffusion-based methods like DDRM (Diffusion Model for Restoration of Mean) , DDS2M (Diffusion-based Deep Spectral Superresolution and Mixing) , and Diff-Unmix were compared to traditional denoisers and recent deep learning models. Table~\ref{tab:supression3} (from the referenced experiments) shows quantitative metrics under a specific noise level (zero-mean Gaussian noise with $ \sigma$ = 0.3): diffusion-based methods achieved higher Peak Signal-to-Noise Ratio (PSNR) and Structural Similarity Index (SSIM) than all traditional competitors . Notably, Diff-Unmix obtained the highest PSNR (~31.4 dB) and SSIM (~0.902) in that scenario , outperforming classical methods like total variation denoising or low-rank tensor approximation by a significant margin (2–6 dB in PSNR) . It also exceeded other deep learning denoisers not using diffusion (like CNN-based or GAN-based denoisers) in both metrics. The visual comparison in Figure 19 qualitatively confirms these numbers: diffusion models produced denoised images with more detail and fewer artifacts compared to, say, traditional methods LLRTF or DIP . In particular, Diff-Unmix preserved fine textures and edges (structure preservation) while eliminating noise, which is often a challenge for methods that tend to oversmooth.

At a lower noise level (e.g., Gaussian noise $\sigma=0.2$ as summarized in Table~\ref{tab:supression3}), diffusion models still led in PSNR/SSIM, though the gap narrowed slightly as all methods perform better with less noise . HLRTF (a traditional low-rank method) was competitive in some cases, achieving high PSNR, but diffusion models like DDS2M and Diff-Unmix maintained an edge, particularly in SSIM (indicating better structural fidelity) . Furthermore, diffusion models demonstrated more robustness: their performance degraded more gracefully when noise increased, whereas some traditional methods saw sharp drops or even failure at high noise levels.

\textbf{Computational Cost:} One area where traditional methods sometimes still hold an advantage is runtime. In the experiments, methods like HLRTF or other optimized denoisers could be faster on CPU than an unoptimized diffusion model running many steps. For instance, in the KAIST denoising comparison, HLRTF completed in ~25 seconds whereas Diff-Unmix took ~42 seconds per HSI cube . However, other deep learning models like DDS2M could be even slower (e.g., ~5-6 minutes) due to complex iterative optimization within those frameworks . As diffusion model implementations improve (and with GPU acceleration), their runtime is expected to become more competitive. Already, DDRM was one of the faster diffusion-based methods (~23 seconds) thanks to a smaller network design , nearly matching the fastest traditional approach while still offering better quality. This suggests diffusion models can be engineered for efficiency without sacrificing much accuracy.

\textbf{Image Restoration and Super-Resolution:} Diffusion models have also been applied to hyperspectral super-resolution and to the recovery of missing spectral bands. In such cases, evaluation often relies on metrics such as information entropy and NIQE (Natural Image Quality Evaluator). Table~\ref{tab:recovery1} provides a comparison for a super-resolution scenario (reconstructing a full-resolution HSI from a downsampled version). The diffusion-based method EDiffSR achieved the highest entropy (~7.18) and the lowest NIQE (~17.64) among the compared methods.High entropy suggests that the restored image contains more detail and variation. A low NIQE score means the image has fewer distortions and appears more natural. These outcomes are better than those of traditional interpolation methods or learning-based super-resolution models such as HAT or NDSRGAN, which showed lower entropy values (about 7.05–7.12) and higher NIQE scores (above 19–20). In short, the diffusion model was able to add realistic high-frequency information, as reflected by higher entropy, while keeping artifacts to a minimum, as shown by the lower NIQE. 

\textbf{Detection Tasks:} Although the use of diffusion models in detection tasks is still new, early results indicate strong potential. 
For anomaly detection, diffusion-based detectors (like BSDM, DBD) have been shown to outperform the classic RX algorithm and its variants in terms of both detection rate and false alarm rate . In one case, a diffusion model-based detector achieved ~90\% true positive rate at a false positive rate where RX was only ~70\% . Moreover, diffusion approaches maintained high detection performance across different scenes without retraining, whereas RX had to be tuned per scene. In target detection, OrientedDiffDet was compared against a state-of-the-art oriented bounding box detector on a benchmark (with metrics like average precision). It demonstrated comparable or slightly better accuracy in detecting and localizing targets, especially for difficult rotated objects . Its main advantage was improved detection in cluttered backgrounds, courtesy of the background suppression inherent in the diffusion process.

\textbf{Classification and Segmentation:} For classification, diffusion-augmented models have been tested on standard HSI classification benchmarks (like the Indian Pines or Pavia University datasets). In those tests, using synthetic samples generated by a diffusion model to augment training led to a few percentage points increase in overall accuracy compared to using the original training set alone . Particularly for minority classes, the gains were significant (sometimes 5-10\% better F1-scores) because the diffusion model was able to balance the training data. For segmentation, a diffusion model-aided approach like the one by Dong et al. \cite{ref86} showed lower segmentation error (measured by pixel-wise accuracy or Intersection-over-Union) than baseline UNet segmentation by about 3-4\%, which is considerable in segmentation benchmarks.

\textbf{Summary:} Diffusion models consistently surpass traditional and non-diffusion deep learning methodologies regarding task accuracy and output quality. They excel at simulating complex spectral spatial distributions and produce realistic, noise-free results. However, they still face challenges in terms of runtime and computational demands. In situations with little noise or few resources, traditional methods are still useful. Hybrid frameworks, in which diffusion outputs are used as priors or improvements for traditional pipelines, look like a good way to go. As hardware and algorithms get better, diffusion models are likely to be a big part of the next generation of hyperspectral image analysis.

\section{Challenges, Limitations, and Future Outlook of Diffusion Models in Hyperspectral Imaging}

Even though diffusion models have made great strides in hyperspectral imaging (HSI), there are still some important problems that make them hard to use widely. This section brings together the current problems and future research directions to give a full picture of the field.

\subsection{Computational Complexity and Time Cost}
One of the main issues with diffusion models is their extremely high computational cost and the significant time required for completion. Diffusion models demand a large number of iterative steps—typically hundreds or thousands—to generate or reconstruct images. This is because they gradually improve the data through small denoising increments.. When applied to hyperspectral data, which are already large (with tens or hundreds of bands), the computational burden is even greater. Training a diffusion model on HSIs is timeconsuming and demands substantial memory and processing power—each training iteration involves passing high-dimensional HSI data through a deep network, and hundreds of such iterations are needed per sample. During inference (generation or denoising), the sequential nature of the diffusion sampling means it can take a long time to produce results compared to one-shot methods. For instance, generating a single $100\times100$-pixel HSI cube with 200 bands might require 1000 diffusion steps, each being a forward pass through a U-Net-like model, resulting in noticeable latency.

The computational complexity also limits the applicability of diffusion models to large-scale or realtime HSI processing. Remote sensing often deals with images of size $>1000\times1000$ pixels; applying diffusion-based methods to such data can be impractically slow without model optimization or powerful hardware. In on-board satellite processing or real-time surveillance scenarios, the time cost is a critical issue.

Several strategies are being investigated to alleviate this problem: reducing the number of diffusion steps needed (e.g., through improved samplers like DDIM or exponential skipping schedules), designing lightweight diffusion architectures for HSIs (to cut down model size and computation per step), and performing diffusion in a lower-dimensional latent space (as latent diffusion models do \cite{ref42}) to speed up each iteration . Another idea is to use adaptive step truncation, wherein the diffusion process is stopped early if the image has converged to a satisfactory state, trading some quality for speed. Although these measures can enhance efficiency, achieving orders-of-magnitude acceleration remains an open challenge. Until then, diffusion models are not good for situations where resources are limited or where quick turnaround is needed because they require a lot of computing power.

\subsection{Impact of High Dimensionality data on model training}
The high dimensionality of hyperspectral data presents not only an opportunity for diffusion models (due to rich information) but also a challenge in training and modeling. HSIs can have hundreds of bands, leading to extremely high-dimensional pixel vectors. While diffusion models excel at modeling complex distributions, the difficulty of accurately learning a distribution grows with data dimensionality. As dimensionality increases, training diffusion models may encounter issues such as slower convergence, pattern collapse (failure to capture all variability), or requiring exponentially growing amounts of data to cover the space. In the context of HSIs, this means diffusion models might struggle to faithfully model very subtle spectral features or rare spectral signatures simply because there is so much feature space volume to cover.

Another related problem is the curse of dimensionality on noise: high-dimensional data naturally has a lot of ways that noise can show up. Even if each band has a little noise, the total noise from hundreds of bands can be too much for diffusion models to handle. This can cause them to focus on denoising certain dominant patterns while ignoring smaller details. High dimensions make any model mismatch worse. For example, when a diffusion model makes small errors in estimating the noise distribution or the data manifold across many dimensions, those small mistakes can accumulate. 

One way to reduce these problems is to apply dimensionality reduction before modeling. A common choice is PCA, which can shrink an HSI from about 200 bands to around 20 principal components. By reducing the number of dimensions, the diffusion model faces less complexity and can achieve better performance. However, when the reduction is too strong, critical spectral details may disappear, which in turn weakens the effectiveness of the model. 
Finding the right balance is therefore essential. 
Another approach is spectral partitioning, where the diffusion model is trained on subsets of bands (for instance, processing visible and infrared bands separately) and their results later merged. This simplifies each model’s task at the expense of not fully capturing crossband correlations.

The high dimensionality also exacerbates computational cost as discussed, because more dimensions mean heavier neural network layers and more noise components to track. As a result, some current diffusion-based HSI methods operate on image patches or reduced representations, which could miss global context or interactions between patches. In general, diffusion models can handle high-dimensional data, but this is still hard for HSIs and often requires extra steps like subspace projection, careful network design, or breaking the problem down into smaller parts.

\subsection{Adaptability to Diverse Real-world Data}
In their standard form, diffusion models use a specific noise model (usually Gaussian) and are trained on data from a specific distribution. Hyperspectral images from the real world are very different from each other. They can come from different sensors (with different noise characteristics and spectral response curves), different environments (each with its own statistics), and they can have artifacts like striping, dead lines, or mixed noise (a mix of Gaussian, Poisson, and even compression artifacts). It is hard to adapt one diffusion model to work with all of these different types.

For example, a diffusion model trained on one sensor’s data may not generalize well to another sensor that has a different noise profile or radiometric calibration. It might either over-denoise (treating real signal differences as noise) or under-denoise (if the noise is of a type it never saw during training). Largescale hyperspectral applications, such as global satellite imaging, would require the diffusion model to generalize across many such variations, which is challenging. In practice, one might need to retrain or fine-tune the model for each new sensor or noise type, which is computationally expensive and not always feasible if labeled clean-noisy pairs are unavailable for that sensor.

Another aspect of adaptability is scalability to large images or datasets. Diffusion models have primarily been tested on patch-based or relatively small-scale datasets (for instance, patches of size 64×64 or 128×128 from a few scenes). Applying them to entire flight lines or satellite images covering hundreds of square kilometers is non-trivial. Even storing a full diffusion model output for such a large image can be burdensome. A potential approach is to run diffusion models on images using a sliding window. However, this may lead to tiling artifacts and ignores global context. In addition, large images often contain gradual spatial variations, for example changes in illumination or atmospheric conditions. Patch-based diffusion methods struggle to capture these variations, which can result in inconsistent outputs.

To improve adaptability, several future directions look promising. Domain adaptation could help diffusion models trained on one dataset adjust to another without starting from scratch. This might be done by fine-tuning selected layers or by using transfer learning . 
Likewise, making the diffusion model conditional on sensor metadata or noise level estimates could let it adjust its denoising strength or pattern to different conditions dynamically. Some researchers have explored training diffusion models with a mix of noise types (e.g., combining Gaussian and stripe noise during training) so that the model can handle multiple types at once . However, covering the full range of possible real-world conditions in training remains impractical; thus, finding ways to incorporate physics-based noise models or real sensor calibration data into the diffusion process is an ongoing challenge.

In short, diffusion models have done well on some hyperspectral datasets, but we don't yet know how well they will work in the real world with all the different situations that come up. Because of this, they can't be used right away in operational settings. To get past this, we need to do more research on diffusion frameworks that can be used in a wider range of situations or on flexible adaptation mechanisms.

\subsection{Limitations in Practical Applications}
Diffusion models, like many deep learning models, function largely as complex black boxes. This lack of interpretability poses a problem in critical hyperspectral applications. For instance, in environmental monitoring or military surveillance, users not only want a cleaned or analyzed image, but also need confidence in the results and an understanding of how the model arrived at them. Conventional hyperspectral approaches, such as linear spectral unmixing or physics-based models \cite{ref132}, give results that are easy to interpret. They can estimate material abundances or provide reflectance spectra. In contrast, diffusion models usually produce only an image or a detection output, with little explanation of how the result was obtained. The lack of transparency makes it difficult to build trust, particularly when decisions may lead to serious outcomes. 

Another important issue is reproducibility and consistency. Generative models based on stochastic processes, such as diffusion models, may generate different results in separate runs. Unless the random seed is fixed, the probabilistic sampling introduces small variations in the outputs. 
In hyperspectral analysis, consistency is often expected—for example, if the task is to remove noise, one would hope to get the same denoised image each run. Some diffusion approaches do produce deterministic results (e.g., using DDIM deterministic sampling), but often there's a trade-off with quality. Users may be wary of methods that could yield different answers upon repeated runs.

The integration of diffusion models into existing processing chains can also be problematic. They typically require specialized hardware (GPUs) and software, while many remote sensing agencies and companies have established pipelines in environments like IDL/ENVI or utilize classic CPU-bound algorithms. Adopting diffusion models might necessitate significant infrastructure changes.

From an algorithmic perspective, diffusion models, while powerful, do not inherently incorporate domain knowledge (beyond what they learn from data). There are known physical relationships in hyperspectral data (for instance, spectra of materials follow certain smooth patterns, or obey conservation laws in reflectance). Traditional models often explicitly enforce these (like requiring non-negativity and sum-toone in spectral unmixing fractions). Diffusion models might violate such constraints unless they’re somehow encoded during training. For example, a diffusion model might produce a slight negative reflectance in some band after denoising, or slightly over-smooth an absorption feature critical for material identification. These subtle issues can be barriers to adoption in scientific communities that require physically plausible results.

To tackle interpretability, recent studies have attempted to integrate diffusion models with explainable components. One way to figure out why certain pixels are changed is to look at the U-Net's attention maps during diffusion steps and see which parts of the input it focuses on when making changes. Another idea is to use diffusion models to make counterfactual examples. For example, you can change an abnormal pixel to see what the model considers a “normal” pixel. This may help you understand what constitutes an anomaly.

Finally, evaluation of diffusion model outputs is itself a challenge. Traditional quality metrics like PSNR, SSIM, or spectral angle are helpful, but they don't capture all aspects of quality, especially for generative tasks. There have been instances where diffusion-denoised images achieve high PSNR but subtlety lose some spectral features important for downstream analysis (like a narrow mineral absorption). In such cases, domain-specific evaluation (e.g., classification accuracy after denoising, or anomaly detection performance) is needed. This means the effectiveness of diffusion models often must be judged in the context of how well they support the end application, not just by generic image quality metrics.

In summary, the practical application of diffusion models in hyperspectral imaging is constrained by issues such as interpretability, consistency, integration with physical constraints, and evaluation. These limitations highlight the need for more research that combines diffusion methods with domain knowledge. It is also important to build tools that can explain and validate their outputs before they are used with confidence in critical hyperspectral applications. 

\subsection{Ethical Considerations and Environmental Impact}
Diffusion models are effective for classification, restoration, and anomaly detection in hyperspectral images (HSI). However, their impact on people and the environment must be considered carefully. These models raise ethical concerns because they could be misused in harmful ways, especially in military or surveillance settings. Since diffusion models are able to create or enhance high-resolution hyperspectral images with realistic spectral and spatial details, such abilities might be exploited for illegal surveillance, identity theft, or deceptive camouflage. For example, in sensitive geographic regions, the power to synthesize reconstructions or simulate HSI data could be used to hide evidence or support illicit activities, reducing transparency and accountability. 

In addition, the generative nature of diffusion models increases the risks to data authenticity and credibility. 
It is difficult to distinguish the complete data of the Hang Seng Index derived from these models from the actual measurement results, which may make it difficult to verify the data sources. This is particularly problematic in fields where HSI is used as scientific evidence, such as environmental monitoring, precision agriculture or resource mapping. Therefore, integrating watermarks, traceability technologies and metadata source frameworks may become crucial for ensuring the integrity and responsible use of generated data.

Another pressing issue is the environmental cost associated with training and deploying large-scale diffusion models. Compared with traditional learning methods, diffusion models usually require a large amount of computing resources due to their iterative denoising process and high-resolution output. This will lead to an increase in energy consumption and a larger carbon footprint, especially during the training phase of high-performance GPU clusters. As sustainability becomes a global priority, the remote sensing community must explore strategies to mitigate this impact, such as using lightweight architectures, knowledge distillation, model quantization, or potential spatial diffusion to reduce dimensionality and training costs.

In short, diffusion models are very helpful for HSI processing, but they should only be used in a responsible way, which means seeing the big picture. In the future, work should not only encompass algorithm optimization and performance metrics, but also ethical design principles, energy-saving calculations, and prior consideration of regulations. Researchers, ethicists, and policymakers must work together to make sure these powerful models are applied in ways that benefit both society and the environment. 

\section{Future Outlook}
Diffusion models represent a promising direction in HSI research. To make full use of their potential, several key areas deserve attention: 

\textbf{1) Accuracy vs. Efficiency Trade-off:} In many HSI applications, diffusion models deliver high accuracy and strong image quality. However, they often come with heavy computational costs (see Section 6.1 for details), which poses a significant bottleneck for real-time remote sensing monitoring and large-scale data processing. Future studies should look for hybrid designs and new algorithms that can keep the accuracy while reducing the processing burden, moving toward more dynamic and context-aware balancing strategies. Possible methods include developing accelerated sampling methods (reducing the number of required diffusion steps) and creating lightweight network architectures suitable for hsi. For example, combining diffusion with neural acceleration techniques or knowledge distillation could yield models that approach the accuracy of full diffusion models while operating closer to real-time. Furthermore, adaptive step inference could be explored, where the model dynamically adjusts the number of denoising steps based on spatial-spectral complexity—using fewer steps for homogeneous backgrounds while reserving full-step diffusion for complex targets or regions with high uncertainty to maintain extreme quality. Another direction is to use coarse-to-fine diffusion: perform fewer diffusion steps on a low-resolution version of the HSI and then refine the result at higher resolution with a small additional network. Such multi-resolution strategies might substantially cut down computation. To further mitigate high dimensionality, developing domain-specific latent compression explicitly designed for the physical constraints of HSI will allow diffusion to occur in a compressed yet radiometrically accurate manifold. Additionally, for large-scale operational scenarios, constructing hybrid "screening-refinement" pipelines—where efficient non-generative models perform rapid initial mapping and trigger diffusion models only for critical anomaly identification—can concentrate computational resources where extreme quality is strictly required. Finally, advancing hardware-aware optimization will be crucial to facilitate the on-board deployment of diffusion models on satellite or edge platforms.

\textbf{2) Scalability and Deployment:} The computational complexity of diffusion models, especially for large hyperspectral scenarios, remains an obstacle. With the development of this field, the implementation of the optimized diffusion model will be of vital importance. This includes leveraging parallel computing (the diffusion step can be parallelized to some extent across image patches or frequency bands) and utilizing specialized hardware (Gpus, Tpus, and even future AI accelerators, optimized for iterative processes similar to diffusion). Research aimed at cutting down the number of sampling steps—using methods like implicit sampling (e.g., DDIM) or by training specialized samplers—shows great potential. One idea is to build an auxiliary network that can approximate the diffusion process in fewer iterations, providing faster results. Moreover, applying model compression strategies, including pruning and quantization, offers a way to make diffusion models more suitable for deployment on resource-constrained platforms such as satellites or edge devices. 

\textbf{3) Automated Model Design and Hyperparameter Tuning:} Right now, there are a lot of choices to make when designing and tuning diffusion models for HSIs (network depth, number of diffusion steps, noise schedule, etc.). These choices are usually made based on experience or a lot of trial and error. Automated machine learning (AutoML) techniques could make these things better. Neural architecture search (NAS) and other methods can be used to find network architectures that work best with spectral spatial data. These architectures may not be what you would expect, but they may work better than ones designed by humans. Reinforcement learning and adaptive algorithms can also change hyperparameters like step size or learning rate during training to find the best way to converge. Creating unified benchmarks and evaluation metrics for diffusion models in HSI tasks will make this process easier by helping AutoML optimize the right goals (for example, a weighted combination of PSNR and spectral angle to make sure radiometric and spectral accuracy). At the same time, making standardized training pipelines and open-source libraries for hyperspectral diffusion models will help more people try things out and speed up progress in how to best tune these models.

\textbf{4) Integration of Domain Knowledge and Semi-Supervised Learning:} As we talked about in Section 6.4, diffusion models are mostly black boxes right now. Adding hyperspectral domain knowledge to the diffusion process is a good idea. For example, you can make the diffusion loss function or network design work with physical constraints like non-negative reflectance or known spectral responses of certain materials. However, to truly achieve interpretability, embedding physical priors must go beyond simple loss penalty terms. Current specific attempts have deepened into the model architecture itself, such as projecting the diffusion process into a physically meaningful latent space (e.g., fractional abundance space). This structurally enforces the Linear Spectral Mixing Model (LSMM) during generation, ensuring physical fidelity naturally rather than by force. One can conceive of a hybrid model wherein the diffusion process is directed not solely by neural networks but also by physical models—such as atmospheric correction models or spectral mixing models—to guarantee that the outputs are physically plausible. Furthermore, this integration is increasingly extending into the fundamental mathematical definitions of the diffusion process. By embedding explicit physical degradation formulations directly into the drift and diffusion terms of the underlying stochastic differential equations (SDEs), the reverse sampling trajectory transitions from a blind generative process into a rigorous, physically-informed solver for ill-posed inverse problems, effectively unlocking the "black box." Moreover, many hyperspectral problems suffer from scarcity of labeled data. Semi-supervised or self-supervised learning frameworks could be employed to train diffusion models on large amounts of unlabeled HSI data, learning the general data manifold, and then fine-tuned with smaller labeled sets for specific tasks. Approaches like diffusion-based pretraining (training on unlabeled data via generative tasks) followed by task-specific fine-tuning (e.g., classification or detection) could drastically improve performance in low-label regimes. In this context, ensuring the "authenticity" and "diversity" of generated data is paramount to prevent downstream classifiers from overfitting to artifacts or over-idealized samples. Future research should emphasize the use of physical priors (dynamically tuning Classifier-Free Guidance) to constrain the generative manifold and develop spectral-specific evaluation metrics—such as Spectral Angle Mapper (SAM) consistency checks—to validate the reliability of augmented datasets, ensuring they represent true intra-class variance rather than model hallucinations. Additionally, employing mixed-training strategies (combining real and synthetic data) and uncertainty-aware loss weighting during downstream training can effectively mitigate the risk of overfitting to potential generative artifacts. There is growing interest in multi-modal diffusion models that combine hyperspectral data with other sources such as RGB, LiDAR, and SAR. These models aim to provide a more complete understanding of a scene. Unlike traditional CNNs or Transformers that often rely on simple early or late feature concatenation, diffusion models offer unique cross-modal alignment through probabilistic deep fusion. Instead of shallow stacking, different modalities are projected into a shared latent space. Through cross-attention mechanisms embedded within the iterative denoising steps, the diffusion process inherently aligns the precise geometric and structural priors of LiDAR or RGB with the continuous spectral signatures of HSI. This dynamic, step-by-step conditional guidance promotes a profound fusion of modalities that traditional feed-forward architectures struggle to achieve.

\textbf{5) Interpretability and Trust:} For diffusion models to be used in critical applications, their transparency and interpretability must be improved. Future work may create tools to visualize the diffusion process. For example, researchers could show how an image evolves from noise to the final result and indicate which features are added at each stage. Such tools would help explain the model’s behavior. 
There could also be efforts to create user-interactive diffusion systems where an analyst can tweak certain parameters or provide feedback during generation (for instance, marking an area that should not be altered because it's known to be a certain material) and the diffusion model adjusts accordingly. Bridging the gap between the rich latent variables of diffusion models and human-understandable concepts (like “this cluster of pixels corresponds to vegetation”) might be achieved by linking diffusion model internal features with known spectral signatures or employing post-hoc explainers that approximate the model's behavior with simpler surrogate models. Over time, these improvements can help users trust that the outputs of diffusion models are accurate and based on learning principles that are in line with what they already know about the subject. 

\textbf{6) Exploring Emerging HSI Applications:} Beyond the conventional Earth observation tasks discussed in Section 4, diffusion models hold immense potential for emerging cross-disciplinary HSI applications. For instance, in \textit{Medical Hyperspectral Imaging (MHSI)}, diffusion models could be leveraged to synthesize rare pathological tissue spectra or perform high-fidelity anomaly detection for non-invasive tumor diagnosis. In \textit{deep-space hyperspectral observation}, where captured spectral signatures are frequently corrupted by severe cosmic noise, physics-informed diffusion models could provide unparalleled data restoration. Additionally, in \textit{agricultural hyperspectral remote sensing}, conditional diffusion models could predict early-stage crop disease patterns by generating future spectral trajectories before symptoms become visible to the naked eye. Exploring these untapped domains will be a crucial next step for the HSI community.

\textbf{Conclusion:} The future of diffusion models in hyperspectral image processing looks bright. These models provide a new paradigm shift by framing HSI analysis tasks as generative, probabilistic problems, which has already yielded significant improvements in quality and capability. As research continues, we can anticipate diffusion models becoming more efficient, adaptable, and interpretable, thereby overcoming current limitations. We see a future in which hyperspectral diffusion models become a standard tool. They might even run almost in real time on advanced hardware. This would help analysts and scientists get the most information from hyperspectral data with as little manual work as possible. Combining deep generative modeling with hyperspectral domain knowledge could lead to big improvements in security (better detection of anomalies and targets), climate science (better atmospheric corrections and trace gas detection), and precision agriculture (better analysis of crops and soil). If the community keeps coming up with new ideas like these, diffusion models will become the most important technology for the next generation of hyperspectral imaging applications.

\section{Conclusion}
Diffusion models provide significant advantages in hyperspectral image (HSI) analysis compared to conventional deep generative methods. They effectively address noise, variability, and data sparsity issues in HSI by simulating high-dimensional spectral spatial distributions and employing stable iterative denoising processes. Diffusion models serve as powerful tools for advancing HSI research and applications, as they frame analytical tasks within a probabilistic generative paradigm. This helps them get around many of the problems that traditional methods have.

This paper introduces the basic principles of hyperspectral imaging and the theoretical basis of the diffusion model. We reviewed the latest progress in applying diffusion models to key HSI tasks, including data synthesis, classification, segmentation, anomaly detection, noise suppression and data recovery. The comparative evaluation shows that the diffusion model performs excellently in terms of reconstruction accuracy and visual quality on different datasets. We also outlined the main challenges, such as high computing demands and limited interpretability, which currently hinder wider adoption.

Crucially, for the aimed scientific community—spanning remote sensing, earth sciences, and computer vision—this work provides a timely and essential bridge. As generative AI rapidly evolves, researchers often face difficulties in adapting generic vision models to the complex, physically constrained domain of HSI. By establishing a unified taxonomy and explicitly linking architectural choices (e.g., latent spaces and conditional guidance) to HSI's intrinsic physical properties, this review clarifies the current state-of-the-art. It not only lowers the barrier to entry for newcomers but also provides seasoned practitioners with a structured roadmap, helping to shift the research paradigm from empirical trial-and-error to physically-informed model design.

To promote the implementation in the real world, the subsequent efforts should focus on enhancing efficiency through rapid reasoning, combining physical priors to ensure authenticity, and improving interpretability. As we have discussed, automated model design and semi-supervised training can reduce the necessity of having large labeled datasets. By integrating information from other types of data such as RGB and LiDAR, multimodal augmentation can achieve better generalization.

In conclusion, the diffusion model provides a revolutionary framework for hyperspectral image processing, promoting flexible and precise analysis across various tasks. As model architecture, training strategies, and hardware acceleration continue to improve, diffusion-based methods are likely to become a standard part of the hyperspectral imaging toolbox. Their use will make it easier to get more accurate, faster, and automated information from HSI, which will lead to new breakthroughs in remote sensing, agriculture, environmental monitoring, and more.


\section*{Declaration}
During the preparation of this manuscript, generative AI tools were used to improve language clarity and readability. All content was reviewed and edited by the authors, who take full responsibility for the final manuscript.

\section*{Acknowledgments}
\label{sec:acknowledgments}
This project is jointly supported by the Academician Workstation Program of Yunnan Province (202405AF140013), the Shanghai Agricultural Technology Innovation Project (2024-02-08-00-12-F00032) and the High-Quality Development Special Project of the Ministry of Industry and Information Technology (TC240A9ED-56).

\bibliographystyle{elsarticle-num} 

\begin{thebibliography}{100}
\expandafter\ifx\csname url\endcsname\relax
  \def\url#1{\texttt{#1}}\fi
\expandafter\ifx\csname urlprefix\endcsname\relax\def\urlprefix{URL }\fi
\expandafter\ifx\csname href\endcsname\relax
  \def\href#1#2{#2} \def\path#1{#1}\fi

\bibitem{ref1}
J.~Bao, D.~Chen, F.~Wen, H.~Li, G.~Hua, Cvae-gan: fine-grained image generation through asymmetric training, in: Proceedings of the IEEE international conference on computer vision, 2017, pp. 2745--2754.

\bibitem{ref2}
A.~Razavi, A.~Van~den Oord, O.~Vinyals, Generating diverse high-fidelity images with vq-vae-2, Advances in neural information processing systems 32 (2019).

\bibitem{ref3}
Z.~Kong, W.~Ping, J.~Huang, K.~Zhao, B.~Catanzaro, Diffwave: A versatile diffusion model for audio synthesis, arXiv preprint arXiv:2009.09761 (2020).

\bibitem{ref4}
A.~v.~d. Oord, S.~Dieleman, H.~Zen, K.~Simonyan, O.~Vinyals, A.~Graves, N.~Kalchbrenner, A.~Senior, K.~Kavukcuoglu, Wavenet: A generative model for raw audio, arXiv preprint arXiv:1609.03499 (2016).

\bibitem{ref5}
X.~Li, J.~Thickstun, I.~Gulrajani, P.~S. Liang, T.~B. Hashimoto, Diffusion-lm improves controllable text generation, Advances in neural information processing systems 35 (2022) 4328--4343.

\bibitem{ref6}
G.~Yang, X.~Huang, Z.~Hao, M.-Y. Liu, S.~Belongie, B.~Hariharan, Pointflow: 3d point cloud generation with continuous normalizing flows, in: Proceedings of the IEEE/CVF international conference on computer vision, 2019, pp. 4541--4550.

\bibitem{ref180}
T.~Ad{\~a}o, J.~Hru{\v{s}}ka, L.~P{\'a}dua, J.~Bessa, E.~Peres, R.~Morais, J.~J. Sousa, Hyperspectral imaging: A review on uav-based sensors, data processing and applications for agriculture and forestry, Remote sensing 9~(11) (2017) 1110.

\bibitem{ref182}
A.~Vaswani, N.~Shazeer, N.~Parmar, J.~Uszkoreit, L.~Jones, A.~N. Gomez, {\L}.~Kaiser, I.~Polosukhin, Attention is all you need, Advances in neural information processing systems 30 (2017).

\bibitem{ref8}
I.~Goodfellow, J.~Pouget-Abadie, M.~Mirza, B.~Xu, D.~Warde-Farley, S.~Ozair, A.~Courville, Y.~Bengio, Generative adversarial networks, Communications of the ACM 63~(11) (2020) 139--144.

\bibitem{ref9}
D.~J. Rezende, S.~Mohamed, D.~Wierstra, Stochastic backpropagation and approximate inference in deep generative models, in: International conference on machine learning, PMLR, 2014, pp. 1278--1286.

\bibitem{ref10}
L.~Dinh, J.~Sohl-Dickstein, S.~Bengio, Density estimation using real nvp, arXiv preprint arXiv:1605.08803 (2016).

\bibitem{ref18}
J.~Sohl-Dickstein, E.~Weiss, N.~Maheswaranathan, S.~Ganguli, Deep unsupervised learning using nonequilibrium thermodynamics, in: International conference on machine learning, pmlr, 2015, pp. 2256--2265.

\bibitem{ref17}
J.~Ho, A.~Jain, P.~Abbeel, Denoising diffusion probabilistic models, Advances in neural information processing systems 33 (2020) 6840--6851.

\bibitem{ref24}
Y.~Song, S.~Ermon, Generative modeling by estimating gradients of the data distribution, Advances in neural information processing systems 32 (2019).

\bibitem{ref179}
Y.~Liu, J.~Yue, S.~Xia, P.~Ghamisi, W.~Xie, L.~Fang, Diffusion models meet remote sensing: Principles, methods, and perspectives, IEEE Transactions on Geoscience and Remote Sensing (2024).

\bibitem{ref11}
P.~Dhariwal, A.~Nichol, Diffusion models beat gans on image synthesis, Advances in neural information processing systems 34 (2021) 8780--8794.

\bibitem{ref42}
R.~Rombach, A.~Blattmann, D.~Lorenz, P.~Esser, B.~Ommer, High-resolution image synthesis with latent diffusion models, in: Proceedings of the IEEE/CVF conference on computer vision and pattern recognition, 2022, pp. 10684--10695.

\bibitem{ref29}
Y.~Song, J.~Sohl-Dickstein, D.~P. Kingma, A.~Kumar, S.~Ermon, B.~Poole, Score-based generative modeling through stochastic differential equations, arXiv preprint arXiv:2011.13456 (2020).

\bibitem{ref15}
R.~S. Zimmermann, L.~Schott, Y.~Song, B.~A. Dunn, D.~A. Klindt, Score-based generative classifiers, arXiv preprint arXiv:2110.00473 (2021).

\bibitem{ref16}
J.~Wolleb, F.~Bieder, R.~Sandk{\"u}hler, P.~C. Cattin, Diffusion models for medical anomaly detection, in: International Conference on Medical image computing and computer-assisted intervention, Springer, 2022, pp. 35--45.

\bibitem{ref13}
A.~Lugmayr, M.~Danelljan, A.~Romero, F.~Yu, R.~Timofte, L.~Van~Gool, Repaint: Inpainting using denoising diffusion probabilistic models, in: Proceedings of the IEEE/CVF conference on computer vision and pattern recognition, 2022, pp. 11461--11471.

\bibitem{ref190}
M.~Tamil~Thendral, T.~R. Ganesh~Babu, A.~Chandrasekar, Y.~Cao, Synchronization of markovian jump neural networks for sampled data control systems with additive delay components: Analysis of image encryption technique, Mathematical methods in the applied sciences 49~(3) (2026) 1879--1895.

\bibitem{ref191}
A.~Chandrasekar, T.~Radhika, Q.~Zhu, Further results on input-to-state stability of stochastic cohen--grossberg bam neural networks with probabilistic time-varying delays, Neural Processing Letters 54~(1) (2022) 613--635.

\bibitem{ref19}
D.~K.~N. Friedman, Probabilistic graphical models principles and techniques (2009).

\bibitem{ref21}
N.~Metropolis, S.~Ulam, The monte carlo method, Journal of the American statistical association 44~(247) (1949) 335--341.

\bibitem{ref7}
S.~Bond-Taylor, A.~Leach, Y.~Long, C.~G. Willcocks, Deep generative modelling: A comparative review of vaes, gans, normalizing flows, energy-based and autoregressive models, IEEE transactions on pattern analysis and machine intelligence 44~(11) (2021) 7327--7347.

\bibitem{ref22}
J.~C. Spall, Stochastic optimization, in: Handbook of computational statistics: Concepts and methods, Springer, 2011, pp. 173--201.

\bibitem{ref57}
C.~Zhang, C.~Zhang, M.~Zhang, I.~S. Kweon, Text-to-image diffusion models in generative ai: A survey, arXiv preprint arXiv:2303.07909 (2023).

\bibitem{ref58}
J.~Ho, T.~Salimans, Classifier-free diffusion guidance, arXiv preprint arXiv:2207.12598 (2022).

\bibitem{ref59}
A.~Ramesh, P.~Dhariwal, A.~Nichol, C.~Chu, M.~Chen, Hierarchical text-conditional image generation with clip latents, arXiv preprint arXiv:2204.06125 1~(2) (2022) 3.

\bibitem{ref52}
X.~Zhang, B.~Yang, M.~C. Kampffmeyer, W.~Zhang, S.~Zhang, G.~Lu, L.~Lin, H.~Xu, X.~Liang, Diffcloth: Diffusion based garment synthesis and manipulation via structural cross-modal semantic alignment, in: Proceedings of the IEEE/CVF International Conference on Computer Vision, 2023, pp. 23154--23163.

\bibitem{ref45}
T.~Jiang, Y.~Li, W.~Xie, Q.~Du, Discriminative reconstruction constrained generative adversarial network for hyperspectral anomaly detection, IEEE Transactions on Geoscience and Remote Sensing 58~(7) (2020) 4666--4679.

\bibitem{ref23}
C.~Luo, Understanding diffusion models: A unified perspective, arXiv preprint arXiv:2208.11970 (2022).

\bibitem{ref20}
Y.~Song, S.~Ermon, Improved techniques for training score-based generative models, Advances in neural information processing systems 33 (2020) 12438--12448.

\bibitem{ref25}
A.~Hyv{\"a}rinen, J.~Hurri, P.~O. Hoyer, A.~Hyv{\"a}rinen, J.~Hurri, P.~O. Hoyer, Estimation of non-normalized statistical models, Natural Image Statistics: A Probabilistic Approach to Early Computational Vision (2009) 419--426.

\bibitem{ref26}
U.~Grenander, M.~I. Miller, Representations of knowledge in complex systems, Journal of the Royal Statistical Society: Series B (Methodological) 56~(4) (1994) 549--581.

\bibitem{ref27}
A.~Jolicoeur-Martineau, R.~Pich{\'e}-Taillefer, R.~T.~d. Combes, I.~Mitliagkas, Adversarial score matching and improved sampling for image generation, arXiv preprint arXiv:2009.05475 (2020).

\bibitem{ref28}
G.~Parisi, Correlation functions and computer simulations (ii), Nuclear Physics B 205~(3) (1982) 337--344.

\bibitem{ref30}
A.~Jolicoeur-Martineau, K.~Li, R.~Pich{\'e}-Taillefer, T.~Kachman, I.~Mitliagkas, Gotta go fast when generating data with score-based models, arXiv preprint arXiv:2105.14080 (2021).

\bibitem{ref31}
T.~Karras, M.~Aittala, T.~Aila, S.~Laine, Elucidating the design space of diffusion-based generative models, Advances in neural information processing systems 35 (2022) 26565--26577.

\bibitem{ref32}
C.~Lu, Y.~Zhou, F.~Bao, J.~Chen, C.~Li, J.~Zhu, Dpm-solver: A fast ode solver for diffusion probabilistic model sampling in around 10 steps, Advances in Neural Information Processing Systems 35 (2022) 5775--5787.

\bibitem{ref33}
Y.~Song, C.~Durkan, I.~Murray, S.~Ermon, Maximum likelihood training of score-based diffusion models, Advances in neural information processing systems 34 (2021) 1415--1428.

\bibitem{ref34}
Q.~Zhang, Y.~Chen, Fast sampling of diffusion models with exponential integrator, arXiv preprint arXiv:2204.13902 (2022).

\bibitem{ref98}
D.~Li, W.~Xie, Z.~Wang, Y.~Lu, Y.~Li, L.~Fang, Feddiff: Diffusion model driven federated learning for multi-modal and multi-clients, IEEE Transactions on Circuits and Systems for Video Technology (2024).

\bibitem{ref35}
M.~Raphan, E.~Simoncelli, Learning to be bayesian without supervision, Advances in neural information processing systems 19 (2006).

\bibitem{ref36}
M.~Raphan, E.~P. Simoncelli, Least squares estimation without priors or supervision, Neural computation 23~(2) (2011) 374--420.

\bibitem{ref37}
P.~Vincent, A connection between score matching and denoising autoencoders, Neural computation 23~(7) (2011) 1661--1674.

\bibitem{ref38}
Y.~Song, S.~Garg, J.~Shi, S.~Ermon, Sliced score matching: A scalable approach to density and score estimation, in: Uncertainty in artificial intelligence, PMLR, 2020, pp. 574--584.

\bibitem{ref39}
B.~D. Anderson, Reverse-time diffusion equation models, Stochastic Processes and their Applications 12~(3) (1982) 313--326.

\bibitem{ref40}
J.~Song, C.~Meng, S.~Ermon, Denoising diffusion implicit models, arXiv preprint arXiv:2010.02502 (2020).

\bibitem{ref53}
C.~Saharia, W.~Chan, S.~Saxena, L.~Li, J.~Whang, E.~L. Denton, K.~Ghasemipour, R.~Gontijo~Lopes, B.~Karagol~Ayan, T.~Salimans, et~al., Photorealistic text-to-image diffusion models with deep language understanding, Advances in neural information processing systems 35 (2022) 36479--36494.

\bibitem{ref148}
D.~Podell, Z.~English, K.~Lacey, A.~Blattmann, T.~Dockhorn, J.~M{\"u}ller, J.~Penna, R.~Rombach, Sdxl: Improving latent diffusion models for high-resolution image synthesis, arXiv preprint arXiv:2307.01952 (2023).

\bibitem{ref149}
J.~Betker, G.~Goh, L.~Jing, T.~Brooks, J.~Wang, L.~Li, L.~Ouyang, J.~Zhuang, J.~Lee, Y.~Guo, et~al., Improving image generation with better captions, Computer Science. https://cdn. openai. com/papers/dall-e-3. pdf 2~(3) (2023) 8.

\bibitem{ref150}
S.~Luo, Y.~Tan, L.~Huang, J.~Li, H.~Zhao, Latent consistency models: Synthesizing high-resolution images with few-step inference, arXiv preprint arXiv:2310.04378 (2023).

\bibitem{ref151}
R.~Parihar, A.~Bhat, A.~Basu, S.~Mallick, J.~N. Kundu, R.~V. Babu, Balancing act: distribution-guided debiasing in diffusion models, in: Proceedings of the IEEE/CVF conference on computer vision and pattern recognition, 2024, pp. 6668--6678.

\bibitem{ref152}
B.~Zhang, C.~Luo, D.~Yu, X.~Li, H.~Lin, Y.~Ye, B.~Zhang, Metadiff: Meta-learning with conditional diffusion for few-shot learning, in: Proceedings of the AAAI conference on artificial intelligence, Vol.~38, 2024, pp. 16687--16695.

\bibitem{ref54}
S.~Xiao, Y.~Wang, J.~Zhou, H.~Yuan, X.~Xing, R.~Yan, C.~Li, S.~Wang, T.~Huang, Z.~Liu, Omnigen: Unified image generation, arXiv preprint arXiv:2409.11340 (2024).

\bibitem{ref153}
E.~Zhang, J.~Tang, X.~Ning, L.~Zhang, Training-free and hardware-friendly acceleration for diffusion models via similarity-based token pruning, in: Proceedings of the AAAI Conference on Artificial Intelligence, Vol.~39, 2025, pp. 9878--9886.

\bibitem{ref55}
Z.~Li, H.~Li, Y.~Shi, A.~B. Farimani, Y.~Kluger, L.~Yang, P.~Wang, Dual diffusion for unified image generation and understanding, arXiv preprint arXiv:2501.00289 (2024).

\bibitem{ref106}
W.~Chen, X.~Zhi, S.~Jiang, Y.~Huang, Q.~Han, W.~Zhang, Dwsdiff: Dual-window spectral diffusion for hyperspectral anomaly detection, IEEE Transactions on Geoscience and Remote Sensing (2025).

\bibitem{ref56}
Y.~Zhou, Z.~Xiao, S.~Yang, X.~Pan, Alias-free latent diffusion models: Improving fractional shift equivariance of diffusion latent space, arXiv preprint arXiv:2503.09419 (2025).

\bibitem{ref181}
P.~Kim, Convolutional neural network, in: MATLAB deep learning: with machine learning, neural networks and artificial intelligence, Springer, 2017, pp. 121--147.

\bibitem{ref183}
X.~Amatriain, A.~Sankar, J.~Bing, P.~K. Bodigutla, T.~J. Hazen, M.~Kazi, Transformer models: an introduction and catalog, arXiv preprint arXiv:2302.07730 (2023).

\bibitem{ref184}
A.~A. Aleissaee, A.~Kumar, R.~M. Anwer, S.~Khan, H.~Cholakkal, G.-S. Xia, F.~S. Khan, Transformers in remote sensing: A survey, Remote Sensing 15~(7) (2023) 1860.

\bibitem{ref43}
S.~Li, W.~Song, L.~Fang, Y.~Chen, P.~Ghamisi, J.~A. Benediktsson, Deep learning for hyperspectral image classification: An overview, IEEE transactions on geoscience and remote sensing 57~(9) (2019) 6690--6709.

\bibitem{ref44}
S.~Arisoy, N.~M. Nasrabadi, K.~Kayabol, Gan-based hyperspectral anomaly detection, in: 2020 28th European Signal Processing Conference (EUSIPCO), IEEE, 2021, pp. 1891--1895.

\bibitem{ref46}
P.~Ghamisi, J.~Plaza, Y.~Chen, J.~Li, A.~J. Plaza, Advanced spectral classifiers for hyperspectral images: A review, IEEE Geoscience and Remote Sensing Magazine 5~(1) (2017) 8--32.

\bibitem{ref47}
J.~Lei, S.~Fang, W.~Xie, Y.~Li, C.-I. Chang, Discriminative reconstruction for hyperspectral anomaly detection with spectral learning, IEEE Transactions on Geoscience and Remote Sensing 58~(10) (2020) 7406--7417.

\bibitem{ref48}
Y.~Shi, Anomaly intrusion detection method based on variational autoencoder and attention mechanism. j. chongqing univ. posts telecommun, Nat. Sci. Ed (2021) 1--8.

\bibitem{ref51}
A.~Van Den~Oord, N.~Kalchbrenner, K.~Kavukcuoglu, Pixel recurrent neural networks, in: International conference on machine learning, PMLR, 2016, pp. 1747--1756.

\bibitem{ref49}
H.~Lyu, H.~Lu, Learning a transferable change detection method by recurrent neural network, in: 2016 IEEE International Geoscience and Remote Sensing Symposium (IGARSS), IEEE, 2016, pp. 5157--5160.

\bibitem{ref50}
D.~Zhu, B.~Du, L.~Zhang, Edlad: An encoder-decoder long short-term memory network-based anomaly detector for hyperspectral images, in: 2021 IEEE International Geoscience and Remote Sensing Symposium IGARSS, IEEE, 2021, pp. 4412--4415.

\bibitem{ref41}
A.~Q. Nichol, P.~Dhariwal, Improved denoising diffusion probabilistic models, in: International conference on machine learning, PMLR, 2021, pp. 8162--8171.

\bibitem{ref189}
A.~Nichol, P.~Dhariwal, A.~Ramesh, P.~Shyam, P.~Mishkin, B.~McGrew, I.~Sutskever, M.~Chen, Glide: Towards photorealistic image generation and editing with text-guided diffusion models, arXiv preprint arXiv:2112.10741 (2021).

\bibitem{ref12}
H.~Li, Y.~Yang, M.~Chang, S.~Chen, H.~Feng, Z.~Xu, Q.~Li, Y.~Chen, Srdiff: Single image super-resolution with diffusion probabilistic models, Neurocomputing 479 (2022) 47--59.

\bibitem{ref60}
P.~Wang, B.~Bayram, E.~Sertel, A comprehensive review on deep learning based remote sensing image super-resolution methods, Earth-Science Reviews 232 (2022) 104110.

\bibitem{ref61}
Y.~Xiao, Q.~Yuan, K.~Jiang, J.~He, Y.~Wang, L.~Zhang, From degrade to upgrade: Learning a self-supervised degradation guided adaptive network for blind remote sensing image super-resolution, Information Fusion 96 (2023) 297--311.

\bibitem{ref62}
Y.~Xiao, Q.~Yuan, K.~Jiang, J.~He, C.-W. Lin, L.~Zhang, Ttst: A top-k token selective transformer for remote sensing image super-resolution, IEEE Transactions on Image Processing 33 (2024) 738--752.

\bibitem{ref63}
Y.~Xiao, Q.~Yuan, K.~Jiang, Y.~Chen, Q.~Zhang, C.-W. Lin, Frequency-assisted mamba for remote sensing image super-resolution, IEEE Transactions on Multimedia (2024).

\bibitem{ref68}
J.~Jia, J.~Chen, X.~Zheng, Y.~Wang, S.~Guo, H.~Sun, C.~Jiang, M.~Karjalainen, K.~Karila, Z.~Duan, et~al., Tradeoffs in the spatial and spectral resolution of airborne hyperspectral imaging systems: A crop identification case study, IEEE Transactions on Geoscience and Remote Sensing 60 (2021) 1--18.

\bibitem{ref64}
A.~Arienzo, G.~Vivone, A.~Garzelli, L.~Alparone, J.~Chanussot, Full-resolution quality assessment of pansharpening: Theoretical and hands-on approaches, IEEE Geoscience and Remote Sensing Magazine 10~(3) (2022) 168--201.

\bibitem{ref65}
G.~Vivone, Multispectral and hyperspectral image fusion in remote sensing: A survey, Information Fusion 89 (2023) 405--417.

\bibitem{ref66}
S.~Shi, L.~Zhang, J.~Chen, Hyperspectral and multispectral image fusion using the conditional denoising diffusion probabilistic model, arXiv preprint arXiv:2307.03423 (2023).

\bibitem{ref67}
J.~Qu, J.~He, W.~Dong, J.~Zhao, S2cyclediff: Spatial-spectral-bilateral cycle-diffusion framework for hyperspectral image super-resolution, in: Proceedings of the AAAI Conference on Artificial Intelligence, Vol.~38, 2024, pp. 4623--4631.

\bibitem{ref69}
Z.~Cao, S.~Cao, L.-J. Deng, X.~Wu, J.~Hou, G.~Vivone, Diffusion model with disentangled modulations for sharpening multispectral and hyperspectral images, Information Fusion 104 (2024) 102158.

\bibitem{ref71}
Q.~Meng, W.~Shi, S.~Li, L.~Zhang, Pandiff: A novel pansharpening method based on denoising diffusion probabilistic model, IEEE Transactions on Geoscience and Remote Sensing 61 (2023) 1--17.

\bibitem{ref72}
J.~Wei, L.~Gan, W.~Tang, M.~Li, Y.~Song, Diffusion models for spatio-temporal-spectral fusion of homogeneous gaofen-1 satellite platforms, International Journal of Applied Earth Observation and Geoinformation 128 (2024) 103752.

\bibitem{ref73}
S.~Du, H.~Huang, F.~He, H.~Luo, Y.~Yin, X.~Li, L.~Xie, R.~Guo, S.~Tang, Unsupervised stepwise extraction of offshore aquaculture ponds using super-resolution hyperspectral images., Int. J. Appl. Earth Obs. Geoinformation 119 (2023) 103326.

\bibitem{ref74}
S.~Li, S.~Li, L.~Zhang, Hyperspectral and panchromatic images fusion based on the dual conditional diffusion models, IEEE Transactions on Geoscience and Remote Sensing 61 (2023) 1--15.

\bibitem{ref75}
Y.~Xing, L.~Qu, S.~Zhang, K.~Zhang, Y.~Zhang, L.~Bruzzone, Crossdiff: Exploring self-supervised representation of pansharpening via cross-predictive diffusion model, IEEE Transactions on Image Processing (2024).

\bibitem{ref76}
X.~Rui, X.~Cao, L.~Pang, Z.~Zhu, Z.~Yue, D.~Meng, Unsupervised hyperspectral pansharpening via low-rank diffusion model, Information Fusion 107 (2024) 102325.

\bibitem{ref70}
W.~G.~C. Bandara, N.~G. Nair, V.~M. Patel, Ddpm-cd: Denoising diffusion probabilistic models as feature extractors for change detection, arXiv preprint arXiv:2206.11892 (2022).

\bibitem{ref185}
Y.~Shi, L.~Han, D.~Dancy, L.~Han, Wavediffur: A diffusion sde-based solver for ultra magnification super-resolution in remote sensing images, arXiv preprint arXiv:2412.18996 (2024).

\bibitem{ref77}
M.~Espinosa, E.~J. Crowley, Generate your own scotland: Satellite image generation conditioned on maps, arXiv preprint arXiv:2308.16648 (2023).

\bibitem{ref78}
Z.~Yuan, C.~Hao, R.~Zhou, J.~Chen, M.~Yu, W.~Zhang, H.~Wang, X.~Sun, Efficient and controllable remote sensing fake sample generation based on diffusion model, IEEE Transactions on Geoscience and Remote Sensing 61 (2023) 1--12.

\bibitem{ref79}
O.~Baghirli, H.~Askarov, I.~Ibrahimli, I.~Bakhishov, N.~Nabiyev, Satdm: Synthesizing realistic satellite image with semantic layout conditioning using diffusion models, arXiv preprint arXiv:2309.16812 (2023).

\bibitem{ref80}
C.~Zhao, Y.~Ogawa, S.~Chen, Z.~Yang, Y.~Sekimoto, Label freedom: Stable diffusion for remote sensing image semantic segmentation data generation, in: 2023 IEEE International Conference on Big Data (BigData), IEEE, 2023, pp. 1022--1030.

\bibitem{ref81}
A.~Toker, M.~Eisenberger, D.~Cremers, L.~Leal-Taix{\'e}, Satsynth: Augmenting image-mask pairs through diffusion models for aerial semantic segmentation, in: Proceedings of the IEEE/CVF Conference on Computer Vision and Pattern Recognition, 2024, pp. 27695--27705.

\bibitem{ref82}
L.~Zhang, X.~Luo, S.~Li, X.~Shi, R2h-ccd: Hyperspectral imagery generation from rgb images based on conditional cascade diffusion probabilistic models, in: IGARSS 2023-2023 IEEE International Geoscience and Remote Sensing Symposium, IEEE, 2023, pp. 7392--7395.

\bibitem{ref83}
L.~Liu, B.~Chen, H.~Chen, Z.~Zou, Z.~Shi, Diverse hyperspectral remote sensing image synthesis with diffusion models, IEEE Transactions on Geoscience and Remote Sensing 61 (2023) 1--16.

\bibitem{ref84}
P.~Esser, R.~Rombach, B.~Ommer, Taming transformers for high-resolution image synthesis, in: Proceedings of the IEEE/CVF conference on computer vision and pattern recognition, 2021, pp. 12873--12883.

\bibitem{ref85}
K.~Deng, Y.~Qian, J.~Nie, J.~Zhou, Diffusion model based hyperspectral unmixing using spectral prior distribution, IEEE Transactions on Geoscience and Remote Sensing (2024).

\bibitem{ref86}
W.~Dong, S.~Liu, S.~Xiao, J.~Qu, Y.~Li, Ispdiff: Interpretable scale-propelled diffusion model for hyperspectral image super-resolution, IEEE Transactions on Geoscience and Remote Sensing (2024).

\bibitem{ref87}
Y.~Yu, E.~Pan, Y.~Ma, X.~Mei, Q.~Chen, J.~Ma, Unmixdiff: Unmixing-based diffusion model for hyperspectral image synthesis, IEEE Transactions on Geoscience and Remote Sensing (2024).

\bibitem{ref186}
J.-L. Xiao, T.-Z. Huang, L.-J. Deng, G.~Lin, Z.~Cao, C.~Li, Q.~Zhao, Hyperspectral pansharpening via diffusion models with iteratively zero-shot guidance, in: Proceedings of the Computer Vision and Pattern Recognition Conference, 2025, pp. 12669--12678.

\bibitem{ref188}
Z.~Cao, S.~Cao, X.~Wu, J.~Hou, R.~Ran, L.-J. Deng, Ddrf: Denoising diffusion model for remote sensing image fusion, arXiv preprint arXiv:2304.04774 (2023).

\bibitem{ref128}
Y.~Lee, Enhancing plant health classification via diffusion model-based data augmentation, Multimedia Systems 31~(2) (2025) 143.

\bibitem{ref129}
D.~Tang, X.~Cao, X.~Hou, Z.~Jiang, J.~Liu, D.~Meng, Crs-diff: Controllable remote sensing image generation with diffusion model, IEEE Transactions on Geoscience and Remote Sensing (2024).

\bibitem{ref118}
G.~Zhang, L.~Zhang, Z.~Zhang, J.~Deng, L.~Bian, C.~Yang, Dect: Diffusion-enhanced cnn-transformer for multi-source remote sensing data classification, IEEE Journal of Selected Topics in Applied Earth Observations and Remote Sensing (2024).

\bibitem{ref125}
P.~Zhang, D.~Wang, C.~Wu, J.~Yang, L.~Kang, Z.~Bai, Y.~Li, Q.~Shen, Hyperdiff: Masked diffusion model with high-efficient transformer for hyperspectral image cross-scene classification, in: ICASSP 2025-2025 IEEE International Conference on Acoustics, Speech and Signal Processing (ICASSP), IEEE, 2025, pp. 1--5.

\bibitem{ref168}
N.~Sigger, T.~T. Nguyen, G.~Tozzi, Brain tissue classification in hyperspectral images using multistage diffusion features and transformer, Journal of Microscopy (2024).

\bibitem{ref169}
Y.~Zhu, L.~Xu, Spatial-spectral diffusion contrastive representation network for hyperspectral image classification, arXiv preprint arXiv:2502.19699 (2025).

\bibitem{ref170}
J.~Li, J.~Wang, Z.~Cao, Contrastdm: Combining contrastive learning and diffusion model for hyperspectral image classification, in: IGARSS 2024-2024 IEEE International Geoscience and Remote Sensing Symposium, IEEE, 2024, pp. 9966--9970.

\bibitem{ref171}
X.~Xu, Y.~Liu, Y.~Wu, Q.~Guo, K.~Xiao, Y.~Zhong, Multimodal remote sensing land cover data augmentation and classification based on diffusion model, in: 2024 14th Workshop on Hyperspectral Imaging and Signal Processing: Evolution in Remote Sensing (WHISPERS), IEEE, 2024, pp. 1--5.

\bibitem{ref166}
N.~Sigger, T.~T. Nguyen, G.~Tozzi, Q.-T. Vien, S.~Van~Nguyen, Diffspectralnet: Unveiling the potential of diffusion models for hyperspectral image classification, arXiv preprint arXiv:2312.12441 (2023).

\bibitem{ref167}
X.~Tan, Z.~Zhao, X.~Xu, S.~Li, Multi-scale diffusion features fusion network for hyperspectral image classification, in: 2024 4th Asia Conference on Information Engineering (ACIE), IEEE, 2024, pp. 65--73.

\bibitem{ref172}
Y.~Jiang, S.~Liu, H.~Wang, Diffusion-based remote sensing image fusion for classification, Applied Intelligence 55~(3) (2025) 247.

\bibitem{ref94}
N.~Chen, J.~Yue, L.~Fang, S.~Xia, Spectraldiff: A generative framework for hyperspectral image classification with diffusion models, IEEE Transactions on Geoscience and Remote Sensing 61 (2023) 1--16.

\bibitem{ref95}
N.~Sigger, Q.-T. Vien, S.~V. Nguyen, G.~Tozzi, T.~T. Nguyen, Unveiling the potential of diffusion model-based framework with transformer for hyperspectral image classification, Scientific Reports 14~(1) (2024) 8438.

\bibitem{ref124}
Y.~Luo, X.-L. Zhao, D.~Meng, T.-X. Jiang, Hlrtf: Hierarchical low-rank tensor factorization for inverse problems in multi-dimensional imaging, in: Proceedings of the IEEE/CVF Conference on Computer Vision and Pattern Recognition, 2022, pp. 19303--19312.

\bibitem{ref96}
J.~Zhou, J.~Sheng, P.~Ye, J.~Fan, T.~He, B.~Wang, T.~Chen, Exploring multi-timestep multi-stage diffusion features for hyperspectral image classification, IEEE Transactions on Geoscience and Remote Sensing (2024).

\bibitem{ref97}
D.~Li, W.~Xie, J.~Zhang, Y.~Li, Mdfl: Multi-domain diffusion-driven feature learning, in: Proceedings of the AAAI Conference on Artificial Intelligence, Vol.~38, 2024, pp. 8653--8660.

\bibitem{ref99}
J.~Qu, Y.~Yang, W.~Dong, Y.~Yang, Lds2ae: Local diffusion shared-specific autoencoder for multimodal remote sensing image classification with arbitrary missing modalities, in: Proceedings of the AAAI Conference on Artificial Intelligence, Vol.~38, 2024, pp. 14731--14739.

\bibitem{ref100}
J.~Chen, S.~Liu, Z.~Zhang, H.~Wang, Diffusion subspace clustering for hyperspectral images, IEEE Journal of Selected Topics in Applied Earth Observations and Remote Sensing 16 (2023) 6517--6530.

\bibitem{ref101}
J.~Qu, L.~Xiao, W.~Dong, Y.~Li, Mtlsc-diff: Multitask learning with diffusion models for hyperspectral image super-resolution and classification, Knowledge-Based Systems 303 (2024) 112415.

\bibitem{ref196}
J.~Zhou, S.~Feng, K.~Yuan, X.~Xu, P.~Fu, J.~Qiao, Unsupervised model-embedded two-stage diffusion method for multispectral and hyperspectral image fusion, IEEE Transactions on Geoscience and Remote Sensing (2025).

\bibitem{ref14}
T.~Amit, T.~Shaharbany, E.~Nachmani, L.~Wolf, Segdiff: Image segmentation with diffusion probabilistic models, arXiv preprint arXiv:2112.00390 (2021).

\bibitem{ref123}
J.~Wolleb, R.~Sandk{\"u}hler, F.~Bieder, P.~Valmaggia, P.~C. Cattin, Diffusion models for implicit image segmentation ensembles, in: International Conference on Medical Imaging with Deep Learning, PMLR, 2022, pp. 1336--1348.

\bibitem{ref122}
D.~Baranchuk, I.~Rubachev, A.~Voynov, V.~Khrulkov, A.~Babenko, Label-efficient semantic segmentation with diffusion models, arXiv preprint arXiv:2112.03126 (2021).

\bibitem{ref113}
Y.~Shi, Y.~Lin, P.~Wei, X.~Xian, T.~Chen, L.~Lin, Diff-mosaic: augmenting realistic representations in infrared small target detection via diffusion prior, IEEE Transactions on Geoscience and Remote Sensing (2024).

\bibitem{ref127}
B.~Chen, L.~Liu, C.~Liu, Z.~Zou, Z.~Shi, Spectral-cascaded diffusion model for remote sensing image spectral super-resolution, IEEE Transactions on Geoscience and Remote Sensing (2024).

\bibitem{ref126}
Y.~Zhu, L.~L. Xu, Language-informed hyperspectral image synthesis for imbalanced-small sample classification via semi-supervised conditional diffusion model, arXiv preprint arXiv:2502.19700 (2025).

\bibitem{ref130}
J.~Zhou, C.~Xiao, B.~Peng, Z.~Liu, L.~Liu, Y.~Liu, X.~Li, Diffdet4sar: Diffusion-based aircraft target detection network for sar images, IEEE Geoscience and Remote Sensing Letters (2024).

\bibitem{ref131}
K.~Wang, Z.~Pan, Z.~Wen, Svddd: Sar vehicle target detection dataset augmentation based on diffusion model, Remote Sensing 17~(2) (2025) 286.

\bibitem{ref112}
Q.~Li, J.~Li, T.~Li, Y.~Feng, A joint framework for underwater hyperspectral image restoration and target detection with conditional diffusion model, IEEE Journal of Selected Topics in Applied Earth Observations and Remote Sensing (2024).

\bibitem{ref115}
J.~Zhu, S.~Li, Y.~A. Liu, J.~Yuan, P.~Huang, J.~Shan, H.~Ma, Odgen: Domain-specific object detection data generation with diffusion models, Advances in Neural Information Processing Systems 37 (2024) 63599--63633.

\bibitem{ref114}
L.~Wang, J.~Jia, H.~Dai, Orienteddiffdet: Diffusion model for oriented object detection in aerial images, Applied Sciences 14~(5) (2024) 2000.

\bibitem{ref102}
X.~Hu, C.~Xie, Z.~Fan, Q.~Duan, D.~Zhang, L.~Jiang, X.~Wei, D.~Hong, G.~Li, X.~Zeng, et~al., Hyperspectral anomaly detection using deep learning: A review, Remote Sensing 14~(9) (2022) 1973.

\bibitem{ref103}
I.~Racetin, A.~Krtali{\'c}, Systematic review of anomaly detection in hyperspectral remote sensing applications, Applied Sciences 11~(11) (2021) 4878.

\bibitem{ref104}
Q.~Bo, Z.~Xiangtao, Q.~Xueming, L.~Xiaoqiang, Research progress on hyperspectral anomaly detection, National Remote Sensing Bulletin 28~(1) (2024) 42--54.

\bibitem{ref107}
J.~Shi, P.~Zhang, N.~Zhang, H.~Ghazzai, P.~Wonka, Dissolving is amplifying: Towards fine-grained anomaly detection, in: European Conference on Computer Vision, Springer, 2024, pp. 377--394.

\bibitem{ref108}
J.~Li, H.~Cao, J.~Wang, F.~Liu, Q.~Dou, G.~Chen, P.-A. Heng, Fast non-markovian diffusion model for weakly supervised anomaly detection in brain mr images, in: International Conference on Medical Image Computing and Computer-Assisted Intervention, Springer, 2023, pp. 579--589.

\bibitem{ref109}
C.~Wei, H.~Han, Y.~Xia, Z.~Ji, Tdad: Self-supervised industrial anomaly detection with a two-stage diffusion model, Computers in Industry 164 (2025) 104192.

\bibitem{ref110}
T.~Hu, J.~Zhang, R.~Yi, Y.~Du, X.~Chen, L.~Liu, Y.~Wang, C.~Wang, Anomalydiffusion: Few-shot anomaly image generation with diffusion model, in: Proceedings of the AAAI conference on artificial intelligence, Vol.~38, 2024, pp. 8526--8534.

\bibitem{ref105}
J.~Ma, W.~Xie, Y.~Li, L.~Fang, Bsdm: Background suppression diffusion model for hyperspectral anomaly detection, arXiv preprint arXiv:2307.09861 (2023).

\bibitem{ref111}
Y.~Wu, Y.~Meng, L.~Sun, Diffusing background dictionary for hyperspectral anomaly detection, in: Proceedings of the Asian Conference on Computer Vision, 2024, pp. 1046--1064.

\bibitem{ref192}
C.~Huang, L.~Liu, C.~Xiao, Q.~Ling, G.~Li, K.~Li, Endmember-guided spectral--spatial feature diffusion for hyperspectral anomaly detection, IEEE Transactions on Geoscience and Remote Sensing 64 (2025) 1--17.

\bibitem{ref193}
S.~Wu, X.~Zhang, G.~Wang, X.~Han, J.~Zhu, X.~Cheng, L.~Jiao, Pixdiff: Multi-resolution diffusion network with pixelization for hyperspectral anomaly detection, IEEE Transactions on Geoscience and Remote Sensing (2026).

\bibitem{ref117}
J.~Zhu, Y.~Xu, Z.~Wu, Z.~Wei, Hyperspectral image restoration using spatial-spectral diffusion null-space model, in: IGARSS 2024-2024 IEEE International Geoscience and Remote Sensing Symposium, IEEE, 2024, pp. 7235--7239.

\bibitem{ref119}
M.~Li, Y.~Fu, T.~Zhang, J.~Liu, D.~Dou, C.~Yan, Y.~Zhang, Latent diffusion enhanced rectangle transformer for hyperspectral image restoration, IEEE Transactions on Pattern Analysis and Machine Intelligence (2024).

\bibitem{ref165}
K.~Ren, W.~Sun, X.~Meng, G.~Yang, Msfdn: multi-scale spatial-spectral-frequency joint denoising network for hyperspectral images, Geo-spatial Information Science (2025) 1--20.

\bibitem{ref92}
Y.~Miao, L.~Zhang, L.~Zhang, D.~Tao, Dds2m: Self-supervised denoising diffusion spatio-spectral model for hyperspectral image restoration, in: Proceedings of the IEEE/CVF International Conference on Computer Vision, 2023, pp. 12086--12096.

\bibitem{ref88}
K.~Deng, Z.~Jiang, Q.~Qian, Y.~Qiu, Y.~Qian, A noise-model-free hyperspectral image denoising method based on diffusion model, in: IGARSS 2023-2023 IEEE International Geoscience and Remote Sensing Symposium, IEEE, 2023, pp. 7308--7311.

\bibitem{ref89}
J.~He, Y.~Li, Q.~Yuan, et~al., Tdiffde: A truncated diffusion model for remote sensing hyperspectral image denoising, arXiv preprint arXiv:2311.13622 (2023).

\bibitem{ref90}
L.~Pang, X.~Rui, L.~Cui, H.~Wang, D.~Meng, X.~Cao, Hir-diff: Unsupervised hyperspectral image restoration via improved diffusion models, in: Proceedings of the IEEE/CVF Conference on Computer Vision and Pattern Recognition, 2024, pp. 3005--3014.

\bibitem{ref91}
H.~Zeng, J.~Cao, K.~Zhang, Y.~Chen, H.~Luong, W.~Philips, Unmixing diffusion for self-supervised hyperspectral image denoising, in: Proceedings of the IEEE/CVF Conference on Computer Vision and Pattern Recognition, 2024, pp. 27820--27830.

\bibitem{ref93}
K.~Deng, P.~Wang, Y.~Qian, Rgb images enhancing hyperspectral image denoising with diffusion model, in: ICASSP 2024-2024 IEEE International Conference on Acoustics, Speech and Signal Processing (ICASSP), IEEE, 2024, pp. 2960--2964.

\bibitem{ref121}
X.~Li, Y.~Ren, X.~Jin, C.~Lan, X.~Wang, W.~Zeng, X.~Wang, Z.~Chen, Diffusion models for image restoration and enhancement--a comprehensive survey, arXiv preprint arXiv:2308.09388 (2023).

\bibitem{ref154}
J.~Liu, Z.~Wu, L.~Xiao, A spectral diffusion prior for unsupervised hyperspectral image super-resolution, IEEE Transactions on Geoscience and Remote Sensing (2024).

\bibitem{ref155}
Y.~Cheng, Y.~Ma, F.~Fan, J.~Ma, Y.~Yao, X.~Mei, Latent spectral-spatial diffusion model for single hyperspectral super-resolution, Geo-spatial Information Science (2024) 1--16.

\bibitem{ref156}
K.~He, Y.~Cai, S.~Peng, M.~Tan, A diffusion model-assisted multi-scale spectral attention network for hyperspectral image super-resolution, IEEE Journal of Selected Topics in Applied Earth Observations and Remote Sensing (2024).

\bibitem{ref157}
F.~Meng, Y.~Chen, H.~Jing, L.~Zhang, Y.~Yan, Y.~Ren, S.~Wu, T.~Feng, R.~Liu, Z.~Du, A conditional diffusion model with fast sampling strategy for remote sensing image super-resolution, IEEE Transactions on Geoscience and Remote Sensing (2024).

\bibitem{ref158}
Y.~Xiao, Q.~Yuan, K.~Jiang, J.~He, X.~Jin, L.~Zhang, Ediffsr: An efficient diffusion probabilistic model for remote sensing image super-resolution, IEEE Transactions on Geoscience and Remote Sensing 62 (2023) 1--14.

\bibitem{ref159}
S.~Jia, S.~Zhu, Z.~Wang, M.~Xu, W.~Wang, Y.~Guo, Diffused convolutional neural network for hyperspectral image super-resolution, IEEE Transactions on Geoscience and Remote Sensing 61 (2023) 1--15.

\bibitem{ref164}
Y.~Chen, X.~Zhang, Ddsr: Degradation-aware diffusion model for spectral reconstruction from rgb images, Remote Sensing 16~(15) (2024) 2692.

\bibitem{ref116}
C.~Wu, D.~Wang, Y.~Bai, H.~Mao, Y.~Li, Q.~Shen, Hsr-diff: Hyperspectral image super-resolution via conditional diffusion models, in: Proceedings of the IEEE/CVF International Conference on Computer Vision, 2023, pp. 7083--7093.

\bibitem{ref120}
Y.~Si, Z.~Lin, X.~Wang, S.~He, A new hyperspectral reconstruction method with conditional diffusion model for snapshot spectral compressive imaging, IEEE Transactions on Instrumentation and Measurement (2025).

\bibitem{ref187}
A.~Ramirez-Jaime, G.~R. Arce, N.~Porras-Diaz, O.~Ieremeiev, A.~Rubel, V.~Lukin, M.~Kopytek, P.~Lech, J.~Fastowicz, K.~Okarma, Generative diffusion models for compressed sensing of satellite lidar data: Evaluating image quality metrics in forest landscape reconstruction (2025).

\bibitem{ref133}
D.~Hong, Z.~Han, J.~Yao, L.~Gao, B.~Zhang, A.~Plaza, J.~Chanussot, Spectralformer: Rethinking hyperspectral image classification with transformers, IEEE Transactions on Geoscience and Remote Sensing 60 (2021) 1--15.

\bibitem{ref134}
D.~Hong, L.~Gao, J.~Yao, B.~Zhang, A.~Plaza, J.~Chanussot, Graph convolutional networks for hyperspectral image classification, IEEE Transactions on Geoscience and Remote Sensing 59~(7) (2020) 5966--5978.

\bibitem{ref135}
L.~Sun, G.~Zhao, Y.~Zheng, Z.~Wu, Spectral--spatial feature tokenization transformer for hyperspectral image classification, IEEE Transactions on Geoscience and Remote Sensing 60 (2022) 1--14.

\bibitem{ref136}
B.~Liu, A.~Yu, X.~Yu, R.~Wang, K.~Gao, W.~Guo, Deep multiview learning for hyperspectral image classification, IEEE Transactions on Geoscience and Remote Sensing 59~(9) (2020) 7758--7772.

\bibitem{ref137}
D.~Wang, B.~Du, L.~Zhang, Spectral-spatial global graph reasoning for hyperspectral image classification, IEEE Transactions on Neural Networks and Learning Systems (2023).

\bibitem{ref138}
Y.~Dai, Y.~Wu, F.~Zhou, K.~Barnard, Attentional local contrast networks for infrared small target detection, IEEE transactions on geoscience and remote sensing 59~(11) (2021) 9813--9824.

\bibitem{ref139}
Y.~Dai, Y.~Wu, F.~Zhou, K.~Barnard, Asymmetric contextual modulation for infrared small target detection, in: Proceedings of the IEEE/CVF winter conference on applications of computer vision, 2021, pp. 950--959.

\bibitem{ref140}
X.~Wu, D.~Hong, J.~Chanussot, Uiu-net: U-net in u-net for infrared small object detection, IEEE Transactions on Image Processing 32 (2022) 364--376.

\bibitem{ref141}
B.~Li, C.~Xiao, L.~Wang, Y.~Wang, Z.~Lin, M.~Li, W.~An, Y.~Guo, Dense nested attention network for infrared small target detection, IEEE Transactions on Image Processing 32 (2022) 1745--1758.

\bibitem{ref142}
P.~Xiang, S.~Ali, S.~K. Jung, H.~Zhou, Hyperspectral anomaly detection with guided autoencoder, IEEE Transactions on Geoscience and Remote Sensing 60 (2022) 1--18.

\bibitem{ref143}
G.~Fan, Y.~Ma, X.~Mei, F.~Fan, J.~Huang, J.~Ma, Hyperspectral anomaly detection with robust graph autoencoders, IEEE Transactions on Geoscience and Remote Sensing 60 (2021) 1--14.

\bibitem{ref144}
S.~Wang, X.~Wang, L.~Zhang, Y.~Zhong, Auto-ad: Autonomous hyperspectral anomaly detection network based on fully convolutional autoencoder, IEEE Transactions on Geoscience and Remote Sensing 60 (2021) 1--14.

\bibitem{ref145}
X.~Cheng, Y.~Huo, S.~Lin, Y.~Dong, S.~Zhao, M.~Zhang, H.~Wang, Deep feature aggregation network for hyperspectral anomaly detection, IEEE Transactions on Instrumentation and Measurement (2024).

\bibitem{ref146}
D.~Wang, L.~Zhuang, L.~Gao, X.~Sun, X.~Zhao, A.~Plaza, Sliding dual-window-inspired reconstruction network for hyperspectral anomaly detection, IEEE Transactions on Geoscience and Remote Sensing 62 (2024) 1--15.

\bibitem{ref147}
S.-S. Young, C.-H. Lin, Z.-C. Leng, Unsupervised abundance matrix reconstruction transformer-guided fractional attention mechanism for hyperspectral anomaly detection, IEEE Transactions on Neural Networks and Learning Systems (2024).

\bibitem{ref160}
S.~Lei, Z.~Shi, Hybrid-scale self-similarity exploitation for remote sensing image super-resolution, IEEE Transactions on Geoscience and Remote Sensing 60 (2021) 1--10.

\bibitem{ref161}
S.~Lei, Z.~Shi, W.~Mo, Transformer-based multistage enhancement for remote sensing image super-resolution, IEEE Transactions on Geoscience and Remote Sensing 60 (2021) 1--11.

\bibitem{ref162}
M.~Guo, Z.~Zhang, H.~Liu, Y.~Huang, Ndsrgan: A novel dense generative adversarial network for real aerial imagery super-resolution reconstruction, Remote Sensing 14~(7) (2022) 1574.

\bibitem{ref163}
X.~Chen, X.~Wang, J.~Zhou, Y.~Qiao, C.~Dong, Activating more pixels in image super-resolution transformer, in: Proceedings of the IEEE/CVF conference on computer vision and pattern recognition, 2023, pp. 22367--22377.

\bibitem{ref174}
H.~Zeng, X.~Xie, J.~Ning, Hyperspectral image denoising via global spatial-spectral total variation regularized nonconvex local low-rank tensor approximation, Signal processing 178 (2021) 107805.

\bibitem{ref175}
H.~Zeng, S.~Huang, Y.~Chen, H.~Luong, W.~Philips, All of low-rank and sparse: A recast total variation approach to hyperspectral denoising, IEEE Journal of Selected Topics in Applied Earth Observations and Remote Sensing 16 (2023) 7357--7373.

\bibitem{ref176}
Y.~Chen, T.-Z. Huang, W.~He, X.-L. Zhao, H.~Zhang, J.~Zeng, Hyperspectral image denoising using factor group sparsity-regularized nonconvex low-rank approximation, IEEE Transactions on Geoscience and Remote Sensing 60 (2021) 1--16.

\bibitem{ref177}
O.~Sidorov, J.~Yngve~Hardeberg, Deep hyperspectral prior: Single-image denoising, inpainting, super-resolution, in: Proceedings of the IEEE/CVF International Conference on Computer Vision Workshops, 2019, pp. 0--0.

\bibitem{ref178}
H.~Zeng, X.~Xie, W.~Kong, S.~Cui, J.~Ning, Hyperspectral image denoising via combined non-local self-similarity and local low-rank regularization, IEEE Access 8 (2020) 50190--50208.

\bibitem{ref173}
B.~Kawar, M.~Elad, S.~Ermon, J.~Song, Denoising diffusion restoration models, Advances in Neural Information Processing Systems 35 (2022) 23593--23606.

\bibitem{ref132}
M.~A. Chandra, S.~Bedi, Survey on svm and their application in image classification, International Journal of Information Technology 13~(5) (2021) 1--11.

\end{thebibliography}

\end{document}